# Benefits, Challenges, and Research Topics: A Multi-vocal Literature Review of Kubernetes

**Shazibul Islam Shamim · Jonathan Alexander Gibson · Patrick Morrison · Akond Rahman**

**Abstract** *Context*: Kubernetes is an open source software that helps in automated deployment of software and orchestration of containers. With Kubernetes, IT organizations, such as IBM, Pinterest, and Spotify have experienced an increase in release frequency. As the adoption of Kubernetes increases amongst IT organizations, understanding the challenges of using Kubernetes becomes a necessity so that the software engineering community is well-equipped to support practitioners who use Kubernetes. *Objective*: *The goal of this paper is to inform practitioners and researchers on benefits and challenges of Kubernetes usage by conducting a multi-vocal literature review of Kubernetes. Methodology*: We conduct a multi-vocal literature review (MLR) where we use 321 Kubernetes-related Internet artifacts to identify benefits and challenges perceived by practitioners. In our MLR, we also analyze 105 peer-reviewed publications to identify the research topics addressed by the research community. *Findings*: We find 8 benefits that include service level objective (SLO)-based scalability and self-healing containers. Our identified 15 challenges related to Kubernetes include unavailability of diagnostics and security tools and attack surface reduction. We observe researchers to address 14 research topics related to Kubernetes, which includes efficient resource uti-

Shazibul Islam Shamim
Department of Computer Science, Tennessee Tech University, Cookeville, TN, USA E-mail: mshamim42@tntech.edu

Jonathan Alexander Gibson

Department of Computer Science, Tennessee Tech University, Cookeville, TN, USA E-mail: jagibson44@tntech.edu

Patrick Morrison
IBM, Research Triangle Park, NC, USA
E-mail: pjmorris@us.ibm.com

Akond Rahman
Department of Computer Science, Tennessee Tech University, Cookeville, TN, USA
E-mail: arahman@tntech.edu

lization. We also identify 9 challenges that are under-explored in research publications, which include cultural change, hardware compatibility, learning curve, maintenance, and testing. *Conclusion*: Our MLR shows that while certain topics, such as efficient resource utilization, are well-investigated by the research community, other topics for which practitioners face challenges, such as learning curve, maintenance, and testing remain under-explored. Furthermore, we observe a nuanced perspective of practitioners for Kubernetes, which has implications for practitioners as well as for researchers.

## 1 Introduction

Kubernetes is an open source software system for automatically deploying, scaling, and managing containerized applications (Kubernetes, 2021). Kubernetes has removed repetitive manual processes in container deployment, and is perceived as one of the most popular container management tools (G2, 2021). Information technology (IT) organizations, such as IBM, Pinterest, and the U.S. Department of Defense (DoD) use Kubernetes to manage their containers and automate the software deployment process (Kubernetes, 2020). Practitioners have reported benefits of Kubernetes, e.g., using Kubernetes the U.S. DoD decreased their release time from 3∼8 months to 1 week (CNCF, 2020).

While other software systems, such as Apache Mesos (Kakadia, 2015) and Openshift (Pousty and Miller, 2014), exist for automated management of container-based applications, certain characteristics of Kubernetes necessitates research focus. These characteristics are:

– *Rapid adoption*: Evidence from multiple practitioner surveys attests to the rapid adoption of Kubernetes for automated container management. Examples include: (i) according to the Stackrox survey, 91% of the surveyed practitioners use Kubernetes for container management (RedHat, 2021); (ii) according to another survey from Portworx (2021), 89% of the survey respondents anticipate Kubernetes to play a big role in their organizations for managing computing infrastructures; (iii) in the State of Enterprise Open Source survey, 78% respondents identified Kubernetes to "*clearly become the* *go-to choice*" for automated management of container-based applications; (iv) recently, Twitter moved from Apache Mesos to Kubernetes for automated management of container-based application (Bhartiya, 2019). According to a technical expert from Alibaba Cloud, "*Twitter's switch from its initial adoption of Mesos to the use of Kubernetes today proves again the assertion that Kubernetes has become an industry standard for container orchestration*" (Bhartiya, 2019); and (v) Bayern (2020) reports Kubernetesrelated job searches to increase by 2,125% in four years, which indicates more and more organizations are adopting or planning to adopt Kubernetes, for which they need professionals with Kubernetes expertise.

The above-mentioned evidence shows how Kubernetes is being rapidly adopted in industry. Similar views were echoed by the co-founder of SaltStack (Bhartiya, 2019), an organization that focuses on automated configuration management, who said "*Kubernetes has become the standard way to run and test applications, almost overnight. The proliferation of the technology is so universal that it has become standard for DevOps team*". According to Mondal et al. (2021), "*Kubernetes is the leader of the container orchestration tool because it makes a series of standards and realizes many fundamental and useful features*".

- *Economics*: Kubernetes has economics-related characteristics as well. According to T4 (2020), Kubernetes has a 77% market share when it comes to container orchestrations, which is higher compared to that of Apache Mesos, Docker Swarm, and Openshift that respectively has a market share of 4.0%, 5.0%, and 9.0%.
- *Diverse use cases*: Kubernetes supports a diverse set of applications. According to Mondal et al. (2021) compared to Kubernetes, other tools, such as Apache Mesos has a narrower audience for high performance computing applications. According to the 2021 Kubernetes adoption survey (2021), 89% of 500 respondents mentioned AI-based applications to be deployed using Kubernetes. Researchers have already used Kubernetes to conduct cutting edge research in the domains of smart cities (Muralidharan et al, 2019), autonomous vehicles (Khakimov et al, 2020), and agriculture (Kim et al, 2021). All of these examples show Kubernetes's relevance to the practitioner and researcher community. All of these examples show Kubernetes's relevance to both, the practitioner and researcher community.
- *Vibrant community*: Kubernetes serves practitioners who come from diverse professional backgrounds. According to another survey (Enlyft, 2021) with 24,441 companies, 37% are small (< 50 employees), 43% are mediumsized and 20% are large (> 1000 employees). This evidence shows that not all users of Kubernetes stem from large-sized companies. Small-sized organizations lack in knowledge resources (Brodman and Johnson, 1994; Majchrowski et al, 2016; Tuape and Ayalew, 2019). Our MLR can benefit these users, as our MLR synthesizes the practitioners' experiences on benefits and challenges for Kubernetes.

Along with a large user base as discussed above, Kubernetes also benefits from having a vibrant community. According to Novoseltseva (2021a), Kubernetes has an active community of more than 2,000 contributors from Fortune 500 companies. Furthermore, the community organizes a series of events that have attracted 21,000 participants (CNCF, 2019). All of the above-mentioned statistics show that Kubernetes has a wide range of users and contributors who come from diverse professional backgrounds. Our MLR could be of interest to both. Kubernetes users can learn from the synthesized benefits and challenges, as well as from the peer-reviewed publications on how to reap benefits from Kubernetes. Contributors can learn from the identified challenges, and use these findings to develop features to improve Kubernetes.

In short, Kubernetes has established itself as the de-facto container management tool, and is perceived to remain dominant in the next few years. Thus, findings generated from our MLR will remain relevant for practitioners and researchers alike. Our MLR also lays the groundwork for studying the Kubernetes community. Community-focused research, where researchers focus on a certain software project or a tool, is commonplace in software engineering. Mockus et al. (2002) studied how developers contribute to Apache projects. Tan et al. (2020) conducted an empirical study on how multiplecommitter model is adopted amongst the Linux contributor community. Hirao et al. (2020) studied divergent code review characteristics for the OpenStack community. With the help of a MLR identified challenges expressed by practitioners can also be triangulated with existing research. Furthermore, as practitioners prefer to learn from other practitioners on new technologies (Moore and McKenna, 1999; Rahman et al, 2015), identification of benefits can help practitioners understand how Kubernetes can satisfy their needs.

*The goal of this paper is to inform practitioners and researchers on benefits and challenges of Kubernetes usage by conducting a multi-vocal literature review of Kubernetes.*

We answer the following research questions:

- **RQ1**: What are the perceived benefits of using Kubernetes as reported by practitioners?
- **RQ2**: What are the perceived challenges of using Kubernetes as reported by practitioners?
- **RQ3**: What research topics have been investigated in Kubernetes-related publications that are peer-reviewed?

We take motivation from prior MLR studies in software engineering related to DevOps (Myrbakken and Colomo-Palacios, 2017), software testing (Garousi and M¨antyl¨a, 2016), and software education (Calder´on et al, 2018). We conduct an MLR following the guidelines of Garousi et al. (2019) to investigate benefits, challenges, and research topics related to Kubernetes. We conduct our MLR by systematically applying inclusion and exclusion criteria to collect 105 peer-reviewed publications and 321 Internet artifacts. We apply a qualitative analysis technique called open coding (Saldana, 2015) to identify benefits and challenges of Kubernetes usage from 321 Internet artifacts, and research topics from the collected 105 peer-reviewed publications. **Contributions**: We list our contributions as follows:

- A list of perceived benefits of Kubernetes usage reported by practitioners;

- A list of perceived challenges of Kubernetes usage reported by practitioners;
- A list of research topics investigated in Kubernetes-related publications that are peer-reviewed;
- A mapping of identified challenges to that with identified research topics; and
- A mapping of identified benefits to that with studied research publications.

We organize the rest of the paper as follows: we provide necessary background and related work in Section 2. We provide our research methodology in Section 3. We provide answers to RQ1, RQ2, and RQ3 in Section 4. In Section 5, we discuss potential research avenues based on our findings. We describe the limitation of our MLR in Section 6. We conclude our paper in Section 7.

## 2 Background and Related Work

In this section, we *first*, provide background on Kubernetes and MLRs. *Next*, we discuss peer-reviewed publications on MLR.

### 2.1 Background

Kubernetes is an open source software system for automating software deployment and management of containerized applications (Kubernetes, 2021). Kubernetes originated off the Borg project, which was developed and managed by Google (Burns et al, 2016). A Kubernetes installation is also referred as a Kubernetes cluster (Miles, 2020). Each Kubernetes cluster contains a set of worker machines defined as nodes. As shown in Figure 1, two types of nodes exist for Kubernetes: Control plane nodes and worker nodes.

Each control plane has the following components: 'kube-apiserver', 'kubescheduler', 'kube-controller-manager', and 'etcd' (Miles, 2020). Kubernetes serves its functionality through an application program interface from the 'kube-api-server'. The 'kube-api-server' orchestrates all the operations within the Kubernetes cluster. The 'kube-controller-manager' is a component on the control plane that watches the state of the cluster through the 'kube-api-server' and changes the current state towards the desired state. The 'kube-scheduler' is the component in the control plane responsible for scheduling pods across multiple nodes. A key-value based database, 'etcd', stores all configuration information for the Kubernetes cluster. A command-line tool called 'Kubectl' is used to communicate with the 'kube-apiserver' in the control plane.

The worker nodes host the applications that run on Kubernetes (Miles, 2020). The following components are included in the worker node: 'kubeproxy', 'kubelet' and 'pod'. 'kube-proxy' maintains the network rules on nodes. 'kubelet' is an agent that ensures containers are running inside a pod. The pod is the smallest Kubernetes entity, which includes at least one active container. A container is a standard software unit that packages the code and associated dependencies to run in any computing environment (Miles, 2020).

Practitioners can provide configuration-related information in forms of configurations files, such as YAML files and JSON files to use Kubernetes. These files hold information on what kind of CPU and memory settings will be used by the Kubernetes installation. For example, using the cpu:2 tag in a YAML file, a practitioner can specify that the Kubernetes installation will use 2 CPUs. Listing 1 shows an example Kubernetes manifest. To specify configurations, Kubernetes uses declarative programming, i.e., the paradigm of structuring computer programs without describing the control flow (Lloyd, 1994).

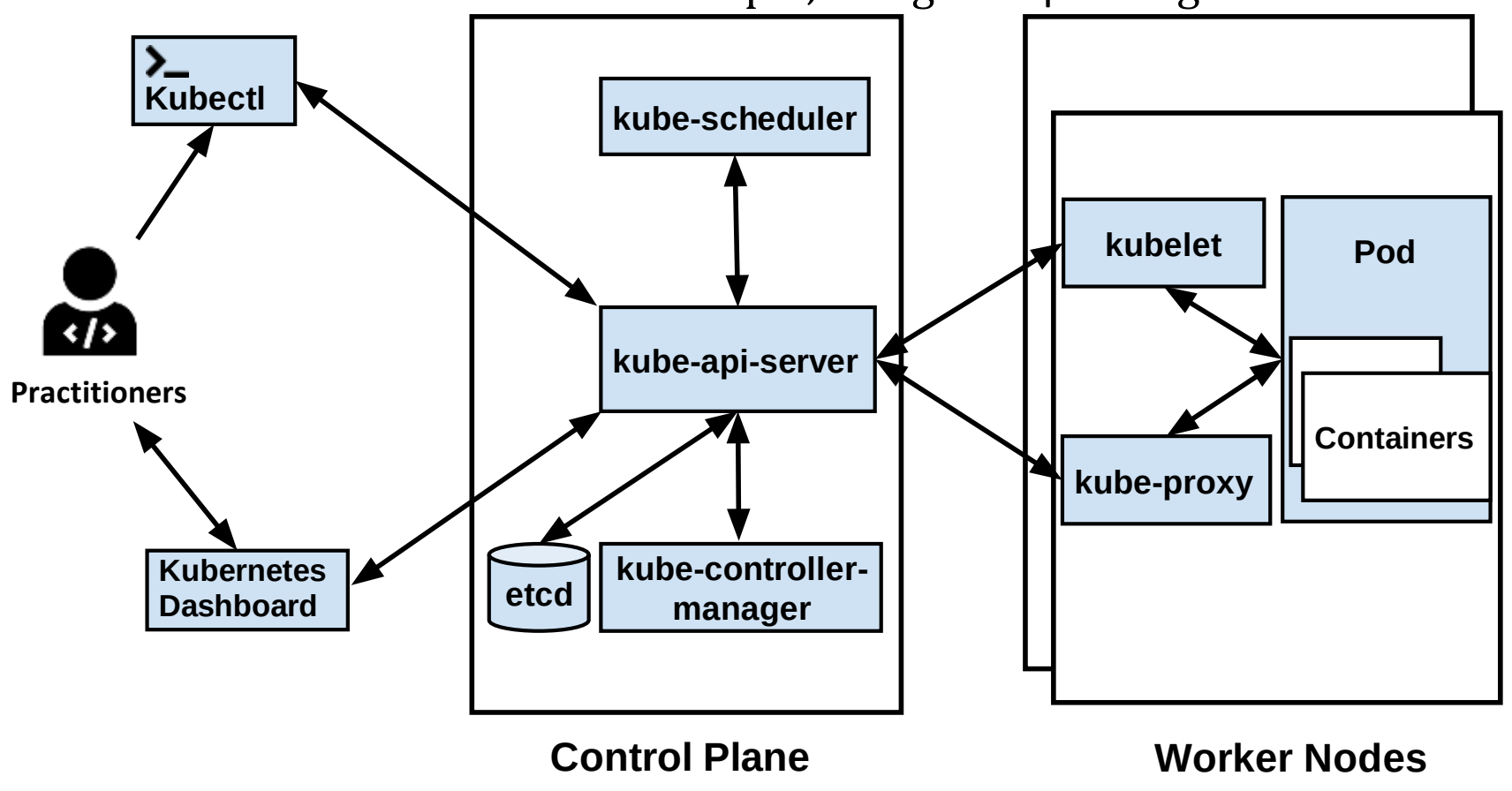

Fig. 1: A brief overview of Kubernetes. Users interact with the control plane using the Kubernetes dashboard and 'kubectl'. The purpose of control plane node is to maintain the desired cluster state and manage worker nodes. Worker nodes are used to run containerized applications inside the pod.

```
kind: Pod metadata:
    name: cpu-demo namespace: cpu-
    example
spec:
    containers:
    - name: cpu-demo-ctr image:
        vish/stress resources:
            limits: cpu:2
```

Listing 1: An example Kubernetes manifest

## 2.2 Related Work

**Brief Overview of Multi-vocal Literature Reviews (MLRs)**: According to Ogawa and Malen (1991), the term 'Multivocal literature' comprises of all accessible writings on a common, contemporary topic from a diverse set of authors, such as academics and industry practitioners. The authors (Ogawa and Malen, 1991) stated that the writings may appear in a variety of forms, and present a diverse set of perspectives by incorporating peer-reviewed and non peer-reviewed insights.

Unlike a systematic literature review, an MLR includes grey literature, i.e., Internet artifacts used by practitioners to share their perceptions and experiences about a certain topic. In software engineering, practitioners tend to express their perceptions and experiences in Internet artifacts, such as blog posts, industrial conference presentations, video demonstrations, and online forums, which are not peer-reviewed. According to Garousi et al. (2016), inclusion of Internet artifacts with peer-reviewed publications can adequately highlight the 'state of practice' in software engineering.

**MLRs in Software Engineering** Our paper is related with reviews that have been conducted in the domains of software engineering. MLRs are commonplace in DevOps-related research, for example, Myrbakken and ColomoPalacios (2017) conducted an MLR study on DevSecOps, and discussed the benefits and challenges of adopting security in DevOps with 2 peer-reviewed publications and 50 Internet artifacts. Sanchez-Gordon et al. (2018) performed an MLR on the evolution and use of DevOps for e-learning systems, and reported growing interest in DevOps adoption for developing e-learning systems. Garousi and Mantyla (2016) performed an MLR study on software test automation decisions, using 78 publications including both peer-reviewed publications and Internet artifacts reported the factors that affect software test automation and summarized a checklist of practical advice for practitioners. Garousi et al. (2017) performed an MLR with 130 peer-reviewed publications and 51 Internet artifacts, and reported 58 software test maturity models, 5 driving factors, 3 benefits and 8 challenges for conducting successful test maturity assessment and test process improvement. Fogarty et al. (2020) performed an MLR and reported challenges in Agile software development to include organizational change and job satisfaction.

**Literature Reviews for Container-related Topics**: Our paper is also related with prior publications that have conducted literature reviews related to containerization. Sadaqat et al. (2018) conducted an MLR study on serverless computing and reported developers perspective, benefits, challenges of adopting serverless computing. Scheuner and Leitner (2020) performed an MLR with 51 peer-reviewed publications and 61 Internet artifacts on performance measurements of function-as-a-service in serverless computing, and reported publication trends and benchmark platforms. Pereira-Vale et al. (2021) performed an MLR with 36 peer-reviewed and 34 Internet artifacts, and observed security mechanisms are inadequate to detect security attacks for microservices. Kumara et al. (2021) conducted a grey literature review for infrastructure as code (IaC), and identified practices that should be followed while developing IaC scripts. Rahman et al. (2018) conducted a systematic mapping study to identify existing gaps in IaC research. In another work, Rahman et al. (2021) conducted a grey literature review to identify secret management practices for IaC. Shamim et al. (2020) conducted a grey literature review to identify security best practices for Kubernetes. Koskinen et al. (2019)

advocated for deeply focused research for containers by reviewing a collection of 56 publications that use containerization for software engineering. Casalicchio and Iannucci (2020) surveyed 97 publications related to containers, and reported performance prediction, multi-layer monitoring, isolation as unresolved challenges for containers. Tyresson (2020) conducted a SLR of 13 publications related to Docker container security, and advocated for application of static and dynamic analysis for Docker containers. Soldani et al. (2018) conducted a grey literature review to identify the technical pain points of developing microservice applications. Li et al. (2020) conducted a SLR with 72 publications to identify the quality attributes of the microservice architecture. Pahl and Jamshidi (2016) conducted a systematic mapping study with 21 publications related to microservices, and reported maturity of microservice-related publication to be low.

The above-mentioned discussion demonstrates the use of MLR in diverse domains, such as software engineering and cloud computing. We take motivation from these studies and conduct an MLR on Kubernetes, an emerging yet under-explored research topic in software engineering.

## 3 Methodology

As shown in Figure 2, we conduct our MLR in five phases that we describe in the following subsections:

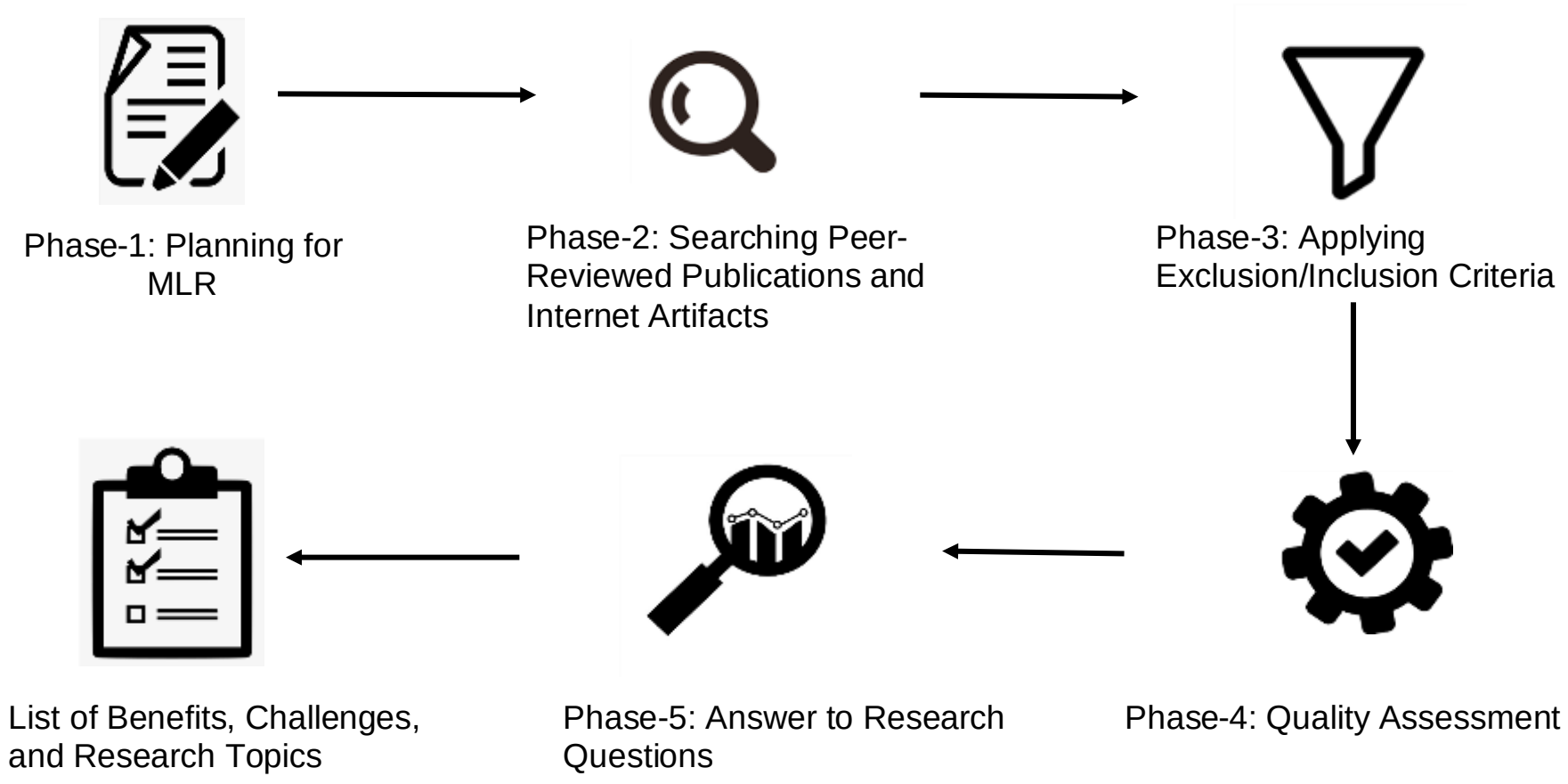

Fig. 2: Our methodology for an MLR related to Kubernetes.

Table 1: Criteria to Plan Our MLR for Kubernetes

| Criteria | First Author | Second Author |
|---|---|---|
| 1. Is the subject complex and not solvable by considering only the formal literature? | **Yes**. Currently, available peer-reviewed publications is not adequate to address the challenges of using, adopting Kubernetes into the production system. | **Yes**. The subject of Kubernetes is complex considering the non-exhaustive research details addressed in the formal literature. |
| 2. Is there a lack of volume or quality of evidence or a lack of consensus of outcome measurement in the formal literature? | **Yes**. Internet artifacts, such as blog posts, tutorials, videos, and white papers are prevalent compared to peer-reviewed publications in Kubernetes. | **Yes**. Peer-reviewed research lacks quality evidence and consensus in Kubernetes. |
| 3. Is the contextual information important to the subject under study? | **Yes**. The practices practitioners adopt for continuous delivery, continuous deployment, configuration management, scaling, load balancing for their production system is crucial for conducting this study. | **Yes**. Kubernetes is a highly complex piece of software that needs investigation from every angle to ensure maximum security and to understand its benefits and flaws. |
| 4. Is it the goal to validate or corroborate scientific outcomes with practical experiences? | **No**. The goal is to identify the benefits, challenges of industry practitioners and research opportunities in Kubernetes research. | **No**.The goal of this research is to recognize the contributions and issues put forth by both fields. |
| 5. Is it the goal to challenge assumptions or falsify results from practice using peerreviewed research or vice versa? | **No**. The goal is not to challenge current assumptions rather identifying the challenges related to Kubernetes adoption in the industry. | **No**. The goal of this research is not to challenge the industry practices or peerreviewed research of Kubernetes. |
| 6. Would a synthesis of insights and evidence from the industrial and academic community be useful to one or even both communities? | **Yes**. A synthesis of insights and evidence from the industrial and academic community for Kubernetes will help both the communities. | **Yes**. Considering the open source nature of Kubernetes, industry and academia would benefit from the combination of industry knowledge and academic knowledge. |
| 7. Is there a large volume of practitioner sources indicating high practitioner interest in a topic? | **Yes**. The prevalence of Internet artifacts and the Kubernetes Github repository's current state shows active contributions from the developers. | **Yes**. The Kubernetes project on GitHub has around 94,000 commits, 25,000 forks, and 2,800 contributors that suggests practitioner interest in Kubernetes. |

### 3.1 **Phase-1: Planning for MLR**

We follow Garousi et al. (2019)'s recommendations to conduct our MLR where the authors provide a list of seven criterion to plan an MLR. According to Garousi et al. (2019) before conducting an MLR the researchers need to evaluate if the seven criterion are satisfied. If all researchers involved in MLR agree for majority of the criteria, then researchers can move forward with the MLR.

From Table 1 we observe the two researchers, i.e., the first author and second author of the paper responded with 'Yes' for five of the seven criteria, and responded with 'No' for criterion#4 and #5.

### 3.2 **Phase-2: Searching Peer-Reviewed Publications and Internet Artifacts**

For our MLR, we use both peer-reviewed publications, and Internet artifacts, such as, blog posts and industrial white papers. To identify necessary peerreviewed publications, we use five scholar databases, namely, (i) ACM Digital Library, (ii) IEEE Xplore, (iii) Springer Link, (iv) ScienceDirect, and (v) Wiley Online Library. We use these five scholar databases for our MLR study because Kuhrmann et al. (2017) recommend these databases to use in systematic mapping studies and systematic literature reviews. We use the Google search engine using our browser in incognito mode to identify Internet artifacts. We collect all Internet artifacts and peer-reviewed publications in September 2020. We describe our search process as follows:

We derive search strings with a technique called snowballing (Wohlin, 2014) following the guidelines of Garousi et al. (2019). To derive initial search strings, we first start with the search string 'Kubernetes usage' and 'Kubernetes challenges'. We do not start with 'Kubernetes benefit' as our assumption is that with the search string 'Kubernetes usage', we will not only get use cases of Kubernetes, but also Kubernetes the contexts of using Kubernetes. The two search strings are used to collect the most relevant 200 Internet artifacts (100 search results for each search string) where relevance is determine by the Google search engine. Our assumption is that by using a set of 200 Internet artifacts we will get the set of search keywords necessary to conduct our MLR. By reading each of these 200 Internet artifacts the first author observes that while describing Kubernetes-related perceptions and experiences practitioners also use other terms, namely, security challenges, deployment challenges, use cases, lesson learned, benefits, in cloud, production, flaws, and trade-offs. Using these terms we derive additional search strings. For validation, the last author inspects each of the search strings, and identifies each of the following search strings to be included to identify Internet artifacts and peer-reviewed publications:

1. 'kubernetes usage'
2. 'kubernetes challenges'
3. 'kubernetes flaws'
4. 'kubernetes security'
5. 'kubernetes benefits'
6. 'kubernetes production'
7. 'kubernetes use cases'
8. 'kubernetes deployment challenges'
9. 'kubernetes security challenges'
10. 'kubernetes adoption challenges'
11. 'kubernetes lesson learned'

12. ‘kubernetes trade-off’
13. ‘kubernetes in cloud’

For each search string we collect first 100 Internet artifacts provided by the Google search engine. From the five scholar databases we obtain 27,398 search results for the 13 search strings.

The high amount of search results necessitates an explanation. In our methodology, we follow a snowball-based approach where we start with two search strings and in the end, use 13 search strings. Each of these search strings included words, such as ‘flaws’, ‘security’, and ‘in the cloud’, which resulted in additional search results. Use of only ‘Kubernetes’ would have reduced the search process, but in that manner, we would have skipped the snowballing process, which is strongly recommended by Garousi et al. (2019)’s guidelines to conduct multi-vocal literature reviews in software engineering.

### 3.3 Phase-3: Applying Exclusion & Inclusion Criteria

The Google search engine and the five scholar databases are susceptible to generate search results that are not relevant to the topic of interest. Following Garousi et al. (2019)’s guidelines, we apply an inclusion and exclusion criteria to filter irrelevant search results that we describe as follows:

**Exclusion Criteria:** We exclude peer-reviewed publications and Internet artifacts that satisfy the following criteria:

- The artifact/publication is not written in English.
- The artifact/publication is published before 2014. We use 2014, as in 2014 the first version of Kubernetes was released. (Miles, 2020).

For publication names returned by scholar databases we apply an additional exclusion criterion: we exclude publications that are indexed in scholar databases but not peer-reviewed, such as keynote abstracts, call for papers, and presentations.

**Inclusion Criteria:** We set the inclusion criteria for peer-reviewed publications and Internet artifacts as follows:

- The artifact/publication is available for reading.
- The artifact/publication is not a duplicate. We determine an Internet artifact to be a duplicate of another if the title, content, and author is exactly the same as another Internet artifact. For duplicates, we randomly pick one of the Internet artifacts, and include it in our set.
- The content of the artifact/publications discusses a topic related with Kubernetes. In the case of Internet artifacts the first and second authors individually read the title and the content to determine this criterion. In the case of peer-reviewed publications, the first and second author individually read the title, abstract, and introduction to determine this criterion.

In Figure 3, we summarize the process of collecting peer-reviewed publications from the five scholar databases. The first author filters search results, identifies publications that are peer-reviewed, written in English, and available for reading. The first author retrieves 27,398 peer-reviewed publications from five scholar databases. The publications were collected on September 15, 2020. Using the exclusion and inclusion criteria, the first author identifies 19,692 peer-reviewed publications. For title and content-based filtering we use two authors: the first and second author. The first and second author reads the titles of identified publications and respectively identifies 457 and 566 publications. After reading the abstract and introduction of each paper, the first author and second author respectively, identifies 181 and 239 publications. At this stage both author inspect each other's set of publications, remove duplicates, and respectively, identify 61 and 112 publications for which they agree and disagree upon. The disagreements are resolved by the last author, and the last author's decision on the disagreed upon publications is final. Upon resolving all disagreements, we obtain a set of 105 peer-reviewed publications that we use in our MLR. Table A1 in the Appendix lists the 105 publication titles.

For Internet artifacts, both the first author and second author of the paper read each of the first 100 results from the Google Search for 13 Search Strings. The search results are collected on September 25, 2020. The first and second author individually reads each of the 1,400 artifacts and respectively, identifies a set of 192 Internet artifacts and 320 Internet artifacts. The first and second author respectively, agreed and disagreed on 85 and 276 Internet artifacts on their relationship with Kubernetes. The disagreements between the authors are resolved by the last author, whose decision is considered final. The last author is given a list of Internet artifacts for which the first and second authors disagreed. By reading the title and the content for each of the 276 Internet artifacts, the last author determines a set of 321 Internet artifacts that we use in our MLR. Table A3 in the Appendix lists the 321 Internet artifact URLs.

### 3.4 **Phase-4: Quality Assessment**

Following guidelines from prior work (Kitchenham et al, 2012; Garousi et al, 2019) we conduct a quality assessment of the collected peer-reviewed publications and Internet artifacts respectively, in Sections 3.4.1 and 3.4.2.

#### *3.4.1 Quality Assessment of Research Articles*

We follow the criteria provided by Kitchenham et al. (2012) to assess the quality of a peer-reviewed publication. A higher-quality score indicates that the publication clearly describes the goal, contains actionable results, clearly discusses the limitations, and contains a clear presentation structure. The criteria set that we use for our set of peer-reviewed publications is listed in Table 2:

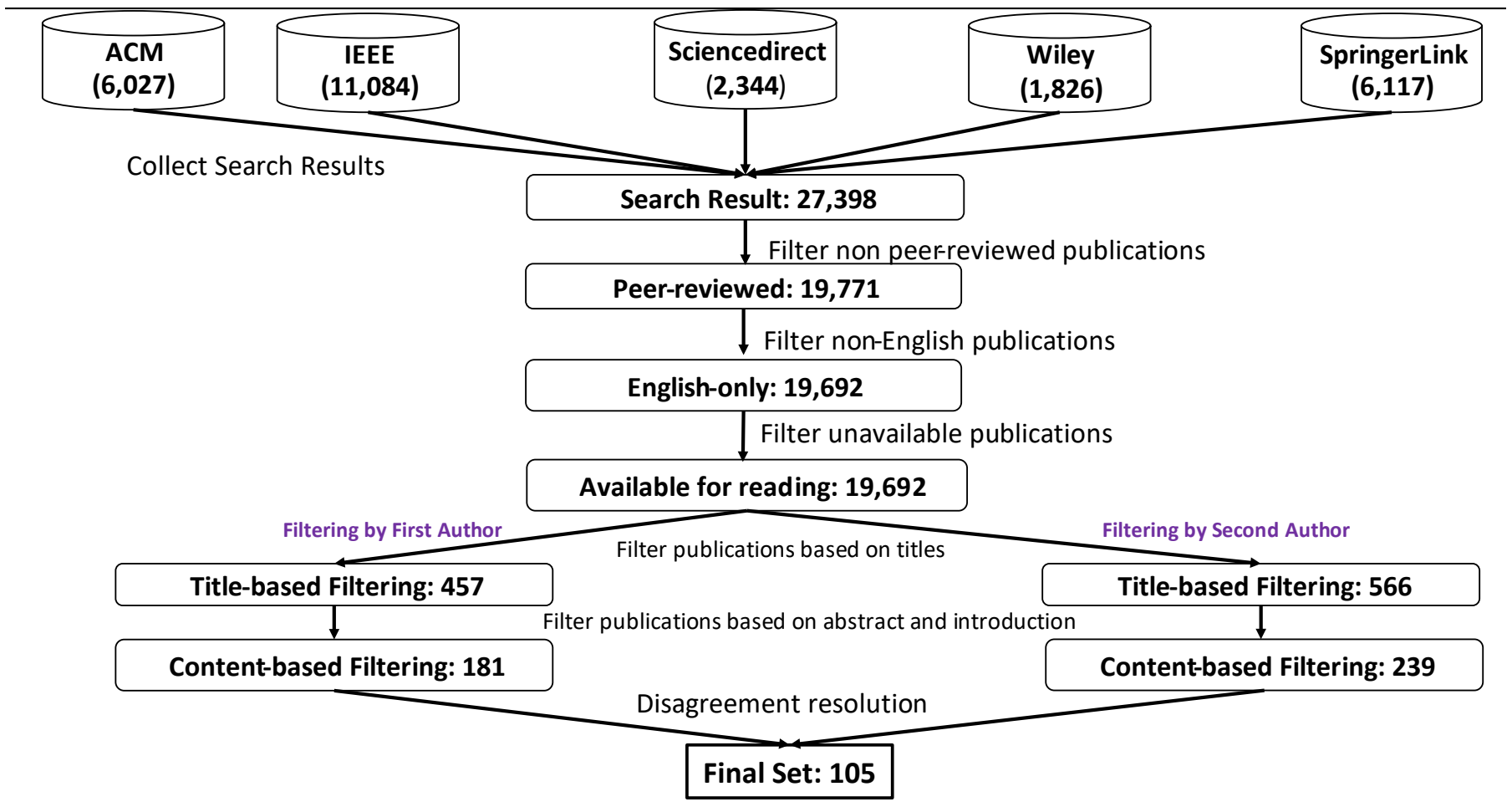

Fig. 3: Summary of the filtering process to obtain the set of 105 peer-reviewed publications in our MLR.

Table 2: Quality Assessment Criteria for Peer-reviewed Publications

| **Criterion** | **Description** |
|---|---|
| **Q1 (Aim)** | Do the authors clearly state the aim of the research? |
| **Q2 (Units)** | Do the authors describe the sample and experimental units? |
| **Q3 (Design)** | Do the authors describe the design of the experiment? |
| **Q4 (Data Collection)** | Do the authors describe the data collection procedures and define the measures? |
| **Q5 (Data Analysis)** | Do the authors define the data analysis procedures? |
| **Q6 (Bias)** | Do the authors discuss potential experimenter bias? |
| **Q7 (Limitations)** | Do the authors discuss the limitations of their study? |
| **Q8 (Clarity)** | Do the authors state the findings clearly? |
| **Q9 (Usefulness)** | Is there evidence that the Experiment/QuasiExperiment can be used by other researchers/practitioners? |

After we answer each of the above nine questions, we provide a rating score associated with each of the answers between 1 to 4. The rating 1 implies 'not at all', 2 implies 'somewhat', 3 implies 'mostly,' and 4 implies 'fully'. As the rating process of the research articles is subjective, we assign two raters, i.e., the first and the second author, who independently provide a rating to each publication. We report the average rating score of both raters for each publication. We summarize the average rating of the quality assessments for Table 3: Quality Assessment for 105 Publications

| Criterion | Average Rating | Min, Max |
|---|---|---|
| **Q1 (Aim)** | 2.95 | 1.0, 4.0 |
| **Q2 (Units)** | 2.42 | 1.0, 4.0 |
| **Q3 (Design)** | 2.92 | 1.0, 4.0 |
| **Q4 (Data Collection)** | 2.56 | 1.0, 4.0 |
| **Q5 (Data Analysis)** | 2.36 | 1.0, 4.0 |
| **Q6 (Bias)** | 1.35 | 1.0, 4.0 |
| **Q7 (Limitations)** | 1.86 | 1.0, 4.0 |
| **Q8 (Clarity)** | 2.60 | 1.0, 4.0 |
| **Q9 (Usefulness)** | 2.83 | 1.5, 4.0 |

the 105 publications in Table 3 and the quality assessment rating for each of the peer-reviewed publications is described in Table A2 of the Appendix. From Table 3, we observe Kubernetes-related publications to score low (< 3.0) on average for all 9 criteria. Furthermore, we observe Kubernetes-related publications to score < 2.0 for with respect to discussing limitations. The range of scores for each criterion is presented as minimum and maximum in the 'Min, Max' column.

### *3.4.2 Quality Assessment of Internet Artifacts*

For quality assessment of identified 321 Internet artifacts, we use the assessment criteria provided by Garousi et al. (2019). Each of the assessment criterion is listed in Table 4:

We use a 3-point scale where '1.0' refers to 'yes', 0.5 refers to 'partially', 0.0 refers to 'no' for Q1-Q11. For Q12 we also use a 3-point scale of 1.0, 0.5, and 0.0, which respectively refers to high, moderate, and low credibility. The first and second author individually read each of the 321 Internet artifacts to determine a value for Q1-Q12. We report the average of the scores reported by the first and second author.

A summary of the average rating for each of the questions for the 321 Internet artifacts given in Table 5. The detailed rating for each of the Internet artifacts is available in Table A4 of the Appendix where each cell represents the of ratings obtained by the first and second author. From Table 5 we observe Kubernetes-related Internet artifacts to score > 0.7 for Q3 and Q8, which respectively corresponds to the aim and date of the Internet artifact. With respect to reputation (Q2), impact (Q11), and credibility (Q12), the average rating for the 321 Internet artifacts is < 0.25. The range of scores for each criterion is presented as minimum and maximum in the 'Min, Max' column.

One of the criteria for quality evaluation is Q7 ('Vested Interest'), where the first and second author individually inspects each Internet artifact, and determine if there is a vested interest to promote a competing tool that also can be used for container orchestration. As shown in Table 5, the average value related to 'Q7 (Vested Interest)' is 0.08. which shows on average our studied Internet artifacts to not have a vested interest in promoting another competing tool. As shown in Table A4 in the Appendix, 276 out of 321 Internet artifacts score a zero (0.0) for Q7, where a value of 0 indicates 'no vested interest' as determined by the first and second author. For 8 of the 321 Internet artifacts, the first and second author record 1.0, which indicates a vested interest to promote a tool that competes with Kubernetes. Our analysis related to vested interest shows the objectivity of our studied Internet artifacts, and further attests the importance of studying the quality of Internet artifacts, as documented by the Garousi et al. (2019).

Table 4: Quality Assessment Criteria for Internet Artifacts

| Criterion | Question |
|---|---|
| **Criterion-1: Reputation** | **Q1:** Is the publishing organization reputable? **Q2:** Is an individual author associated with a reputable organization? |
| **Criterion-2: Methodology (Aim, Reference, Coverage)** | **Q3:** Does the source have a clearly stated aim? **Q4:** Is the source supported by authoritative, contemporary references? **Q5:** Does the work cover a specific question? |
| **Criterion-3: Objectivity** | **Q6:** Is the statement in the sources as objective as possible? Or, is the statement a subjective opinion? **Q7:** Is there a vested interest? For example, a tool comparison by authors that are working for particular tool vendor. |
| **Criterion-4: Date** | **Q8:** Does the item have a clearly stated date? |
| **Criterion-5: Position with respect to related sources** | **Q9:** Have key related Internet artifacts or peer-reviewed publications been linked to / discussed? |
| **Criterion-6: Novelty** | **Q10:** Does it strengthen or refute a current position? |
| **Criterion-7: Impact** | **Q11:** What is the impact of the Internet artifact? The raters apply subjective evaluation to determine impact of an Internet artifact. The rater considers the following concepts to determine impact: count of back links, count of comments, count of views, and count of shares. |
| **Criterion-8: Credibility** | **Q12:** What is the credibility of the Internet artifact? (i): High credibility: Books, magazines, theses, government reports, white papers; (ii) Moderate credibility: Annual reports, news articles, presentations, videos, Q/A sites (e.g. StackOverflow), Wikipedia articles; (iii) Low credibility: Blogs, emails, tweets. |

### 3.5 **Phase-5: Methodology to Answer Research Questions**

We describe our methodology to answer the three research questions in this section. We use the content of Internet artifacts to answer RQ1 and RQ2. Practitioners are likely to expresses their perceptions in non-academic venues, such as in blog posts instead of peer-reviewed journals (Garousi et al, 2020). Our assumption is that by applying a qualitative analysis technique we can identify the benefits and challenges that practitioners face while using Kubernetes. We use a qualitative analysis technique called open coding (Saldana, 2015) with 321 Internet artifacts to answer RQ1 and RQ2. Open coding is a technique for qualitative analysis that can summarize the latent theme from Table 5: Quality Assessment for 321 Internet Artifacts

| Criterion | Average Rating | Min, Max |
|---|---|---|
| **Q1 (Reputation of Publishing Organization)** | 0.25 | 0.0, 1.0 |
| **Q2 (Reputation of Author's Organization)** | 0.22 | 0.0, 1.0 |
| **Q3 (Clearly Stated Aim)** | 0.79 | 0.0, 1.0 |
| **Q4 (References)** | 0.61 | 0.0, 1.0 |
| **Q5 (Coverage)** | 0.27 | 0.0, 1.0 |
| **Q6 (Content Objectivity)** | 0.67 | 0.0, 1.0 |
| **Q7 (Vested Interest)** | 0.08 | 0.0, 1.0 |
| **Q8 (Clearly Stated Date)** | 0.72 | 0.0, 1.0 |
| **Q9 (Links to Important Literature)** | 0.59 | 0.0, 1.0 |
| **Q10 (Strengthen/Refute Position)** | 0.37 | 0.0, 1.0 |
| **Q11 (Impact)** | 0.11 | 0.0, 1.0 |
| **Q12 (Credibility)** | 0.22 | 0.0, 1.0 |

unstructured text data (Saldana, 2015). Our hypothesis is that by analyzing Internet artifacts we will have an understanding of benefits and challenges related to Kubernetes as reported by practitioners. We use open coding as this qualitative analysis technique can be used to extract patterns from Internet artifacts, as done in prior work (Hasan et al, 2020; Ur Rahman and Williams, 2016). In the case of RQ3, we use peer-reviewed publications that we identify from Section 3.3.

### *3.5.1 Methodology to Answer RQ1*

While applying open coding for RQ1 the rater inspects each Internet artifact for content that expresses a benefit of using Kubernetes. To conduct open coding for RQ1 we first extract text features i.e., sentences that express a benefit of using Kubernetes from an Internet artifact. From the initial text features, the rater extracts 'initial code'. From the one or multiple initial codes, the raters determine 'codes', from which we categorize benefits related to Kubernetes.

In Figure 4 we demonstrate a hypothetical example of identifying a benefit related to Kubernetes from our set of 321 Internet artifacts. According to Figure 4, in the left-most column, we select the initial text features from Internet artifacts. We summarize from the initial text feature to generate the initial code *"Kubernetes can run in public, private or hybrid cloud environments"*, *"Enterprises can run Kubernetes on any combination of public and private clouds"*, and *"Applications written in Kubernetes can be deployed in any cloud provider"*. From these three initial codes we derive two initial categories: (i) *'Ease in provisioning multi-cloud environments'*, and (ii) *'Ease in using cloud provided utilities'*. We combine these two initial categories to derive the category *'Ease in Cloud-based Interfacing'* as both initial categories relate to Kubernetes's benefit related to cloud-based interfacing.

The first and second authors are raters who individually applied open coding to identify benefits of using Kubernetes. Upon determining the benefits the authors compared their agreements and disagreements. The Cohen's Kappa (Cohen, 1960) is 0.68 for benefits. The first author disagree on 3 benefits. For benefits that are disagreed upon, the last author acts as the resolver. To resolve the disagreements the last author read the definitions and examples for benefits that are disagreed upon. The last author's decision is final for resolving disagreements.

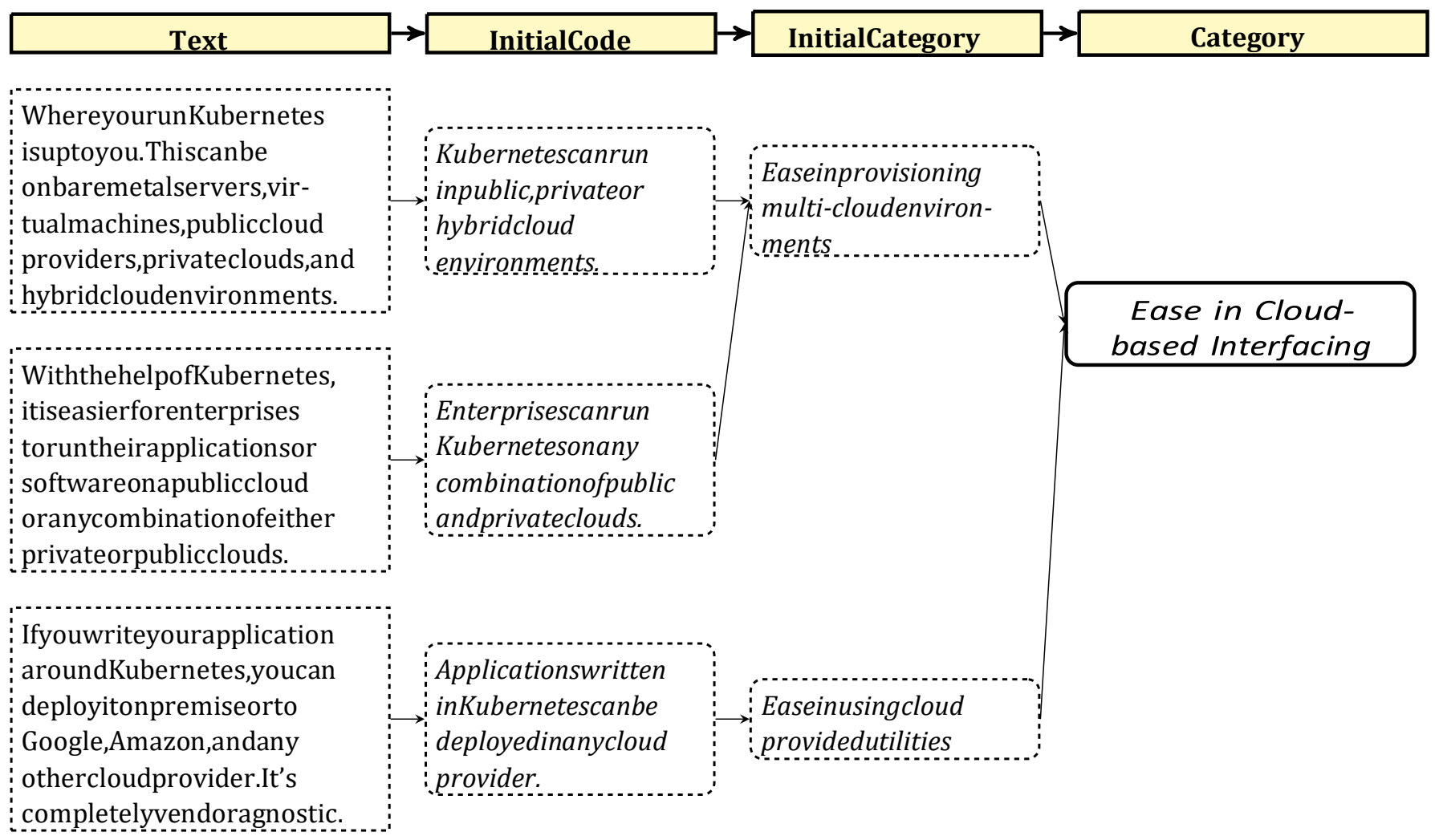

Fig. 4: An example to demonstrate how we used open coding to determine benefits of using Kubernetes.

### *3.5.2 Methodology to Answer RQ2*

For RQ2 we repeat the process of open coding where the rater inspects each Internet artifact for content that expresses a challenge of using Kubernetes. We use Figure 5 to demonstrate how we use open coding to identify challenges from Internet artifacts. Similar to the open coding procedure described in Section 3.5.1, we derive initial code, then codes, and finally from code we identify challenges.

For challenge identification we follow the guidelines proposed by Tornatzky et al. (1990), similar to prior work (Batubara et al, 2018) that used the same set of guidelines to identify challenges in using blockchain technologies. According to Tornatzky et al. (1990), a challenge describes a characteristic of a technology or its surrounding ecosystem, which prevent users from adopting a technology or not reap the full benefits from the technology. Tornatzky et al. (1990) considers both: technical and non-technical such as organizational challenges. While identifying the challenges both the first and second author inspected for statements in Internet artifacts that indicate one or multiple characteristics of Kubernetes and its ecosystem, which can hinder adoption of Kubernetes or prevent existing users to reap full benefits from the tool.

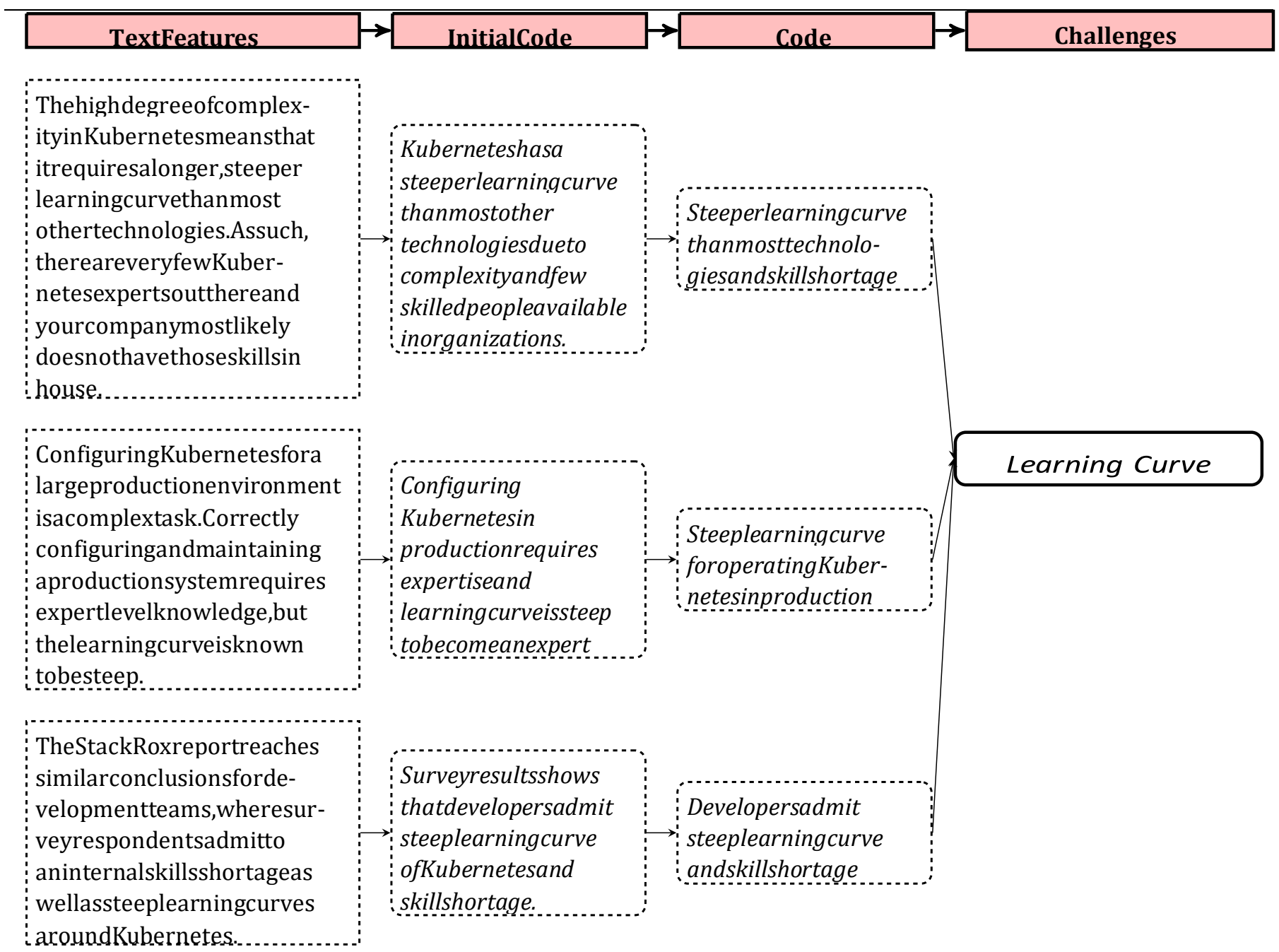

Fig. 5: An illustrative example to demonstrate the methodology of applying open coding to determine challenges of using Kubernetes.

The first and second authors are raters who individually applied open coding to identify challenges of using Kubernetes. Upon determining the challenges the authors compared their agreements and disagreements. The Cohen's Kappa (Cohen, 1960) is 0.71 for challenges identification. The first and second author disagree on 4 challenges. For challenges that are disagreed upon, the last author acts as the resolver. To resolve the disagreements the last author read the definitions and examples for challenges that are disagreed upon. The last author's decision is final for resolving disagreements.

**Mapping of Challenges to Product Types**: The Cloud Native Computing Foundation (CNCF), uses the following classification for product types related to Kubernetes (CNCF, 2021):

- Distribution: "*A distribution is software based on Kubernetes that can be installed by an end user on to a public cloud or bare metal and includes patches to the upstream codebase*".
- Hosted: "*A hosted platform is a Kubernetes service provided and managed by a vendor*".

– Installer: "*An installer downloads and then installs vanilla upstream Kubernetes*".

---

The first author uses the description of challenge and applies closed coding (Saldana, 2015) to identify which of the identified challenges are applicable for any or multiple of the three product types. As closed coding is a qualitative analysis procedure, we use a PhD student in the department to perform rater verification for all challenges and their mapping with the three product types. We record a Cohen's Kappa (Cohen, 1960) of 0.84 between the first author and the PhD student.

### *3.5.3 Methodology to Answer RQ3*

Similar to RQ1 and RQ2, we also apply open coding to determine research topics that are investigated in peer-reviewed publications related with Kubernetes. While applying open coding, the rater inspects the content of the entire publication to determine what research topic the publication is addressing. The first and second author individually applies open coding and, respectively, identified 15 and 21 research topics. The Cohen's Kappa (Cohen, 1960) is 0.67. The raters disagree on 5 research topics, which necessitated the last author to act as a resolver. After reading the name, description, and example of each topic the last author identifies 14 research topics. For disagreement resolution the last author's decision is final.

**Mapping of Benefits and Challenges with Publication Content**: Upon deriving the research topics we also map each of the identified benefits and challenges, respectively from Sections 3.5.1 and 3.5.2. For mapping the benefits, the first author reads the content for each of the 105 publications, and identifies if any or multiple benefits are mentioned in a publication for using Kubernetes. In the case of challenges, the first author read the content of each publication, and identified if the research topic addressed in the publication addresses any or multiple of the challenges.

## 4 Results

We provide answers to our research questions in this section. We identify 8 benefits, 15 challenges, and 14 research topics from our MLR of Kubernetes, as summarized in Table 6.

### 4.1 *RQ1*: What are the perceived benefits of using Kubernetes as reported by practitioners?

From our analysis of 321 Internet artifacts we identify 8 benefits. A complete mapping of Internet artifacts and identified benefits is provided in Table A5 of the Appendix. We describe the identified benefits below, where the challenges are sorted based on the count of Internet artifacts that mention the challenge.

Table 6: Summary of Our MLR with 321 Internet Artifacts and 105 Peerreviewed Publications

| Summary | Name |
|---|---|
| **8 Benefits** | 'Deployment of Software & Services', 'Ease in Cloud-based Interfacing', 'Configuration Management', 'Community Support', 'Service Level Objective (SLO)-based Scalability', 'Resource Limit Specification', 'Availability of Software', 'Self-healing Containers & Pods' |
| **15 Challenges** | 'Lack of Security Practices & Tools', 'Attack Surface Reduction', 'Lack of Diagnostics Tools', 'Maintenance-related Challenges', 'Learning Curve', 'Networking', 'Migration Cost', 'Storage', 'System Environment Configurations', 'Testing', 'Failure Troubleshooting', 'Performance', 'Cultural Change', 'Idempotency', 'Hardware Compatibility' |
| **14 Research Topics** | 'Performance Evaluation', 'Resource Allocation', 'Internet of Things', 'Networking', 'Data Mining & Machine Learning', 'Microservice Orchestration', 'Security', 'Fault Tolerance', 'High Performance Computing', 'Logging & Monitoring', 'Configuration Abstraction', 'Database Management', 'Electronic Vehicle', 'Discrete Time System Simulation' |

The count of internet artifacts that mention the challenge is expressed in parenthesis.

**Deployment of Software & Services (126):** We observe practitioners to report deployment of software and services to be a benefit. Kubernetes provides unique utilities, such as 'kubernetes deployment' (Kubernetes, 2021) and Helm (2021) to provision the required deployment infrastructure using pods. Furthermore, Kubernetes automates the manual process of software deployment, which further helps practitioners to scale deployment infrastructures (VMWare, 2021). We identify four sub-categories of deployment-related benefits below:

(i) *Automation of deployments:* Practitioners perceive Kubernetes's ability to automate software deployments, as a benefit. According to a practitioner, *"The Kubernetes API is a great tool when it comes to automating a deployment pipeline. Deployments are not only more reliable, but also much faster, because we're no longer dealing with VMs"* (IA233).

(ii) *Diversity in deployment strategies:* Kubernetes has facilitated practitioners to experiments and use deployment strategies, namely, canary deployment and blue/green deployment. A canary deployment is a deployment strategy that allows deployment of new software changes to a subset of end-users as an initial test, instead of deploying to all end-users (Humble and Farley, 2010). In blue/green deployment, practitioners deploy the 'blue' version i.e., the new version of software changes along with the 'green' version, i.e., the latest working version of the software so that if quality concerns arise due to the blue version, practitioners can roll back to the green version of software changes (Humble and Farley, 2010). While blue/green deployment focuses on safe and reliable software releases, alpha /beta testing focuses functionality of a software feature as perceived by end-users (Humble and Farley, 2010). In IA213 a practitioner discussed how Kubernetes utilities, such as pods and controllers can be used to adopt above-mentioned diverse deployment strategies.

(iii) *Ease of deployment:* Kubernetes has removed complexities when deploying software changes in container-based deployment platforms. A practitioner discusses how Kubernetes can resolve deployment-related challenges in a team with multiple developers who develop a diverse set of products: "*Pods, or groups of containers can group together container images developed by different teams into a single deployable unit*"(IA60). Using Kubernetes-based utilities, such as Helm practitioners can mitigate the complexity of deploying a Hadoop-based software across multiple servers stating "*It's harder to deploy Hadoop across servers than to run a Ruby on Rails application on my laptop. With Kubernetes, this is going to change*"(IA76).

(iv) *High velocity of deployments:* Kubernetes is helpful to improve the velocity of software deployments, which has lead to rapid deployment of software and services. For example, a practitioner mentioned "*Rather than a team of developers spending their time wrapping their heads around the development and deployment lifecycle, Kubernetes (along with Docker) can effectively manage it for you so the team can spend their time on more meaningful work that gets products out the door*" (IA27).

**Ease in Cloud-based Interfacing (111):** Kubernetes's ability to interface with cloud-based utilities is perceived as a benefit of using Kubernetes. According to practitioners, Kubernetes provides flexibility for practitioners to deploy containers in on-premise, public, or private cloud infrastructure, or in hybrid multi-cloud environment. We observe practitioners to report two types of benefits:

(i) *Ease in using cloud provided utilities:* Practitioners find Kubernetes to be easily interfaced with cloud-provided utilities, such as IBM Cloud (IBM, 2021). One practitioner mentions 'kops' (Kubernetes, 2021), a Kubernetesprovided utility, and how that is helpful to interface with Azure and AWS-based cloud instances "*If you find and use a tool, such as kops for managing your Kubernetes, you can get a lot of the same benefits Azure provides*" (IA255).

(ii) *Ease in provisioning multi-cloud environments:* A hybrid cloud-based resource becomes multi-cloud when there is more than one public cloud service combined with private cloud resources. A hybrid cloud is a combination of public and private clouds in order to orchestrate a single IT solution between both. Companies, such as Amazon (AWS, 2021) and Microsoft (Azure, 2021) provide hybrid multi-cloud resources.

Practitioners report Kubernetes's ability to provision deployment environments in hybrid cloud-based resources as a benefit. For example, a practitioner states *"What's great about Kubernetes is that it's built to be used anywhere so you can deploy to public/private/hybrid clouds, enabling you to reach users where they're at, with greater availability and security"* (IA27). Another practitioner states that using Kubernetes one can "*move workloads without having to redesign your applications or completely rethink your infrastructure—which lets you to standardize on a platform and avoid vendor lock-in*" (IA2). In IA291, a practitioner describes that the same Kubernetes configuration can be transferred from one cloud-based infrastructure (e.g. AWS) to another (e.g. Azure), instantly.

**Configuration Management(86):** Deployment of software involves specifying and managing a wide range of configurations. According to our analysis of Internet artifacts, Kubernetes helps practitioners in configuration management. For example, in IA176 a practitioner argues that the separation of configuration from orchestration is why Kubernetes is popular. A practitioner from GitHub describes the complexities in configuration management before adopting Kubernetes *"Over the last several months, engineers have already deployed dozens of applications to [our] cluster. Each of these applications would have previously required configuration management and provisioning support from SREs. With a self-service application provisioning workflow in place, SREs [at GitHub] can devote more of our time to delivering infrastructure products to the rest of the engineering organization in support of our best practices, building toward a faster and more resilient GitHub experience for everyone*"(IA24). Kubernetes uses declarative programming to manage configurations, which is also perceived as a benefit by a practitioner in IA160.

**Community Support (75):** The fact that Kubernetes is an open source community-driven tool is perceived as a benefit by practitioners. According to Novoseltseva, Kubernetes has an active community of more than 2,000 contributors from Fortune 500 companies (Novoseltseva, 2021b). The Kubernetes community pro-actively engages in mitigation of bugs, as well development of new features. According to the official Kubernetes Security and disclosure information page (Kubernetes, 2021), Kubernetes community volunteers thoroughly investigate and respond to each of the vulnerability reported by security researchers and security practitioners. Kubernetes is built with 15 years of Google's experience of running production workload and best practices from the community (Miles, 2020; Kubernetes, 2021).

**Service Level Objective (SLO)-based Scalability (75):** Kubernetes allows its users to scale their deployment infrastructure based on the concept of service level objective (SLO), which practitioners perceive as a benefit. SLO is the concept of providing constraints for the computing infrastructure of interest based on certain a metric, such as latency. For example, a user can specify a SLO where the latency for a HTTP-based web software can never be > 1.0*ms* within a time window of 60 seconds. If this SLO is specified, then while adding containers and pods Kubernetes will periodically check if the latency is > 1.0*ms* while adding containers or pods as the user of the web software increases. In this manner, with Kubernetes the user does not have to be concerned about performance-related metrics, such as latency while provisioning more infrastructure. Kubernetes provides SLO-based scalability using utilities, such as cluster auto-scaler, horizontal pod auto-scaler (HPA), and vertical pod auto-scaler (VPA).

Practitioners report that SLO-based scalability has solved the problem for IT organizations to handle the growing number of users. For example, in IA196 a practitioner describes companies, such as Tinder, Airbnb, and Pinterest have used Kubernetes's SLO-based scalability to solve challenges related to a high volume of users.

**Resource Limit Specification (72):** Kubernetes allows end-users to limit resources, such as CPU, memory, and network bandwidth as configurations with ranges, which practitioners perceive as a benefit. In IA51, a practitioner describes that Kubernetes has resource quota configurations for CPU, memory, and disk and Kubernetes components such as pods, services, and volumes. As mentioned in IA11, the vertical pod auto-scaler (VPA) (Kubernetes, 2021; Miles, 2020) provided by Kubernetes is also perceived as beneficial by practitioners as it can automatically optimize resources consumed by containers. In this manner, practitioners do not have to manage allocate and de-allocate resources by themselves. With respect to resource allocation, practitioners from GitLab observed benefits of using Kubernetes: *"We knew then that running GitLab.com on Kubernetes would benefit the SaaS platform for scaling, deployments, and efficient use of compute resources"* (IA152).

**Availability of Software (53):** We observe practitioners to report availability of software to be a benefit of using Kubernetes. In the context of continuous deployment, availability refers to the feature of provisioning a software application so that the software is resilient against single point of failures (Humble and Farley, 2010). Practitioners report that Kubernetes provides utilities, such as 'kube-scheduler' (Kubernetes, 2021) and ReplicaSet (Kubernetes, 2021) to make software applications resilient against single point of failures. Using these utilities Kubernetes automatically searches for pods that are not functioning, and automatically provisions a new pod to ensure software availability for end-users. The benefits related to availability can be further divided into two sub-categories:

(i) *Availability through automated orchestration:* Practitioners find Kubernetes to eliminate manual work related to container-based orchestration. For software provisioning "*Kubernetes does the work itself and distributes the workloads to make sure everything runs as the operator envisioned*" (IA79). One practitioner states "*anyone interested in designing large scale applications that remain highly available ... should consider using Kubernetes for container orchestration, combined with an engineering team that has expert knowledge of configuring, maintaining and updating*

*Kubernetes*" (IA57). According to one practitioner, "*Kubernetes' ReplicaSet helps developers solve this problem by ensuring a specified number of Pods are kept alive continuously*" (IA220).

(ii) *Multi-zone availability:* In cloud computing, multi-zone availability refers to the concept of ensuring availability of cloud computing utilities in geographically isolated locations from which public cloud services originate and operate (Xie et al, 2019). Using cloud-based providers, such as Amazon Web Services (AWS) practitioners can deploy their software applications in cloud-based deployment environments that have multizone availability. In this manner, provisioned software applications are resilient against single point of failures.
In IA51, a practitioner describes that Kubernetes provides high availability by starting a new instance of a service if the service instance fails. Another practitioner says, *"Today, our Kubernetes infrastructure fleet consists of over 400 virtual machines spread across multiple data-centres. The platform hosts highly-available mission-critical software applications and systems, to manage a massive live network with nearly four million active devices"* (IA58).

**Self-healing Containers & Pods (26):** Kubernetes can periodically check the status of containers or pods, and restart containers or pods that are malfunctioning. This feature of checking functionality-related status of containers or pods, and restarting them if needed is called 'self-healing' (Kubernetes, 2021). Practitioners find Kubernetes's self-healing utility to be beneficial for resilient deployment infrastructure *"Kubernetes can self-heal containerized applications, making them resilient to unexpected failures"*. A practitioner perceives Kubernetes to be helpful to mitigate cybersecurity-related threats "*We're often stuck in a balancing act where we're asked to prioritize growth and innovation or security and compliance. With Kubernetes, these goals are no longer mutually exclusive*" (IA159).

## 4.2 *RQ2*: What are the perceived challenges of using Kubernetes as reported by practitioners?

While we have identified 8 types of benefits of using Kubernetes, practitioners face challenges too. A complete mapping of Internet artifacts and identified challenges is provided in Table A6 of the Appendix. We describe the 15 challenges that practitioners face with Kubernetes as follows:
**Lack of Security Practices & Tools (132)**: This category describes the lack of practices and tools for Kubernetes to increase the security of Kubernetes installations and Kubernetes-based deployments. We observe practitioners to seek guidance on (i) securing Kubernetes configurations and components, such as containers and pods (IA93), (ii) securing network traffic (IA269/IA271), and (iii) applying role-based access control (RBAC) policies (IA302/IA303). The lack of security practices and tools impact software deployments: according to the Stackrox security survey, 44% survey participants describe that they delay Kubernetes-based software deployments due to lack of confidence that relates with Kubernetes security (Stackrox, 2021).

**Attack Surface Reduction (126):** This category describes challenges that practitioners encounter while reducing the attack surface that stems from Kubernetes usage. Attack surface corresponds to the set of ways in which an adversary can enter the system and potentially cause damage (Theisen et al, 2018). This challenge is different to lack of security practices and tools because the challenge includes the challenge of assessing and reducing the attack surface of Kubernetes. The components and the software projects that use Kubernetes for deployment increase the attack surface of Kubernetes, making Kubernetes-related deployments susceptible to security attacks (Fei Huang, 2018). This view was echoed by a practitioner who stated "*the more apps or worker nodes that you expose publicly [using Kubernetes], the more steps you must take to prevent external malicious attacks*" (IA13). Practitioners face challenges with the following Kubernetes-related entities to reduce the attack surface: databases (IA20), Kubernetes configurations (IA54/IA70/IA89), bugs and vulnerabilities in packages used by Kubernetes (IA92), and unstable Kubernetes releases (IA160).

**Lack of Diagnostics Tools (121):** This category describes the lack of diagnostics tools, i.e., monitoring and logging tools for Kubernetes as expressed by practitioners. While a wide range of utilities, such as self-healing and SLOs exist to facilitate scalable deployment, Kubernetes does not provide any builtin logging or monitoring utilities (Kubernetes, 2021). Practitioners mitigate this challenge by using open source tools, such as Prometheus (prometheus, 2021), but find these tools inadequate. A practitioner in IA70 reports that relying on open source third party diagnostics utilities make Kubernetes-based deployments susceptible to unauthorized accesses. Another practitioner finds existing open source diagnostics tools to not be a good fit for mid or smallsized companies due to lack of resources *"smaller organizations generally have less need to create a formal monitoring process, while larger ones have the resources to create a more robust, customized monitoring system. Stuck in the middle are those organizations with 100 to 999 employees"* (IA91).

While describing this challenge practitioners also mention the areas where availability of diagnostic tools are pivotal. For example, in IA1 practitioners describe that monitoring is necessary to prevent deployments from consuming too many resources. Diagnostics tools can help practitioners gain visibility in Kubernetes-based deployments (IA119), and also troubleshoot root causes of failures (IA199/IA203). Security is also another use case for Kubernetesrelated diagnostics tools: a practitioner states "*At some point, you will have to implement centralized log management for your Kubernetes logs in order to meet security and quality requirements for your organization*"(IA158). **Maintenance-related Challenges (109):** This category describes challenges that are related to managing Kubernetes-related installations and deployments. Our review of Internet artifacts show that even though Kubernetes helps in automated deployment of software and services, dedicated computing and human resources are required to adequately maintain Kubernetes installations.

According to one practitioner, *"Kubernetes management is a largely manual exercise because all you get with Kubernetes is Kubernetes. The platform doesn't come with anything to run it, so you need to figure out how to deliver resources to Kubernetes itself no easy feat"* (IA283). A practitioner states "*Kubernetes is not a set-and-forget platform; managed Kubernetes services from cloud providers help get you started faster, but even perfectly built infrastructure needs to be monitored, configured, and managed by dedicated DevOps resources*"(IA265). Practitioners mention topics for which maintenance is required: policy upgrading (IA14/IA261), security-related maintenance (IA51/IA269), database operations (IA69), cluster management as user base grows (IA142/IA77/IA121), and interfacing with cloud-based platforms, such as AWS (IA273).

**Learning Curve (91):** This category describes challenges related to learning Kubernetes-related concepts. From our set of Internet artifacts we observe practitioners to perceive Kubernetes as a complex tool, which comes with a steep learning curve. According to a practitioner in IA66, the Kubernetes codebase is large, complex, contains minimal documentation, and has a wide range of external dependencies. According to a practitioner in IA119, Kubernetes has a steeper learning curve than most other technologies, and the number of Kubernetes experts out there are few. A practitioner further comments "*Indeed, Kubernetes has a steep learning curve, a bewilderingly vast ecosystem, and a seemingly infinite landscape of third-party providers attempting to sell you some kind of 'easy Kubernetes' (which in itself is a telltale sign of its complexity). At the end of the day, dealing with Kubernetes can be mindnumbing*" (IA166). A practitioner suggests the complexities associated with learning Kubernetes configurations stating "*Kubernetes is very complicated, and it's very easy to make a mistake on how you configure it*" (IA19). Training team members on Kubernetes requires time and effort, which was echoed by a practitioner in IA57 *"Many teams that tried to abruptly switch to Kubernetes without proper training, faced difficulties and often switched back to their old solution. Any new team, before choosing to use Kubernetes, should first allocate time to learn the fundamentals required to work with Kubernetes"* (IA27). Practitioners perceive the steep learning curve of Kubernetes to cause security-related weaknesses (IA18/IA141) and hard to debug run-time failures (IA235).

**Networking (84):** This category describes challenges of using network-related utilities in Kubernetes. Practitioners face challenges when setting up networkbased utilities in Kubernetes as overlay networking is handled dynamically by Kubernetes (IA62). In IA72, a practitioner perceives containers in Kubernetes to pose unique network-related challenges, such as network segmentation because how containers communicate across nodes. According to IA194, understanding the trade-offs related to network-related features between cloud vendors, such as AWS and Azure, is challenging and necessitates derivation of best practices.

**Migration Cost (82):** This category describes challenges related to the cost that practitioners encounter while migrating to a Kubernetes-based deployment or installation. Migration-related costs include investment in time and team resource to experiment with Kubernetes (IA2), changes in existing codebase (IA52), and monetary costs incurred due to provisioning of multiple services using cloud-based utilities, such as AWS (IA55) as well as setting up virtual networks (IA194). In IA233, a practitioner discusses the trade-offs between deployment cost and deployment size stating money-wise Kubernetes is cost effective for larger deployments compared to that of smaller deployments. While describing migration-related challenges practitioners recommend potential adoptees to reflect on why they want to use Kubernetes: *"Kubernetes transformation is not cheap. The price you pay for it must really justify 'your' use case and how it leverages the platform"* (IA58).

**Storage (52):** This category describes challenges related to storage while using Kubernetes. From our analysis, we observe practitioners to face challenges with (i) encrypting Kubernetes storage, (ii) defining storage volume types, (iii) creating persistent layers for storage, and (iv) implementing backup and restore process for storage. For example, in IA13 a practitioner describes Kubernetes default storage to not be encrypted, and perceives encryption of storage to be challenging.

**System Environment Configurations (46):** This category describes challenges related to managing configurations of system environments. In Section 4.1 we have discussed that practitioners perceive configuration management of Kubernetes as a benefit. Yet, while setting up and managing configurations for system environments we observe practitioners to face challenges. For example as discussed in IA11, we observe practitioners face challenges on how to isolate system configurations for dev/stage deployment environments. **Testing (36):** This category refers to testing Kubernetes-based deployments and installations. Practitioners mention differences in system environment configurations to create challenges for testing: *"Because infrastructure is tightly coupled to the concept of environment, testing infrastructure changes is challenging, as differences tend to exist between the infrastructures of different environments"* (IA10). Testing a functionality with Kubernetes can require spinning up the entire Kubernetes installation, which is time consuming according to IA66. Kubernetes releases can also create testing challenges as one practitioner mentions *"Although rare, there are times when new changes may not be backward compatible and could cause production outages. Thus, one should be diligent and design a testing mechanism for detecting application breaking changes, whenever they update their Kubernetes version"*(IA57). **Failure Troubleshooting (32):** This category describes challenges related to failure troubleshooting in Kubernetes. A practitioner discusses how fixing failures is complex "*With Kubernetes ... it's difficult to fix if something goes wrong, and you can't automate upgrades*" (IA20). We observe practitioners to face challenges when troubleshooting networking (IA30) and componentrelated failures (IA167). A practitioner emphasized the need for in-depth knowledge of Kubernetes to troubleshoot Kubernetes-related failures stating "*Kubernetes requires experience and extensive training for its debugging and troubleshooting in due time*" (IA113).

**Performance (26):** This category describes challenges related to performance of the deployed software on Kubernetes as well as the performance of the Kubernetes installation itself. Performance metrics that practitioners have mentioned include network bandwidth, CPU usage, and disk read/write rate. For example, in IA13, a practitioner mentioned storage encryption to impact disk read/write rate of Kubernetes. According to IA30, etcd-related failures can cause reduced performance for Kubernetes as etcd stores the state of all jobs running on Kubernetes. In IA262, a practitioner reports workload-related configurations to be correlated with performance for Kubernetes.

**Cultural Change (23):** This category describes challenges related to team culture while using Kubernetes. Our analysis shows that cultural issues, such as mindset of team members can pose challenges while adopting and using Kubernetes. In IA27, a practitioner warns that Kubernetes adoption could be challenging for teams who are unwilling to experiment and take risks. Furthermore, as stated in IA257, building teams for adopting Kubernetes is challenging due to a 'legacy mindset'. According to IA320, corporate culture complicates Kubernetes usage by creating a contrary view between developers and upper-level executives.

**Idempotency (16):** This category describes challenges related to idempotency while using Kubernetes. For Kubernetes, idempotency is the property, which ensures that even after $n$ executions, where $n > 1$, a Kubernetes-based environment is exactly the same as it was after the first provisioning of the Kubernetes-based infrastructure. While using pods in Kubernetes practitioners can modify pod configurations manually, which can result in a Kubernetesbased deployment that is different to what is desired. For example in IA288, a practitioner observes a lack of homogeneity in Kubernetes, which can contribute to idempotency-related bugs in Kubernetes.

**Hardware Compatibility (4):** This category describes challenges related to hardware compatibility when using Kubernetes. For example, as described in IA274, a practitioner faced memory-related challenges when trying use Kubernetes on a Raspberry PI device. The incompatibility was attributed to the inavailability of container images that support Raspberry PI devices with a 32-bit Advanced RISC Machine (ARM) architecture.

**Yearly Trends of Identified Challenges**: We further report the temporal trends of the identified challenges with Figure 6. The y-axis presents the proportion of Internet artifacts that report a challenge for a certain year. Except for 'Hardware Compatibility', we observe challenges to increase with time. The implication of this finding is that the identified challenges are relevant amongst practitioners, and the trends are increasing for 14 of the 15 identified challenges.

**Mapping of Challenges to Product Types**: We further map each challenge to a product type as defined by CNCF. The mapping is provided in Table 7. We observe majority of the challenges to be applicable for the three product types. 'Hardware Compatibility' is not applicable for distribution-type products.

### 4.3 *RQ3*: What research topics have been investigated in Kubernetes-related publications that are peer-reviewed?

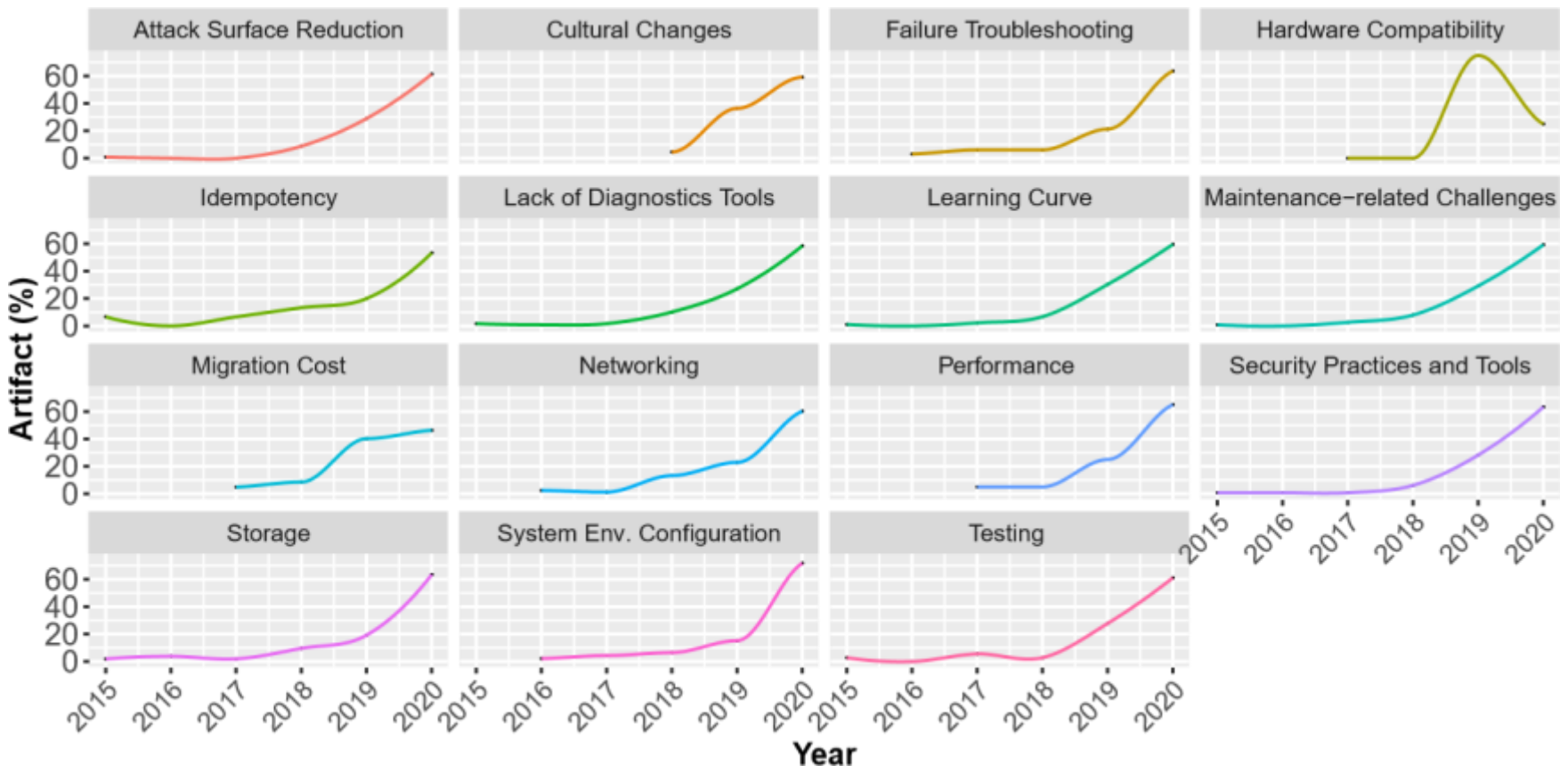

Fig. 6: Yearly trends in identified challenges. Out of 15, 14 challenges recur and show an increasing trend.

Table 7: Mapping of Identified Challenges to Product Types

| Product Type | Challenge |
|---|---|
| **Distribution** | ‘Attack Surface Reduction’, ‘Cultural Changes’, ‘Failure Troubleshooting’, ‘Idempotency’, ‘Lack of Diagnostics Tools’, ‘Learning Curve’, ‘Maintenance-related Challenges’, ‘Migration Cost’, ‘Networking’, ‘Performance’, ‘Security Practices and Tools’, ‘Storage’, ‘System Environment Configuration’, ‘Testing’ |
| **Hosted** | ‘Attack Surface Reduction’, ‘Cultural Changes’, ‘Failure Troubleshooting’, ‘Hardware Compatibility’, ‘Idempotency’, ‘Lack of Diagnostics Tools’, ‘Learning Curve’, ‘Maintenance-related Challenges’, ‘Migration Cost’, ‘Networking’, ‘Performance’, ‘Security Practices and Tools’, ‘Storage’, ‘System Environment Configuration’, ‘Testing’ |
| **Installer** | ‘Attack Surface Reduction’, ‘Cultural Changes’, ‘Failure Troubleshooting’, ‘Hardware Compatibility’, ‘Idempotency’, ‘Lack of Diagnostics Tools’, ‘Learning Curve’, ‘Maintenance-related Challenges’, ‘Migration Cost’, ‘Networking’, ‘Performance’, ‘Security Practices and Tools’, ‘Storage’, ‘System Environment Configuration’, ‘Testing’ |

We answer RQ3 by describing the research topics that we identify from our analysis. Next, we map the identified benefits and challenges to the identified research publications.

#### *4.3.1 Description of Identified Research Topics*

In our qualitative analysis one publication can discuss one or more identified topics, and can belong to multiple topics. We provide a mapping between the research topics and publications in Table 8. The description of each of the research topic is given below:

**Performance Evaluation (50):** Performance evaluation is the category of peer-reviewed publications that investigates performance issues in Kubernetesbased deployments. We observe this category of publication to include two sub-categories:

(i) *Technique for performance improvement*: Publications that belong to this category proposes and evaluates techniques that can improve a Kubernetesbased deployment. For example, in P18, the authors propose a technique called 'AlloX', and evaluated the performance improvement obtained by Allox for TensorFlow (Abadi et al, 2016). In P19, the authors proposed and evaluated a configuration tuning tool called 'Accordia' that generates configurations so that performance overhead is reduced for resourceintensive software. Similarly, in P53, the authors propose 'ConfAdvisor' to improve container performance. In P23, the authors propose a tool called 'KubeShare' that allows graphics processing units (GPU)based deployments using Kubernetes. In P52, the authors propose a technique to minimize CPU consumption when the CPU resources are shared among co-located containerized software. In P91, the authors propose a technique that uses the non-dominated sorting genetic algorithm II (NSGA II) to optimize container CPU and memory.

(ii) *Performance benchmarks*: Publications that belong to this category investigate and compare performance of Kubernetes-based deployments using curated data benchmarks. For example, in P37, the authors compare five deployment tools including Kubernetes, based on the criteria of provisioning, packaging, monitoring, reconfiguration. Authors of P37 report that Kubernetes provide robust deployment features. In P41, the authors use a benchmark to compare the performance of Kubernetes with AWS, Azure, and Google Compute Engine (GCE), and reports when interfaced with GCE, Kubernetes to show best overall performance. In P43, the authors use a benchmark to compare performance between Docker Swarm and Kubernetes, and observe Kubernetes to do better. On the contrary to P43, authors of P67 observed Docker Swarm to perform better than Kubernetes when used on their benchmark.

**Resource Allocation (44):** Resource allocation is the category of peerreviewed publications that proposes and evaluates techniques on how Kubernetes can be configured so that resources are efficiently allocated for one our multiple software deployments using Kubernetes. We observe prior research to apply a diverse set of algorithms, such as search-based algorithms, graph algorithms, and machine learning algorithms to efficiently allocate resources. For example, in P8, the authors use search-based algorithms, namely, the ant colony algorithm (Dorigo et al, 2006), and the particle swarm optimization algorithm (Shi et al, 2001) to develop a scheduling model for Kubernetes-based deployments. In P21, authors use BestConfig algorithm (Zhu et al, 2017) and Bayesian optimization (Alipourfard et al, 2017) to find cost-effective resource allocation policies for SLOs in Kubernetes. In P26, the stable marriage algorithm (Morizumi et al, 2011) is used to find compatible hosts and containers in order to achieve the best deployment with respect to deployment speed. In P55, the authors provide an effective resource allocator for containers running on the Kubernetes cluster. In P80, authors use deep reinforcement learning (Li, 2017) to allocate resources for deployments in Kubernetes. In P105, authors use a graph algorithm called the minimum cost flow algorithm (Goldberg and Kharitonov, 1992) where resources are allocated by representing each container request with a graph.

**Internet of Things (16):** Internet of things (IoT) is the category of peerreviewed publications that investigates how Kubernetes can be used for IoTbased software applications. Peer-reviewed publications that belong to this category focus on improving network latency, scheduling, and fault tolerance of IoT applications. For example, In P38, the authors propose a fault-tolerant architecture for IoT applications in the cloud. In P45, the authors propose an extension to Kubernetes called 'KubeEdge' architecture with a network protocol stack called 'KubeBus' for IoT applications. In P62, the authors propose a custom Kubernetes scheduler where the nodes decide scheduling for IoT agents.

**Networking (14):** Networking is the category of peer-reviewed publications that investigates networking-related challenges in Kubernetes-based deployments. For example, in P64, the authors propose a remote direct memory access (RDMA) architecture to control network bandwidth in Kubernetes. In P65, the authors propose a framework to automatically configure virtualized networks with Kubernetes. In P85, the authors propose a solution for monitoring vehicular networks provisioned using Kubernetes. In P87, the authors analyze performance bottlenecks for container network interface (CNI) plugins used in Kubernetes.

**Data Mining & Machine Learning (11):** Data mining & machine learning is the category of peer-reviewed publications that investigates how software projects that use data mining and machine learning algorithms can be deployed in Kubernetes. For example, in P71, the author uses Kubernetes to design and deploy experiments for a data mining application used in particle imaging (Bobkov et al, 2018). In P98, the authors propose 'JOVIAL' a cloudbased data mining platform that can be used for astronomical data analysis with JupyterHub and Kubernetes.

**Microservice Orchestration (8):** Microservice orchestration is the category of peer-reviewed publications that investigates techniques on how to orchestrate microservice-based software applications while maintaining availability. For example, in P59, the authors propose a strategy to improve availability of microservices that relies on the state of the service by implementing state controller support for Kubernetes. In P72, the authors propose a new framework to support synchronization among microservices in Kubernetes/Openstack and test various use cases. In P74, the authors compare the deployment of microservices in CI/CD pipelines with Rundeck, Docker, Kubernetes and report that Kubernetes provides the most efficient way to achieve highly available and scalable microservices.

**Security (7):** Security is the category of peer-reviewed publications that investigates techniques to mitigate security weaknesses for Kubernetes. Anomaly detection is one security-related topic that has been addressed by researchers. In P3, the authors propose an anomaly detection tool for detecting anomalies in astronomy data analysis tools that are deployed with Kubernetes. In P93, the authors implement 'KubeAnomaly' a tool for anomaly detection in the Kubernetes cluster, using neural network approaches. Security-focused frameworks have also garnered interest: in P33, the authors propose an automated threat mitigation architecture for Kubernetes that continuously scan containers for vulnerabilities to quarantine and isolate vulnerable containers. In P92, the authors built a security framework for integrity protection for microservices-based systems. In P100, the authors propose a zero-trust secure design for a Kubernetes-based data center. Zero-trust refers to the concept that requires all users to be authenticated, authorized, and continuously validated before being granted or keeping access to software and data (Kindervag et al, 2010).

**Fault Tolerance (6):** Fault tolerance is the category of peer-reviewed publications that proposes frameworks to increase reliability for Kubernetes. For example in P96, the authors propose a Kubernetes Multi-Master Robust (KMMR) platform to facilitate robust fault tolerance of Kubernetes.

**High Performance Computing (2):** High performance computing (HPC) is the category of peer-reviewed publications that investigates techniques on how to efficiently provision HPC applications on Kubernetes. For example in P68, the authors discuss how Kubernetes can be used to deploy HPC applications. The authors further compare Kubernetes-based deployments with Docker Swarm, and bare metal deployments with respect to memory and network bandwidth. The authors of P68 observe Docker Swarm to outperform Kubernetes.

**Logging & Monitoring (2):** Logging & monitoring is the category of peer-reviewed publications that investigates how logging can integrated in Kubernetes-based deployments. For example, in P30, the author proposes a technique to mitigate challenges related to logging in pods and containers.

**Configuration Abstraction (1):** Configuration abstraction is the category of peer-reviewed publications that investigates how novel configuration abstractions can be conducted for Kubernetes. The only publication belonging to this category is P20, where the authors propose 'Isopod' that directly identifies and abstracts Kubernetes objects using the Kubernetes API instead of using Kubernetes manifests. The authors of P20 reported that YAML-based Kubernetes manifests are untyped, can contain wrong indents, and miss important fields, which necessitates abstractions of Kubernetes objects using the Kubernetes API.

**Database Management (1):** Database management is the category of peer-reviewed publications that investigates how database management tools can be provisioned using Kubernetes. The only publication belonging to this category is P12, where authors propose the Greenplum Database for Kubernetes (GP4K) tool to aid database administrators in automatically deploying databases in Kubernetes.

**Electronic Vehicle (1):** Electronic vehicle is the category of peer-reviewed publications that investigates how Kubernetes can be used to simulate behaviors of electronic vehicles. The only publication belonging to this category is P49, where the authors use Kubernetes to simulate electric vehicle fleet behavior in a distributed manner.

**Discrete Time System Simulation (1):** Discrete time system simulation is the category of peer-reviewed publications that investigates how Kubernetes can be used to simulate discrete time systems. The only publication belonging to this category is P36, where the authors use Kubernetes to simulate a linear multi-variable discrete time system. A discrete-time system is a system that takes a discrete time signal as input and generates a discrete time signal as output (Oppenheim, 1999).

Table 8: Mapping Between Publications and Research Topics

| **Topic** | **Publication Index** | **Count** |
|---|---|---|
| Performance Evaluation | P1, P2, P6, P7, P14, P17, P18, P19, P21, P22, P23, P24, P27, P35, P37, P39, P41, P42, P43, P46, P47, P51, P52, P53, P55, P57, P58, P59, P60, P61, P67, P68, P69, P70, P73, P75, P77, P78, P79, P84, P86, P87, P91, P95, P96, P99 ,P101, P103, P104, P105 | 50 |
| Resource Allocation | P1, P2, P4, P5, P6, P8, P16, P18, P19, P21, P22, P23, P24, P26, P27, P28, P32, P41, P50, P51, P52, P54, P55, P57, P58, P62, P70, P73, P77, P78, P79, P80, P81, P82, P83, P84, P86, P88, P91, P95, P99, P102, P104, P105 | 44 |
| Internet of Things (IoT) | P5, P14, P34, P35, P38, P45, P46, P47, P54, P57, P62, P63, P76, P86, P102, P104 | 16 |
| Networking | P6, P7, P15,P45, P47, P54, P56, P64, P65, P76, P85, P86, P87, P100 | 14 |
| Data Mining & Machine Learning | P9, P11, P14, P18, P25, P51, P71, P79, P80, P93, P98 | 11 |
| Microservice Orchestration | P42, P59, P66, P72, P74, P90, P92, P94 | 8 |
| Security | P3, P33, P47, P66, P92, P93, P100 | 7 |
| Fault tolerance | P10, P17, P44, P59, P96, P97 | 6 |
| High Performance Computing | P68,P71 | 2 |
| Logging & Monitoring | P30, P60 | 2 |
| Configuration Abstraction | P20 | 1 |
| Database Management | P12 | 1 |
| Electronic Vehicle | P49 | 1 |
| Discrete Time System Simulation | P36 | 1 |

#### *4.3.2 Mapping of Identified Benefits and Challenges with Research Publications*

We provide a mapping between the identified benefits and challenges with our set of research publications in the following subsections.

*Mapping of Identified Benefits with Research Publications*: We use this section to describe which of the identified benefits have been discussed by existing research publications. We provide a mapping between identified benefits and identified research topics in Table 9. For each benefit we report if a benefit is mentioned by a publication as a motivation for using Kubernetes, then we list that publication. For example, the benefit 'Deployment of Software & Services' has been reported as a motivation for 55 publications out of 105 (52.4%). The most frequently-reported benefit in our set of 105 publications is 'Deployment of Software & Services'.

Table 9: Mapping Between the Publications and Benefits of Kubernetes

| **Benefit** | **Publication Index** | **Count** |
|---|---|---|
| Deployment of Software & Services | P1, P2, P4, P6, P8, P9, P11, P12, P14, P27, P29, P31, P32, P35, P36, P42, P43, P44, P46, P48, P49, P50, P52, P53, P54, P56, P57, P58, P59, P60, P62, P64, P66, P67, P70, P71, P72, P74, P75, P76, P78, P79, P82, P85, P88, P91, P92, P93, P94, P95, P96, P98, P100, P102, P105 | 55 |
| Service Level Objective (SLO)-based Scalability | P1, P3, P4, P5, P8, P9, P11, P16, P17, P20, P21, P22, P27, P29, P31, P32, P41, P42, P43, P48, P49, P50, P52, P54, P55, P56,P57, P60, P63, P67, P68, P69, P70, P72, P74, P75, P76, P78, P80, P81, P83, P84, P87, P88, P90,P94, P98, P99, P102, P104, P105 | 51 |
| Ease in Cloud-based Interfacing | P1, P2, P4, P5, P6, P7, P8, P11, P12, P16, P20, P21, P23, P25, P27, P28, P29, P31, P32, P33, P34, P36, P37, P40, P41, P42, P44, P45, P46, P47, P48, P52, P56, P57, P61, P62, P66, P67, P80, P82, P93, P94, P95, P96, P100, P101, P102, P103, P104 | 49 |
| Resource Limit Specification | P1, P4, P5, P6, P8, P13, P17, P19, P21, P22, P23, P24, P25, P26, P27, P28, P29, P31, P32, P39, P41, P44, P51, P52, P54, P56, P57, P58, P59, P66, P67, P68, P70, P73, P77, P78, P79, P81, P84, P86, P88, P90, P91, P95, P96, P97, P98, P99, P105 | 49 |
| Availability of Software | P8, P17, P38, P42, P43, P44, P54, P56, P59, P64, P67, P68, P74, P82, P87,P88, P89, P92, P94, P96, P97, P98 | 22 |
| Configuration Management | P1, P4, P6, P9, P12, P14, P18, P23, P38, P43, P45, P48, P49, P50, P56, P71, P91, P92, P93, P94, P96 | 21 |
| Self-healing Containers & Pods | P1, P3, P4, P8, P33, P38, P42, P43, P48, P53, P54, P56, P70, P72, P76, P86, P87 | 17 |
| Community Support | P13, P23, P47, P64, P86, P89 | 6 |

*Mapping of Identified Challenges with Research Publications*: We use this section to describe which of the identified challenges have been discussed by existing research publications that are related with Kubernetes. We provide a mapping between identified challenges and identified research topics in Table 10. For each challenge we report if there is a direct mapping between a research topic and a challenge, then we report the topic name. We also report how many research publications belong to the topic that has a mapping to the identified challenge. For example, the challenge 'Lack of Security Practices & Tools', mentioned in 41.1% of the 321 Internet artifacts, have been addressed by 6.7% of the 105 publications in our set. According to Table 10, the challenges for which no mapping exits with a research topic are: 'Attack Surface Reduction', 'Maintenance-related Challenges', 'Learning Curve', 'Migration Cost', 'Storage', 'Testing', 'Cultural Change', 'Idempotency', and 'Hardware Compatibility'.

Table 10: Mapping Between Identified Challenges and Research Topics

| **Identified Challenge** | **Addressed in Research** |
|---|---|
| Lack of Security Practices & Tools (41.1%) | Security (6.7%) |
| Attack Surface Reduction (39.2%) | None (0.0%) |
| Lack of Diagnostics Tools (37.7%) | Logging & Monitoring (1.9%) |
| Maintenance-related Challenges (33.9%) | None (0.0%) |
| Learning Curve (28.3%) | None (0.0%) |
| Networking (26.2%) | Networking (13.3%) |
| Migration Cost (26.2%) | None (0.0%) |
| Storage (16.2%) | None (0.0%) |
| System Environment Configurations (14.3%) | Configuration Abstraction (0.9%) |
| Testing (11.2%) | None (0.0%) |
| Failure Troubleshooting (9.9%) | Logging & Monitoring (1.9%) |
| Performance (8.1%) | Performance Evaluation (47.6%) |
| Cultural Change (7.5%) | None (0.0%) |
| Idempotency (5.0%) | None (0.0%) |
| Hardware Compatibility (1.2%) | None (0.0%) |

# 5 Discussion

In this section, we discuss the implications of our findings.

## 5.1 Research opportunities in Kubernetes

Our MLR synthesizes the challenges faced by practitioners while using Kubernetes that are not adequately addressed in research. We conjecture that researchers can benefit from our synthesis of reported challenges to identify new research directions, which we describe below:

**Idempotency, Maintenance, Migration, and Testing: Low-hanging research fruits?** According to our MLR, idempotency, maintenance, migration, and testing are domains for which practitioner face challenges. As shown in Table 8 no peer-reviewed publications address any of these topics.

These topics provide researchers to conduct research in these areas for Kubernetes, and pinpoint the uniqueness of Kubernetes. For example, researchers may discover that idempotency in infrastructure as code scripts (Hummer et al, 2013) have similar characteristics to that with Kubernetes. Existing empirical research methodologies related to migration (Lenarduzzi et al, 2020; Olsson et al, 2012) and maintenance (Rahman and Gao, 2015) in DevOps can be applied for Kubernetes to reveal new insights. Furthermore, researchers can investigate the differences between traditional software testing and Kubernetes deployment testing, and derive novel techniques to test Kubernetes deployments automatically.

**Education-related Research:** Learning curve is a frequently-mentioned challenge: in 91 out of 317 artifacts, practitioners mention how difficult it is to learn Kubernetes as new tool. Practitioners find Kubernetes to have a steep learning curve, yet, we do not find any of the collected peer-reviewed publications to address the issue of educating practitioners or students on Kubernetes. Researchers can explore the challenges of educating students or professionals on Kubernetes. Educators can integrate education materials on Kubernetes into computer science curriculum, and share their experiences in form of research reports. Such initiatives can also help future educators on how to design DevOps-related course materials so that the existing workforce shortage related to Kubernetes in industry (VMware, 2020) is mitigated.

**Logging & Monitoring:** We observe a lack of research in the domain of Kubernetes logging and monitoring. According to a survey result (Pemmaraju, 2019) 49% respondents reported logging and monitoring in Kubernetes as their biggest challenge. Practitioner perceptions are supported by our review of Internet artifacts: 121 of the studied 321 Internet artifacts in our MLR mention the challenge of logging and monitoring for Kubernetes. Based on our findings, we conjecture that monitoring and logging is important and challenging for Kubernetes-related software deployments, which necessitates systematic derivation of techniques and tools that will mitigate monitoring-related challenges. As described in Section 4, we only found 2 out of 105 publications (P30, P60) that investigate the topic of logging and monitoring. This suggests monitoring in Kubernetes to be an under-explored area that researchers can take advantage of.

**Security:** We have observed the need of security-related tools and practices for Kubernetes from our MLR. According to the StackRox Survey in 2020, 67% of surveyed 400 IT practitioners have reported security-related misconfigurations that further necessitates of mitigating security weaknesses in Kubernetes (Stackrox, 2021). The software engineering community has started to respond to this need by developing open source tools, such as Skan (alcideio, 2021). We welcome this initiative, and advocate for further development of static and dynamic analysis tools to mitigate security weaknesses in Kubernetes deployments. Of our studied 105 peer-reviewed publications, only 7 publications investigate security issues, suggesting a lack of research in the domain of Kubernetes security, e.g., identifying, detecting, and repairing Kubernetes security misconfiguration categories. Similar initiatives have been pursued in the domain of infrastructure as code (IaC), which has resulted in tools, such as SLIC and SLAC that identifies security weaknesses in IaC scripts. However, configuration specification in Kubernetes manifests is different to that of IaC scripts, namely Ansible, Chef, and Puppet scripts. Therefore, we advocate for novel static analysis tools that can identify security weaknesses, such as security misconfigurations in Kubernetes manifests.

As described in Table 1, our goal of conducting this MLR is not to challenge results derived from practice or research. Rather, our MLR complements existing research related to Kubernetes, by identifying which of the research topics have a mapping with the identified challenges. As discussed in Section 4.3.2, we observe 9 of the 15 challenges to not be addressed by existing research related to Kubernetes. We hope our paper will be helpful for researchers to pursue exciting research problems in the domain of Kubernetes.

## 5.2 Nuanced Perspectives of Practitioners

Our results related to RQ1 and RQ2 provide a nuanced perspective on how practitioners perceive Kubernetes. For example, while ease in cloud-based interfacing is reported as a benefit, practitioners also view Kubernetes to increase the attack surface. As attack surface corresponds to the set of ways in which an adversary can enter the system and potentially cause damage (Theisen et al, 2018), interfacing with cloud-based utilities can increase the attack surface for the software that is deployed using Kubernetes. As another example, practitioners find Kubernetes to be helpful in dealing with configurations that are necessary for container-based software deployments. Yet, they also report that setting up and using Kubernetes necessitates interacting with system configurations that pose challenges. Along with these examples, we notice practitioners to report challenges related to cultural issues, learning curve, and migration costs. Furthermore, anecdotally practitioners have reported that Kubernetes require a lot of investment with respect to time and effort, which may not be worthwhile for an IT organization that do not deploy software and services rapidly.

The implication of this discussion is that practitioners should account the challenges that come with Kubernetes usage. Our findings reported in Sections 4.1 and 4.2 can be a good source to learn from other practitioners who already are using this Kubernetes. Another implication of our discussion is related to research. While we observe well-known IT organizations from Section 4 to benefit from Kubernetes usage, the trade-offs and relationships between process metrics, such as team size and deployment frequency with Kubernetes benefits remain inconclusive. We advocate researchers to conduct empirical studies that will tease out the process metrics that correlate with Kubernetes benefits. Such empirical studies can help practitioners to decide on the tradeoffs related to Kubernetes usage.

### 5.3 The Need for Better Reporting of Research Results

Findings from Table 3 show that on average research related to Kubernetes score low when it comes to the criteria provided by Kitchenham et al. (2012). The implication of this findings is that while presenting research techniques and results, researchers miss important content that helps other researchers as well as practitioners to contextualize their research techniques and results. For example, we observe Kubernetes-related papers to score < 2.0, when discussing and mitigating biases. As another example, the average score for limitationrelated discussion is < 2.0, indicating there is a lack in discussing limitations for publications related to Kubernetes.

While majority of the Kubernetes-related publications scored low based on our publication review criteria, some scored high and could be used as guiding examples of how to better present research related to Kubernetes. For example, P46 scored 3.5 out of 4.0, and received a full score (4.0) for 4 of the 9 quality criteria, including bias discussion. Researchers can read this paper to get suggestions on how Kubernetes-related research can be reported. If researchers want are interested in publications that score well on specific criteria, such as limitation reporting, then they can use Table A2 of our Appendix. For example, P2, which has an average score of 3.2 out of 4.0, scores 4.0 for Q7, which is related to limitation discussion.

We also hope that future research will take the findings of Table 3 into account while reporting research related to Kubernetes. We advocate researchers to follow the recommended practices provided by Kitchenham et al. (2002) to report research results. Existing surveys (Ampatzoglou et al, 2020; Feldt and Magazinius, 2010; Ampatzoglou et al, 2019; Kitchenham et al, 2010) of primary and secondary studies, as well as guidelines reported by other researchers (Runeson and Host, 2009; Tuma et al, 2018) could also be of interest to researchers who plan to engage in Kubernetes-related research.

## 6 Threats to Validity

In this section, we discuss the limitation of our MLR:

– **Conclusion Validity:** We apply a set of inclusion and exclusion criteria for selecting the peer-reviewed publications and Internet artifacts related to Kubernetes using 13 search strings as described in Section 3.2. We acknowledge the fact that the process of selecting these peer-reviewed publications and Internet artifacts can be subjective. Our selection process may miss peer-reviewed publications and Internet artifacts that are related to Kubernetes. To mitigate this limitation, we use two raters who independently search peer-reviewed publications and Internet artifacts.
  Our identified benefits and challenges are limited to the Internet artifacts that we have analyzed. We do not include peer-reviewed publications for identifying benefits and challenges.

We apply open coding to determine identified benefits, challenges, and research topics. We acknowledge that our open coding process can be subjective. We use two raters to mitigate this limitation. Furthermore, for resolution of disagreements we use the last author of the paper as another rater.

- **Internal Validity:** We acknowledge that our search process for collecting Internet artifacts and research publications may not be comprehensive. We use five scholar databases and the Google search engine respectively, to collect our set of peer-reviewed publications and Internet artifacts. Our constructed search strings may not yield all publications related to Kubernetes. Furthermore, the selected scholar databases may not include all Kubernetes-related publications.
- **Construct Validity:** We use raters in our MLR to identify benefits, challenges and research topics. The process can be susceptible to mono-method bias, where a rater's bias can influence the findings. We use two raters to mitigate this threat. In this study, one rater has professional experience of using Kubernetes for one year, and another rater has no prior professional experience in Kubernetes. We acknowledge that the experience of the raters may bias the process. Moreover, we have a resolver with seven years of experience in DevOps research that can also affect the identified benefits, challenges, and research topics.
- **External Validity:** Our MLR is susceptible to external validity because the identified benefits, challenges, and research topics are limited to the collected 105 peer-reviewed publications and 321 Internet artifacts. Findings from our Internet artifacts may not generalize to another grey literature review, which uses another set of Internet artifacts. In such a grey literature review practitioners may discuss certain benefits and challenges that are not included in our paper. Also, with time research trends related to Kubernetes can evolve. As a result, a potential future review of Kubernetesrelated publications can identify research topics that are not included in our paper.

## 7 Conclusion

Kubernetes is a popular container orchestration tool among practitioners for automatically manage and deploy containers. As adoption of Kubernetes steadily increases amongst IT organizations, gaining an understanding of practitionerreported benefits and challenges for Kubernetes could be of relevance to the software engineering research community. A systematic investigation of benefits and challenges of Kubernetes reported by practitioners can help researchers understand the needs of practitioners, and engage in novel research directions. We conduct an MLR with 105 peer-reviewed publications and 321 Internet artifacts that discuss Kubernetes-related concepts. From our MLR we identify 8 benefits and 15 challenges of using Kubernetes reported by practitioners. We also identify 14 topics investigated in peer-reviewed publications. We observe the mostly frequently mentioned benefit of Kubernetes is related to automated software deployment, whereas the most frequently mentioned challenge to be related with unavailability of security best practices and tools. The frequently investigated research topic for Kubernetes is performance evaluation. We also observe a gap between reported challenges and investigated research topics for Kubernetes. For example, even though logging and monitoring has been reported by practitioners as a challenge in 121 of the studied 131 Internet artifacts, only 1.9% of the peer-reviewed publications addresses the topic of logging and monitoring.

Based on our findings, we recommend the software engineering research community to purse research efforts in novel directions, such as idempotency, testing, and education for Kubernetes in order to mitigate practitioner-reported challenges. Furthermore, researchers can use our findings to understand the pain points of current Kubernetes adoptees, and pursue research efforts that will resolve industry's need related to Kubernetes. We hope our MLR study will facilitate further research in Kubernetes.

# A Appendix

Table A1: List of 105 Publications for the Multi-vocal Literature Review

| Index | Publication |
|---|---|
| P1 | Medel, V´ıctor, Omer Rana, Jos´e Angel Ban˜ares, and Unai Arronategui. “Modelling´ performance & resource management in kubernetes.” In Proceedings of the 9th International Conference on Utility and Cloud Computing, pp. 257-262. 2016. |
| P2 | Takahashi, Kimitoshi, Kento Aida, Tomoya Tanjo, Jingtao Sun, and Kazushige Saga. “A Portable Load Balancer with ECMP Redundancy for Container Clusters.” IEICE TRANSACTIONS on Information and Systems 102, no. 5 (2019): 974-987. |
| P3 | Hariri, Sahand, and Matias Carrasco Kind. “Batch and online anomaly detection for scientific applications in a Kubernetes environment.” In Proceedings of the 9th Workshop on Scientific Cloud Computing, pp. 1-7. 2018. |
| P4 | Sarajlic, Semir, Julien Chastang, Suresh Marru, Jeremy Fischer, and Mike Lowe. “Scaling JupyterHub using Kubernetes on Jetstream cloud: Platform as a service for research and educational initiatives in the atmospheric sciences.” In Proceedings of the Practice and Experience on Advanced Research Computing, pp. 1-4. 2018. |
| P5 | Li, Qiankun, Gang Yin, Tao Wang, and Yue Yu. “Building a Cloud-Ready Program: A highly scalable Implementation based on Kubernetes.” In Proceedings of the 2nd International Conference on Advances in Image Processing, pp. 159-164. 2018. |
| P6 | Xu, Cong, Karthick Rajamani, and Wesley Felter. “Nbwguard: Realizing network qos for kubernetes.” In Proceedings of the 19th International Middleware Conference Industry, pp. 32-38. 2018. |
| P7 | Liu, Haifeng, Shugang Chen, Yongcheng Bao, Wanli Yang, Yuan Chen, Wei Ding, and Huasong Shan. “A High Performance, Scalable DNS Service for Very Large Scale Container Cloud Platforms.” In Proceedings of the 19th International Middleware Conference Industry, pp. 39-45. 2018. |
| P8 | Wei-guo, Zhang, Ma Xi-lin, and Zhang Jin-zhong. “Research on Kubernetes’ Resource Scheduling Scheme.” In Proceedings of the 8th International Conference on Communication and Network Security, pp. 144-148. 2018. |
| P9 | Zhuang, Jinfeng, and Yu Liu. “PinText: A Multitask Text Embedding System in Pinterest.” In Proceedings of the 25th ACM SIGKDD International Conference on Knowledge Discovery & Data Mining, pp. 2653-2661. 2019. |
| P10 | Tu, Tengfei, Xiaoyu Liu, Linhai Song, and Yiying Zhang. “Understanding real-world concurrency bugs in Go.” In Proceedings of the Twenty-Fourth International Conference on Architectural Support for Programming Languages and Operating Systems, pp. 865-878. 2019. |
| P11 | Govind, Yash, Pradap Konda, Paul Suganthan GC, Philip Martinkus, Palaniappan Nagarajan, Han Li, Aravind Soundararajan et al. “Entity matching meets data science: A progress report from the magellan project.” In Proceedings of the 2019 International Conference on Management of Data, pp. 389-403. 2019. |
| P12 | Patel, Jemish, Goutam Tadi, Oz Basarir, Lawrence Hamel, David Sharp, Fei Yang, and Xin Zhang. “Pivotal Greenplum© for Kubernetes: Demonstration of Managing Greenplum Database on Kubernetes.” In Proceedings of the 2019 International Conference on Management of Data, pp. 1969-1972. 2019. |
| P13 | Carcassi, Gabriele, Joe Breen, Lincoln Bryant, Robert W. Gardner, Shawn Mckee, and Christopher Weaver. “SLATE: Monitoring Distributed Kubernetes Clusters.” In Practice and Experience in Advanced Research Computing, pp. 19-25. 2020. |
| P14 | Huang, Yuzhou, Kaiyu cai, Ran Zong, and Yugang Mao. “Design and implementation of an edge computing platform architecture using docker and kubernetes for machine learning.” In Proceedings of the 3rd International Conference on High Performance Compilation, Computing and Communications, pp. 29-32. 2019. |

**Table A1 – continued from previous page**

| | |
|---|---|
| P15 | Kouchaksaraei, Hadi Razzaghi, and Holger Karl. "Service function chaining across openstack and kubernetes domains." In Proceedings of the 13th ACM International Conference on Distributed and Event-based Systems, pp. 240-243. 2019. |
| | Continued on next page |

| Index | Publication |
|---|---|
| P16 | Thurgood, Brandon, and Ruth G. Lennon. "Cloud computing with Kubernetes cluster elastic scaling." In Proceedings of the 3rd International Conference on Future Networks and Distributed Systems, pp. 1-7. 2019. |
| P17 | Ambati, Pradeep, and David Irwin. "Optimizing the cost of executing mixed interactive and batch workloads on transient vms." Proceedings of the ACM on Measurement and Analysis of Computing Systems 3, no. 2 (2019): 1-24. |
| P18 | Le, Tan N., Xiao Sun, Mosharaf Chowdhury, and Zhenhua Liu. "AlloX: compute allocation in hybrid clusters." In Proceedings of the Fifteenth European Conference on Computer Systems, pp. 1-16. 2020. |
| P19 | Liu, Yang, Huanle Xu, and Wing Cheong Lau. "Accordia: Adaptive cloud configuration optimization for recurring data-intensive applications." In Proceedings of the ACM Symposium on Cloud Computing, pp. 479-479. 2019. |
| P20 | Xu, Charles, and Dmitry Ilyevskiy. "Isopod: An Expressive DSL for Kubernetes Configuration." In Proceedings of the ACM Symposium on Cloud Computing, pp. 483483. 2019. |
| P21 | Kaminski, Matthijs, Eddy Truyen, Emad Heydari Beni, Bert Lagaisse, and Wouter Joosen. "A framework for black-box SLO tuning of multi-tenant applications in Kubernetes." In Proceedings of the 5th International Workshop on Container Technologies and Container Clouds, pp. 7-12. 2019. |
| P22 | Verreydt, Stef, Emad Heydari Beni, Eddy Truyen, Bert Lagaisse, and Wouter Joosen. "Leveraging Kubernetes for adaptive and cost-efficient resource management." In Proceedings of the 5th International Workshop on Container Technologies and Container Clouds, pp. 37-42. 2019. |
| P23 | Yeh, Ting-An, Hung-Hsin Chen, and Jerry Chou. "KubeShare: A Framework to Manage GPUs as First-Class and Shared Resources in Container Cloud." In Proceedings of the 29th International Symposium on High-Performance Parallel and Distributed Computing, pp. 173-184. 2020. |
| P24 | Zhong, Zhiheng, and Rajkumar Buyya. "A Cost-Efficient Container Orchestration Strategy in Kubernetes-Based Cloud Computing Infrastructures with Heterogeneous Resources." ACM Transactions on Internet Technology (TOIT) 20, no. 2 (2020): 1-24. |
| P25 | Lee, Chun-Hsiang, Zhaofeng Li, Xu Lu, Tiyun Chen, Saisai Yang, and Chao Wu. "Multi-Tenant Machine Learning Platform Based on Kubernetes." In Proceedings of the 2020 6th International Conference on Computing and Artificial Intelligence, pp. 5-12. 2020. |
| P26 | Alimudin, Akhmad, and Yoshiteru Ishida. "Service-Based Container Deployment on Kubernetes Using Stable Marriage Problem." In Proceedings of the 2020 The 6th International Conference on Frontiers of Educational Technologies, pp. 164-167. 2020. |
| P27 | Li, Dong, Yi Wei, and Bing Zeng. "A Dynamic I/O Sensing Scheduling Scheme in Kubernetes." In Proceedings of the 2020 4th International Conference on High Performance Compilation, Computing and Communications, pp. 14-19. 2020. |
| P28 | Fan, Dayong, and Dongzhi He. "A Scheduler for Serverless Framework base on Kubernetes." In Proceedings of the 2020 4th High Performance Computing and Cluster Technologies Conference & 2020 3rd International Conference on Big Data and Artificial Intelligence, pp. 229-232. 2020. |

**Table A1 – continued from previous page**

| Index | Publication |
|---|---|
| P29 | Burns, Brendan, Brian Grant, David Oppenheimer, Eric Brewer, and John Wilkes. “Borg, Omega, and Kubernetes: Lessons learned from three container-management systems over a decade.” Queue 14, no. 1 (2016): 70-93. |
| P30 | Singh, Satnam. “Cluster-level Logging of Containers with Containers: Logging Challenges of Container-Based Cloud Deployments.” Queue 14, no. 3 (2016): 83-106. |
| P31 | Bernstein, David. “Containers and cloud: From lxc to docker to kubernetes.” IEEE Cloud Computing 1, no. 3 (2014): 81-84. |
| P32 | Medel, V´ıctor, Omer Rana, Jos´e Angel Ban˜ares, and Unai Arronategui. “Adaptive´ application scheduling under interference in kubernetes.” In 2016 IEEE/ACM 9th International Conference on Utility and Cloud Computing (UCC), pp. 426-427. IEEE, 2016. |

Continued on next page

| Index | Publication |
|---|---|
| P33 | Bila, Nilton, Paolo Dettori, Ali Kanso, Yuji Watanabe, and Alaa Youssef. “Leveraging the serverless architecture for securing linux containers.” In 2017 IEEE 37th International Conference on Distributed Computing Systems Workshops (ICDCSW), pp. 401-404. IEEE, 2017. |
| P34 | Dupont, Corentin, Raffaele Giaffreda, and Luca Capra. “Edge computing in IoT context: Horizontal and vertical Linux container migration.” In 2017 Global Internet of Things Summit (GIoTS), pp. 1-4. IEEE, 2017. |
| P35 | Tsai, Pei-Hsuan, Hua-Jun Hong, An-Chieh Cheng, and Cheng-Hsin Hsu. “Distributed analytics in fog computing platforms using tensorflow and kubernetes.” In 2017 19th Asia-Pacific Network Operations and Management Symposium (APNOMS), pp. 145150. IEEE, 2017. |
| P36 | Sima, Vasile, Alexandru Stanciu, and Florin Hartescu. “New software applications for system identification.” In 2017 21st International Conference on System Theory, Control and Computing (ICSTCC), pp. 106-111. IEEE, 2017. |
| P37 | Coullon, H´el`ene, Christian Perez, and Dimitri Pertin. “Production deployment tools for IaaSes: an overall model and survey.” In 2017 IEEE 5th International Conference on Future Internet of Things and Cloud (FiCloud), pp. 183-190. IEEE, 2017. |
| P38 | Javed, Asad, Keijo Heljanko, Andrea Buda, and Kary Fr¨amling. “Cefiot: A faulttolerant iot architecture for edge and cloud.” In 2018 IEEE 4th world forum on internet of things (WF-IoT), pp. 813-818. IEEE, 2018. |
| P39 | Tosh, Deepak, Sachin Shetty, Peter Foytik, Charles Kamhoua, and Laurent Njilla. “CloudPoS: A proof-of-stake consensus design for blockchain integrated cloud.” In 2018 IEEE 11th International Conference on Cloud Computing (CLOUD), pp. 302309. IEEE, 2018. |
| P40 | Herger, Lorraine M., Mercy Bodarky, and Carlos Fonseca. “Breaking down the barriers for moving an enterprise to cloud.” In 2018 IEEE 11th International Conference on Cloud Computing (CLOUD), pp. 572-576. IEEE, 2018. |
| P41 | Podolskiy, Vladimir, Anshul Jindal, and Michael Gerndt. “Iaas reactive autoscaling performance challenges.” In 2018 IEEE 11th International Conference on Cloud Computing (CLOUD), pp. 954-957. IEEE, 2018. |
| P42 | Vayghan, Leila Abdollahi, Mohamed Aymen Saied, Maria Toeroe, and Ferhat Khendek. “Deploying microservice based applications with Kubernetes: experiments and lessons learned.” In 2018 IEEE 11th international conference on cloud computing (CLOUD), pp. 970-973. IEEE, 2018. |
| P43 | Modak, Arsh, S. D. Chaudhary, P. S. Paygude, and S. R. Ldate. “Techniques to secure data on cloud: Docker swarm or kubernetes?.” In 2018 Second International Conference on Inventive Communication and Computational Technologies (ICICCT), pp. 7-12. IEEE, 2018. |

**Table A1 – continued from previous page**

| | |
|---|---|
| P44 | Netto, Hylson Vescovi, Aldelir Fernando Luiz, Miguel Correia, Luciana de Oliveira Rech, and Caio Pereira Oliveira. "Koordinator: A service approach for replicating Docker containers in Kubernetes." In 2018 IEEE Symposium on Computers and Communications (ISCC), pp. 00058-00063. IEEE, 2018. |
| P45 | Xiong, Ying, Yulin Sun, Li Xing, and Ying Huang. "Extend cloud to edge with kubeedge." In 2018 IEEE/ACM Symposium on Edge Computing (SEC), pp. 373377. IEEE, 2018. |
| P46 | Aly, Mohab, Foutse Khomh, and Soumaya Yacout. "Kubernetes or openShift? Which technology best suits eclipse hono IoT deployments." In 2018 IEEE 11th Conference on Service-Oriented Computing and Applications (SOCA), pp. 113-120. IEEE, 2018.. |
| P47 | Brito, Andrey, Christof Fetzer, Stefan Ko¨psell, Marcelo Pasin, Pascal Felber, Keiko Fonseca, Marcelo Rosa et al. "Cloud challenge: Secure end-to-end processing of smart metering data." In 2018 IEEE/ACM International Conference on Utility and Cloud Computing Companion (UCC Companion), pp. 36-42. IEEE, 2018. |
| P48 | Shah, Jay, and Dushyant Dubaria. "Building modern clouds: using docker, kubernetes & Google cloud platform." In 2019 IEEE 9th Annual Computing and Communication Workshop and Conference (CCWC), pp. 0184-0189. IEEE, 2019. |

Continued on next page

| Index | Publication |
|---|---|
| P49 | Rehman, Kasim, Orthodoxos Kipouridis, Stamatis Karnouskos, Oliver Frendo, Helge Dickel, Jonas Lipps, and Nemrude Verzano. "A cloud-based development environment using hla and kubernetes for the co-simulation of a corporate electric vehicle fleet." In 2019 IEEE/SICE International Symposium on System Integration (SII), pp. 47-54. IEEE, 2019. |
| P50 | Townend, Paul, Stephen Clement, Dan Burdett, Renyu Yang, Joe Shaw, Brad Slater, and Jie Xu. "Improving data center efficiency through holistic scheduling in kubernetes." In 2019 IEEE International Conference on Service-Oriented System Engineering (SOSE), pp. 156-15610. IEEE, 2019. |
| P51 | Bao, Yixin, Yanghua Peng, and Chuan Wu. "Deep learning-based job placement in distributed machine learning clusters." In IEEE INFOCOM 2019-IEEE conference on computer communications, pp. 505-513. IEEE, 2019. |
| P52 | Podolskiy, Vladimir, Michael Mayo, Abigail Koay, Michael Gerndt, and Panos Patros. "Maintaining SLOs of cloud-native applications via self-adaptive resource sharing." In 2019 IEEE 13th International Conference on Self-Adaptive and Self-Organizing Systems (SASO), pp. 72-81. IEEE, 2019. |
| P53 | Chiba, Tatsuhiro, Rina Nakazawa, Hiroshi Horii, Sahil Suneja, and Seetharami Seelam. "Confadvisor: A performance-centric configuration tuning framework for containers on kubernetes." In 2019 IEEE International Conference on Cloud Engineering (IC2E), pp. 168-178. IEEE, 2019. |
| P54 | Santos, Jose, Tim Wauters, Bruno Volckaert, and Filip De Turck. "Towards networkaware resource provisioning in Kubernetes for fog computing applications." In 2019 IEEE Conference on Network Softwarization (NetSoft), pp. 351-359. IEEE, 2019. |
| P55 | Rattihalli, Gourav, Madhusudhan Govindaraju, Hui Lu, and Devesh Tiwari. "Exploring potential for non-disruptive vertical auto scaling and resource estimation in kubernetes." In 2019 IEEE 12th International Conference on Cloud Computing (CLOUD), pp. 33-40. IEEE, 2019. |
| P56 | Gawel, Maciej, and Krzysztof Zielinski. "Analysis and evaluation of kubernetes based nfv management and orchestration." In 2019 IEEE 12th International Conference on Cloud Computing (CLOUD), pp. 511-513. IEEE, 2019. |

**Table A1 – continued from previous page**

| | |
|---|---|
| P57 | Kaur, Kuljeet, Sahil Garg, Georges Kaddoum, Syed Hassan Ahmed, and Mohammed Atiquzzaman. "Keids: Kubernetes-based energy and interference driven scheduler for industrial iot in edge-cloud ecosystem." IEEE Internet of Things Journal 7, no. 5 (2019): 4228-4237. |
| P58 | Kelley, Jaimie, and Nathaniel Morris. "Rapid In-situ Profiling of Colocated Workloads." In IEEE INFOCOM 2019-IEEE Conference on Computer Communications Workshops (INFOCOM WKSHPS), pp. 528-534. IEEE, 2019. |
| P59 | Vayghan, Leila Abdollahi, Mohamed Aymen Saied, Maria Toeroe, and Ferhat Khendek. "Microservice based architecture: Towards high-availability for stateful applications with Kubernetes." In 2019 IEEE 19th International Conference on Software Quality, Reliability and Security (QRS), pp. 176-185. IEEE, 2019. |
| P60 | Astyrakakis, Nikolaos, Yannis Nikoloudakis, Ioannis Kefaloukos, Charalabos Skianis, Evangelos Pallis, and Evangelos K. Markakis. "Cloud-Native Application Validation & Stress Testing through a Framework for Auto-Cluster Deployment." In 2019 IEEE 24th International Workshop on Computer Aided Modeling and Design of Communication Links and Networks (CAMAD), pp. 1-5. IEEE, 2019. |
| P61 | Marathe, Nikhil, Ankita Gandhi, and Jaimeel M. Shah. "Docker swarm and kubernetes in cloud computing environment." In 2019 3rd International Conference on Trends in Electronics and Informatics (ICOEI), pp. 179-184. IEEE, 2019. |
| P62 | Casquero, Oskar, Aintzane Armentia, Isabel Sarachaga, Federico P´erez, Dar´ıo Orive, and Marga Marcos. "Distributed scheduling in Kubernetes based on MAS for Fogin-the-loop applications." In 2019 24th IEEE International Conference on Emerging Technologies and Factory Automation (ETFA), pp. 1213-1217. IEEE, 2019. |

Continued on next page

| Index | Publication |
|---|---|
| P63 | Chen, Hung-Li, and Fuchun Joseph Lin. "Scalable IoT/M2M platforms based on kubernetes-enabled NFV MANO architecture." In 2019 International Conference on Internet of Things (iThings) and IEEE Green Computing and Communications (GreenCom) and IEEE Cyber, Physical and Social Computing (CPSCom) and IEEE Smart Data (SmartData), pp. 1106-1111. IEEE, 2019. |
| P64 | Link, Coleman, Jesse Sarran, Garegin Grigoryan, Minseok Kwon, M. Mustafa Rafique, and Warren R. Carithers. "Container Orchestration by Kubernetes for RDMA Networking." In 2019 IEEE 27th International Conference on Network Protocols (ICNP), pp. 1-2. IEEE, 2019. |
| P65 | Bringhenti, Daniele, Guido Marchetto, Riccardo Sisto, Fulvio Valenza, and Jalolliddin Yusupov. "Towards a fully automated and optimized network security functions orchestration." In 2019 4th International Conference on Computing, Communications and Security (ICCCS), pp. 1-7. IEEE, 2019. |
| P66 | Hussain, Fatima, Weiyue Li, Brett Noye, Salah Sharieh, and Alexander Ferworn. "Intelligent Service Mesh Framework for API Security and Management." In 2019 IEEE 10th Annual Information Technology, Electronics and Mobile Communication Conference (IEMCON), pp. 0735-0742. IEEE, 2019. |
| P67 | Pan, Yao, Ian Chen, Francisco Brasileiro, Glenn Jayaputera, and Richard Sinnott. "A performance comparison of cloud-based container orchestration tools." In 2019 IEEE International Conference on Big Knowledge (ICBK), pp. 191-198. IEEE, 2019. |
| P68 | Beltre, Angel M., Pankaj Saha, Madhusudhan Govindaraju, Andrew Younge, and Ryan E. Grant. "Enabling HPC workloads on cloud infrastructure using Kubernetes container orchestration mechanisms." In 2019 IEEE/ACM International Workshop on Containers and New Orchestration Paradigms for Isolated Environments in HPC (CANOPIE-HPC), pp. 11-20. IEEE, 2019. |

**Table A1 – continued from previous page**

| Index | Publication |
|---|---|
| P69 | Ferreira, Arnaldo Pereira, and Richard Sinnott. "A performance evaluation of containers running on managed kubernetes services." In 2019 IEEE International Conference on Cloud Computing Technology and Science (CloudCom), pp. 199-208. IEEE Computer Society, 2019. |
| P70 | Wu, Qiang, Jiadi Yu, Li Lu, Shiyou Qian, and Guangtao Xue. "Dynamically adjusting scale of a kubernetes cluster under QoS guarantee." In 2019 IEEE 25th International Conference on Parallel and Distributed Systems (ICPADS), pp. 193-200. IEEE, 2019. |
| P71 | Tesliuk, Anton, Sergey Bobkov, Viacheslav Ilyin, Alexander Novikov, Alexey Poyda, and Vasily Velikhov. "Kubernetes container orchestration as a framework for flexible and effective scientific data analysis." In 2019 Ivannikov Ispras Open Conference (ISPRAS), pp. 67-71. IEEE, 2019. |
| P72 | De Iasio, Antonio, and Eugenio Zimeo. "Avoiding Faults due to Dangling Dependencies by Synchronization in Microservices Applications." In 2019 IEEE International Symposium on Software Reliability Engineering Workshops (ISSREW), pp. 169-176. IEEE, 2019. |
| P73 | Fu, Yuqi, Shaolun Zhang, Jose Terrero, Ying Mao, Guangya Liu, Sheng Li, and Dingwen Tao. "Progress-based container scheduling for short-lived applications in a kubernetes cluster." In 2019 IEEE International Conference on Big Data (Big Data), pp. 278-287. IEEE, 2019. |
| P74 | Rajavaram, Harika, Vineet Rajula, and B. Thangaraju. "Automation of Microservices Application Deployment Made Easy By Rundeck and Kubernetes." In 2019 IEEE International Conference on Electronics, Computing and Communication Technologies (CONECCT), pp. 1-3. IEEE, 2019. |
| P75 | Dewi, Lily Puspa, Agustinus Noertjahyana, Henry Novianus Palit, and Kezia Yedutun. "Server Scalability Using Kubernetes." In 2019 4th Technology Innovation Management and Engineering Science International Conference (TIMES-iCON), pp. 1-4. IEEE, 2019. |

Continued on next page

| Index | Publication |
|---|---|
| P76 | Schneider, Stefan, Manuel Peuster, Kai Hannemann, Daniel Behnke, Marcel Muller, Patrick-Benjamin B¨ok, and Holger Karl. ""Producing Cloud-Native": Smart Manufacturing Use Cases on Kubernetes." In 2019 IEEE Conference on Network Function Virtualization and Software Defined Networks (NFV-SDN), pp. 1-2. IEEE, 2019. |
| P77 | Beltre, Angel, Pankaj Saha, and Madhusudhan Govindaraju. "Kubesphere: An approach to multi-tenant fair scheduling for kubernetes clusters." In 2019 IEEE Cloud Summit, pp. 14-20. IEEE, 2019. |
| P78 | Wang, Mingming, Dongmei Zhang, and Bin Wu. "A Cluster Autoscaler Based on Multiple Node Types in Kubernetes." In 2020 IEEE 4th Information Technology, Networking, Electronic and Automation Control Conference (ITNEC), vol. 1, pp. 575-579. IEEE, 2020. |
| P79 | Surya, Rahmad Yesa, and Achmad Imam Kistijantoro. "Dynamic Resource Allocation for Distributed TensorFlow Training in Kubernetes Cluster." In 2019 International Conference on Data and Software Engineering (ICoDSE), pp. 1-6. IEEE, 2019. |
| P80 | Huang, Jiaming, Chuming Xiao, and Weigang Wu. "RLSK: A Job Scheduler for Federated Kubernetes Clusters based on Reinforcement Learning." In 2020 IEEE International Conference on Cloud Engineering (IC2E), pp. 116-123. IEEE, 2020. |
| P81 | Balla, David, Csaba Simon, and Markosz Maliosz. "Adaptive scaling of Kubernetes pods." In NOMS 2020-2020 IEEE/IFIP Network Operations and Management Symposium, pp. 1-5. IEEE, 2020. |
| P82 | Liu, Qingyang, E. Haihong, and Meina Song. "The Design of Multi-Metric Load Balancer for Kubernetes." In 2020 International Conference on Inventive Computation Technologies (ICICT), pp. 1114-1117. IEEE, 2020. |

**Table A1 – continued from previous page**

| Index | Publication |
|---|---|
| P83 | Donca, Ionut-Catalin, Cosmina Corches, Ovidiu Stan, and Liviu Miclea. "Autoscaled RabbitMQ Kubernetes Cluster on single-board computers." In 2020 IEEE International Conference on Automation, Quality and Testing, Robotics (AQTR), pp. 1-6. IEEE, 2020. |
| P84 | Donca, Ionut-Catalin, Cosmina Corches, Ovidiu Stan, and Liviu Miclea. "Autoscaled RabbitMQ Kubernetes Cluster on single-board computers." In 2020 IEEE International Conference on Automation, Quality and Testing, Robotics (AQTR), pp. 1-6. IEEE, 2020. |
| P85 | Botez, Robert, Calin-Marian Iurian, Iustin-Alexandru Ivanciu, and Virgil Dobrota. "Deploying a Dockerized Application With Kubernetes on Google Cloud Platform." In 2020 13th International Conference on Communications (COMM), pp. 471-476. IEEE, 2020. |
| P86 | Eidenbenz, Raphael, Yvonne-Anne Pignolet, and Alain Ryser. "Latency-Aware Industrial Fog Application Orchestration with Kubernetes." In 2020 Fifth International Conference on Fog and Mobile Edge Computing (FMEC), pp. 164-171. IEEE, 2020. |
| P87 | Qi, Shixiong, Sameer G. Kulkarni, and K. K. Ramakrishnan. "Understanding container network interface plugins: design considerations and performance." In 2020 IEEE International Symposium on Local and Metropolitan Area Networks (LANMAN, pp. 1-6. IEEE, 2020. |
| P88 | Nguyen, Nguyen, and Taehong Kim. "Toward Highly Scalable Load Balancing in Kubernetes Clusters." IEEE Communications Magazine 58, no. 7 (2020): 78-83. |
| P89 | Muddinagiri, Ruchika, Shubham Ambavane, and Simran Bayas. "Self-Hosted Kubernetes: Deploying Docker Containers Locally With Minikube." In 2019 International Conference on Innovative Trends and Advances in Engineering and Technology (ICITAET), pp. 239-243. IEEE, 2019. |
| P90 | Rossi, Fabiana, Valeria Cardellini, and Francesco Lo Presti. "Hierarchical scaling of microservices in Kubernetes." In 2020 IEEE International Conference on Autonomic Computing and Self-Organizing Systems (ACSOS), pp. 28-37. IEEE, 2020. |
| P91 | Guerrero, Carlos, Isaac Lera, and Carlos Juiz. "Genetic algorithm for multi-objective optimization of container allocation in cloud architecture." Journal of Grid Computing 16, no. 1 (2018): 113-135. |
| P92 | Ahmadvand, Mohsen, Alexander Pretschner, Keith Ball, and Daniel Eyring. "Integrity protection against insiders in microservice-based infrastructures: From threats to a security framework." In Federation of International Conferences on Software Technologies: Applications and Foundations, pp. 573-588. Springer, Cham, 2018. |
| P93 | Tien, Chin-Wei, Tse-Yung Huang, Chia-Wei Tien, Ting-Chun Huang, and Sy-Yen Kuo. "KubAnomaly: Anomaly detection for the Docker orchestration platform with neural network approaches." Engineering Reports 1, no. 5 (2019): e12080. |
| P94 | Bogo, Matteo, Jacopo Soldani, Davide Neri, and Antonio Brogi. "Component-aware orchestration of cloud-based enterprise applications, from TOSCA to Docker and Kubernetes." Software: Practice and Experience 50, no. 9 (2020): 1793-1821. |
| P95 | Medel, V´ıctor, Rafael Tolosana-Calasanz, Jos´e Angel Ban˜ares, Unai Arronategui, and´ Omer F. Rana. "Characterising resource management performance in Kubernetes." Computers & Electrical Engineering 68 (2018): 286-297. |
| P96 | Diouf, Gor Mack, Halima Elbiaze, and Wael Jaafar. "On Byzantine fault tolerance in multi-master Kubernetes clusters." Future Generation Computer Systems 109 (2020): 407-419. |
| P97 | Netto, Hylson V., Lau Cheuk Lung, Miguel Correia, Aldelir Fernando Luiz, and Luciana Moreira S´a de Souza. "State machine replication in containers managed by Kubernetes." Journal of Systems Architecture 73 (2017): 53-59. |

Continued on next page

**Table A1 – continued from previous page**

| | |
|---|---|
| P98 | Araya, Mauricio, Maximiliano Osorio, Mat´ıas D´ıaz, Carlos Ponce, Mart´ın Villanueva, Camilo Valenzuela, and Mauricio Solar. "JOVIAL: Notebook-based astronomical data analysis in the cloud." Astronomy and computing 25 (2018): 110-117. |
| P99 | Christodoulopoulos, Christos, and Euripides GM Petrakis. "Commodore: fail safe container scheduling in kubernetes." In International Conference on Advanced Information Networking and Applications, pp. 988-999. Springer, Cham, 2019. |
| P100 | Surantha, Nico, and Felix Ivan. "Secure kubernetes networking design based on zero trust model: A case study of financial service enterprise in indonesia." In International Conference on Innovative Mobile and Internet Services in Ubiquitous Computing, pp. 348-361. Springer, Cham, 2019. |
| P101 | Mercl, Lubos, and Jakub Pavlik. "Public Cloud Kubernetes Storage Performance Analysis." In International Conference on Computational Collective Intelligence, pp. 649-660. Springer, Cham, 2019. |
| P102 | Goethals, Tom, Filip De Turck, and Bruno Volckaert. "Fledge: Kubernetes compatible container orchestration on low-resource edge devices." In International Conference on Internet of Vehicles, pp. 174-189. Springer, Cham, 2019. |
| P103 | Kratzke, Nane. "About the complexity to transfer cloud applications at runtime and how container platforms can contribute?." In International Conference on Cloud Computing and Services Science, pp. 19-45. Springer, Cham, 2017. |
| P104 | Zheng, Wei-Sheng, and Li-Hsing Yen. "Auto-scaling in Kubernetes-based fog computing platform." In International Computer Symposium, pp. 338-345. Springer, Singapore, 2018. |
| P105 | Hu, Yang, Huan Zhou, Cees de Laat, and Zhiming Zhao. "Concurrent container scheduling on heterogeneous clusters with multi-resource constraints." Future Generation Computer Systems 102 (2020): 562-573. |

| | | | | | | | | | | |
|---|---|---|---|---|---|---|---|---|---|---|
| P1 | 3 | 3 | 4 | 3.5 | 3 | 1 | 1.5 | 2.5 | 3.5 | 2.78 |
| P2 | 3.5 | 3 | 4 | 3.5 | 2.5 | 1.5 | 4 | 4 | 3 | 3.22 |
| P3 | 3.5 | 2.5 | 2.5 | 2 | 2.5 | 1 | 1.5 | 3.5 | 2.5 | 2.39 |
| P4 | 4 | 2.5 | 2.5 | 3 | 2 | 2.5 | 2.5 | 2.5 | 3 | 2.72 |

**Table A1 – continued from previous page**

Continued on next page

Table A2: Quality Assessments of Each of the 105 Publications for Multi-vocal Literature Review

| PI | Q1 | Q2 | Q3 | Q4 | Q5 | Q6 | Q7 | Q8 | Q9 | Avg. |
|---|---|---|---|---|---|---|---|---|---|---|

**Table A2 – continued from previous page**

| Index | Q1 | Q2 | Q3 | Q4 | Q5 | Q6 | Q7 | Q8 | Q9 | Avg. |
|---|---|---|---|---|---|---|---|---|---|---|
| P5 | 2.5 | 2.5 | 2.5 | 2 | 2 | 1 | 1.5 | 2.5 | 2.5 | 2.11 |
| P6 | 3.5 | 2.5 | 3 | 2 | 2.5 | 1 | 1.5 | 2.5 | 3 | 2.39 |
| P7 | 3 | 2 | 3 | 2.5 | 3 | 1 | 1.5 | 3 | 2.5 | 2.39 |
| P8 | 3 | 2 | 3 | 2.5 | 3 | 1 | 1.5 | 2.5 | 2.5 | 2.33 |
| P9 | 4 | 3.5 | 3.5 | 2.5 | 2.5 | 1.5 | 2.5 | 3 | 3 | 2.89 |
| P10 | 3.5 | 3 | 3.5 | 4 | 3.5 | 1.5 | 3.5 | 4 | 3 | 3.28 |
| P11 | 3.5 | 3 | 3.5 | 3 | 3 | 1 | 1.5 | 3.5 | 3 | 2.78 |
| P12 | 3.5 | 2 | 3.5 | 2 | 1.5 | 1 | 1 | 2.5 | 2.5 | 2.17 |
| P13 | 4 | 2 | 3.5 | 2.5 | 2.5 | 1.5 | 3 | 2.5 | 3.5 | 2.78 |
| P14 | 2.5 | 2 | 3.5 | 2 | 1.5 | 1 | 2 | 1.5 | 2.5 | 2.06 |
| P15 | 3 | 1.5 | 3 | 1 | 1.5 | 1 | 1 | 1.5 | 3 | 1.83 |
| P16 | 2.5 | 2 | 3.5 | 2 | 2 | 1 | 1.5 | 2 | 2.5 | 2.11 |
| P17 | 3.5 | 4 | 4 | 3.5 | 4 | 1 | 1.5 | 4 | 4 | 3.28 |
| P18 | 3 | 3 | 3 | 2 | 2.5 | 1.5 | 2 | 3.5 | 3 | 2.61 |
| P19 | 3 | 2 | 3 | 2.5 | 2.5 | 1 | 1 | 3 | 2.5 | 2.28 |
| P20 | 3 | 1 | 1 | 1 | 1 | 1 | 1 | 2.5 | 3.5 | 1.67 |
| P21 | 3 | 2.5 | 3 | 2 | 2.5 | 1.5 | 1.5 | 2.5 | 2.5 | 2.33 |
| P22 | 4 | 3 | 4 | 2.5 | 2.5 | 2 | 3.5 | 3 | 3 | 3.06 |
| P23 | 3.5 | 3 | 4 | 2.5 | 3 | 1 | 2 | 3 | 4 | 2.89 |
| P24 | 3.5 | 3.5 | 3 | 3.5 | 3 | 2.5 | 3 | 3 | 3 | 3.11 |
| P25 | 3.5 | 3 | 3 | 3 | 2 | 1 | 1.5 | 2 | 3 | 2.44 |
| P26 | 3 | 2.5 | 2.5 | 2 | 1.5 | 1.5 | 1.5 | 2.5 | 2 | 2.11 |
| P27 | 3 | 3 | 3 | 2.5 | 2.5 | 1 | 1.5 | 2 | 2.5 | 2.33 |
| P28 | 3 | 3 | 3 | 2.5 | 2.5 | 1 | 2 | 3 | 2.5 | 2.5 |
| P29 | 3 | 1 | 1 | 2 | 1.5 | 1 | 1.5 | 3 | 3 | 1.89 |
| P30 | 3 | 2 | 2.5 | 3.5 | 3 | 1 | 2 | 2 | 3.5 | 2.5 |
| P31 | 1 | 1.5 | 1.5 | 1 | 1 | 1 | 1.5 | 2 | 2 | 1.39 |
| P32 | 3 | 2 | 2.5 | 2.5 | 1.5 | 1 | 1.5 | 1 | 2.5 | 1.94 |
| P33 | 3 | 2.5 | 3.5 | 2 | 1.5 | 1 | 1.5 | 1 | 2.5 | 2.06 |
| P34 | 3 | 2 | 3 | 2.5 | 1.5 | 1 | 2 | 1 | 2.5 | 2.06 |
| P35 | 2.5 | 3.5 | 3 | 3 | 2.5 | 1.5 | 3.5 | 3.5 | 3.5 | 2.94 |
| P36 | 1.5 | 2.5 | 2 | 2.5 | 2.5 | 1 | 1.5 | 2 | 2.5 | 2 |
| P37 | 3 | 2.5 | 2 | 2 | 2.5 | 1.5 | 1.5 | 2.5 | 3.5 | 2.33 |
| P38 | 3 | 2.5 | 3.5 | 2.5 | 2 | 1 | 2 | 2.5 | 2.5 | 2.39 |
| P39 | 3 | 3.5 | 3.5 | 2 | 2 | 2.5 | 2 | 2.5 | 3 | 2.67 |
| P40 | 3 | 1.5 | 1.5 | 1.5 | 1.5 | 1 | 1.5 | 2 | 2.5 | 1.78 |
| P41 | 3.5 | 2.5 | 3 | 2.5 | 2.5 | 1 | 2 | 3 | 3 | 2.56 |
| P42 | 3.5 | 2.5 | 3 | 3 | 3 | 1.5 | 2 | 2.5 | 2 | 2.56 |
| P43 | 2 | 2 | 1.5 | 1.5 | 1 | 1 | 1.5 | 2 | 2.5 | 1.67 |
| P44 | 2.5 | 2 | 3.5 | 3 | 3 | 1.5 | 1.5 | 2.5 | 3 | 2.5 |
| P45 | 1.5 | 2 | 3 | 2 | 1.5 | 1 | 1.5 | 2 | 2.5 | 1.89 |
| P46 | 4 | 3 | 3 | 4 | 4 | 4 | 3 | 3.5 | 3 | 3.5 |
| P47 | 4 | 2 | 2.5 | 2 | 3 | 2.5 | 2 | 2 | 2.5 | 2.5 |
| P48 | 1 | 2 | 2 | 1.5 | 1.5 | 1 | 1 | 2 | 3 | 1.67 |
| P49 | 4 | 2.5 | 3 | 2.5 | 1.5 | 1 | 2 | 1.5 | 2 | 2.22 |
| P50 | 3 | 3 | 3 | 3.5 | 2.5 | 2 | 2 | 2.5 | 2 | 2.61 |
| P51 | 3.5 | 2.5 | 3.5 | 3 | 2.5 | 1.5 | 2 | 3 | 3 | 2.72 |
| P52 | 4 | 3.5 | 4 | 3.5 | 3 | 2 | 2.5 | 3 | 2.5 | 3.11 |
| P53 | 3.5 | 2.5 | 2.5 | 3 | 2.5 | 2 | 2 | 2.5 | 2.5 | 2.56 |
| P54 | 3 | 2.5 | 3 | 3 | 2 | 1.5 | 1.5 | 3 | 3 | 2.5 |
| P55 | 3 | 3 | 3 | 3.5 | 3.5 | 1 | 2 | 4 | 3.5 | 2.94 |
| P56 | 2 | 2 | 2.5 | 2 | 2 | 1.5 | 1.5 | 3 | 2.5 | 2.11 |
| P57 | 3 | 2.5 | 3 | 3.5 | 3.5 | 2.5 | 1.5 | 3.5 | 3 | 2.89 |
| P58 | 3.5 | 2.5 | 2.5 | 3 | 3 | 1.5 | 3 | 2.5 | 3.5 | 2.78 |
| P59 | 4 | 2.5 | 4 | 3.5 | 4 | 2 | 3.5 | 3.5 | 3 | 3.33 |

Continued on next page

| Index | URL |
|---|---|

**Table A3 – continued from previous page**

| | |
|---|---|
| IA1 | https://azure.microsoft.com/en-us/services/kubernetes-service/ |
| IA2 | https://blog.newrelic.com/engineering/what-is-kubernetes/ |

Continued on next page

**Table A2 – continued from previous page**

| Index | Q1 | Q2 | Q3 | Q4 | Q5 | Q6 | Q7 | Q8 | Q9 | Avg. |
|---|---|---|---|---|---|---|---|---|---|---|
| P60 | 3 | 2.5 | 2.5 | 3 | 3 | 1.5 | 1.5 | 3 | 3 | 2.56 |
| P61 | 1.5 | 2 | 1.5 | 1.5 | 1.5 | 1 | 1 | 1.5 | 2.5 | 1.56 |
| P62 | 3.5 | 2.5 | 3 | 2.5 | 3 | 1 | 2 | 2 | 3 | 2.5 |
| P63 | 2 | 2.5 | 3 | 2.5 | 2.5 | 1 | 2 | 2.5 | 2 | 2.22 |
| P64 | 2.5 | 1.5 | 2 | 2.5 | 1.5 | 1.5 | 1.5 | 1.5 | 2 | 1.83 |
| P65 | 3.5 | 2.5 | 3 | 2.5 | 2.5 | 2 | 2 | 2 | 2.5 | 2.5 |
| P66 | 3 | 3 | 3 | 2.5 | 2 | 2 | 1.5 | 1 | 2.5 | 2.28 |
| P67 | 3 | 3.5 | 3.5 | 3 | 2 | 1 | 1.5 | 3 | 3 | 2.61 |
| P68 | 3 | 2.5 | 3 | 3.5 | 3.5 | 1 | 1.5 | 3 | 2.5 | 2.61 |
| P69 | 3.5 | 2.5 | 3.5 | 3 | 3 | 1.5 | 2 | 3.5 | 3 | 2.83 |
| P70 | 3.5 | 3 | 3 | 3 | 3 | 1 | 2 | 3 | 3 | 2.72 |
| P71 | 2.5 | 2 | 2.5 | 2 | 2.5 | 1 | 1.5 | 2 | 1.5 | 1.94 |
| P72 | 3.5 | 2 | 3 | 2.5 | 2 | 1.5 | 2 | 2.5 | 3 | 2.44 |
| P73 | 3 | 3 | 3 | 3 | 3 | 2 | 1.5 | 4 | 2.5 | 2.78 |
| P74 | 2.5 | 1.5 | 1.5 | 1.5 | 1.5 | 1 | 1 | 1.5 | 2.5 | 1.61 |
| P75 | 3.5 | 2.5 | 3 | 2.5 | 1.5 | 1 | 1.5 | 2.5 | 2 | 2.22 |
| P76 | 2 | 1.5 | 2 | 1.5 | 1.5 | 1.5 | 1.5 | 2 | 2.5 | 1.78 |
| P77 | 3 | 2.5 | 3.5 | 2.5 | 3 | 1 | 1.5 | 4 | 3.5 | 2.72 |
| P78 | 2 | 2 | 2 | 2.5 | 1.5 | 1.5 | 1.5 | 2 | 3 | 2 |
| P79 | 1.5 | 2 | 2.5 | 2.5 | 2 | 1 | 1.5 | 2.5 | 2.5 | 2 |
| P80 | 2.5 | 2 | 3.5 | 3 | 3 | 1.5 | 1.5 | 2 | 3 | 2.44 |
| P81 | 3 | 2 | 3.5 | 2.5 | 1.5 | 1.5 | 2.5 | 2.5 | 3 | 2.44 |
| P82 | 3 | 2 | 2.5 | 2.5 | 2 | 1.5 | 1.5 | 2 | 3 | 2.22 |
| P83 | 2.5 | 2 | 2.5 | 2.5 | 1.5 | 1 | 1.5 | 2 | 3 | 2.06 |
| P84 | 3.5 | 2.5 | 3 | 2.5 | 2 | 1.5 | 1.5 | 2.5 | 3 | 2.44 |
| P85 | 1.5 | 2 | 3 | 2 | 2 | 1.5 | 1.5 | 2.5 | 2 | 2 |
| P86 | 2.5 | 3 | 3 | 2.5 | 2.5 | 1 | 2 | 2.5 | 2.5 | 2.39 |
| P87 | 2.5 | 2.5 | 3.5 | 2.5 | 2.5 | 1 | 1.5 | 3 | 3.5 | 2.5 |
| P88 | 2.5 | 2.5 | 3 | 3 | 1.5 | 2 | 2 | 2 | 2 | 2.28 |
| P89 | 2.5 | 1.5 | 1.5 | 1 | 1 | 1 | 1.5 | 1 | 2.5 | 1.5 |
| P90 | 3.5 | 3 | 4 | 3.5 | 3.5 | 1 | 2 | 3 | 3.5 | 3 |
| P91 | 3 | 2.5 | 3.5 | 2 | 3 | 1 | 1.5 | 3.5 | 3 | 2.56 |
| P92 | 4 | 1 | 2 | 4 | 4 | 1 | 1 | 2 | 4 | 2.56 |
| P93 | 3 | 3.5 | 3.5 | 4 | 3.5 | 1.5 | 4 | 3.5 | 3.5 | 3.33 |
| P94 | 3 | 2.5 | 4 | 3 | 1.5 | 1.5 | 2 | 3 | 3.5 | 2.67 |
| P95 | 3 | 2 | 3.5 | 2.5 | 3.5 | 1.5 | 1.5 | 3.5 | 3.5 | 2.72 |
| P96 | 3 | 3 | 3.5 | 2.5 | 2.5 | 1.5 | 4 | 3 | 3.5 | 2.94 |
| P97 | 2.5 | 3.5 | 3 | 3 | 2.5 | 1 | 2.5 | 2.5 | 4 | 2.72 |
| P98 | 3 | 2.5 | 3 | 2.5 | 2.5 | 1 | 1.5 | 2.5 | 3.5 | 2.44 |
| P99 | 2.5 | 2 | 3 | 2.5 | 1 | 1 | 1.5 | 2.5 | 3 | 2.11 |
| P100 | 3 | 1.5 | 3 | 1.5 | 2 | 1.5 | 1.5 | 2.5 | 2 | 2.06 |
| P101 | 1 | 1.5 | 2 | 2.5 | 2 | 1 | 1.5 | 3 | 2 | 1.83 |
| P102 | 3 | 2.5 | 3 | 2.5 | 2.5 | 2 | 2.5 | 3 | 3.5 | 2.72 |
| P103 | 3 | 3.5 | 3 | 3 | 2.5 | 1.5 | 2.5 | 3.5 | 4 | 2.94 |
| P104 | 2.5 | 2 | 3 | 2.5 | 2.5 | 1 | 1 | 3 | 2 | 2.17 |
| P105 | 3 | 2.5 | 3 | 3 | 3 | 1.5 | 2 | 3.5 | 3 | 2.72 |
| Avg. | 2.95 | 2.42 | 2.92 | 2.56 | 2.36 | 1.35 | 1.86 | 2.6 | 2.83 | 2.43 |

Table A3: Index and URL of 321 Internet Artifacts

| Index | URL |
|---|---|
| IA3 | https://blog.pythian.com/lessons-learned-kubernetes/ |
| IA4 | https://cilium.io/blog/2020/07/27/2020-07-27-multitenancy-network-security/ |
| IA5 | https://cloud.google.com/containers/security |
| IA6 | https://cloud.google.com/kubernetes-engine/docs/concepts/cluster-autoscaler |
| IA7 | https://cloud.google.com/kubernetes-engine/docs/concepts/security-overview |
| IA8 | https://cloud.google.com/kubernetes-engine/docs/security-bulletins |
| IA9 | https://cloud.google.com/learn/what-is-kubernetes |
| IA10 | https://cloud.google.com/solutions/addressing-continuous-delivery-challenges-in-akubernetes-world |
| IA11 | https://cloud.google.com/solutions/prep-kubernetes-engine-for-prod |
| IA12 | https://cloud.ibm.com/docs/containers?topic=containers-cs ov |
| IA13 | https://cloud.ibm.com/docs/containers?topic=containers-security |
| IA14 | https://codefresh.io/kubernetes-tutorial/kubernetes-cloud-aws-vs-gcp-vs-azure/ |
| IA15 | https://codilime.com/harnessing-the-power-of-kubernetes-7-use-cases/ |
| IA16 | https://containerjournal.com/topics/container-ecosystems/extending-kubernetes-withservice-mesh/ |
| IA17 | https://containerjournal.com/topics/container-ecosystems/threatstack-report-highlightscommon-kubernetes-security-issues/ |
| IA18 | https://containerjournal.com/topics/container-management/running-kubernetes-at-scaletop-2020-challenge/ |
| IA19 | https://containerjournal.com/topics/container-security/common-container-and-kubernetesvulnerabilities/ |
| IA20 | https://d2iq.com/blog/the-top-5-challenges-to-getting-started-with-kubernetes |
| IA21 | https://devops.com/kubernetes-adoption-are-you-game-for-it/ |
| IA22 | https://devspace.cloud/blog/2019/10/31/advantages-and-disadvantages-of-kubernetes |
| IA23 | https://enterprisersproject.com/article/2020/4/kubernetes-everything-you-need-know |
| IA24 | https://github.blog/2017-08-16-kubernetes-at-github/ |
| IA25 | https://github.com/cuongnv23/awesome-k8s-lessons-learned |
| IA26 | https://hackernoon.com/lessons-learned-from-moving-my-side-project-to-kubernetesc28161a16c69 |
| IA27 | https://hackernoon.com/why-and-when-you-should-use-kubernetes-8b50915d97d8 |
| IA28 | https://jaxenter.com/kubernetes-practical-implications-171647.html |
| IA29 | https://jvns.ca/blog/2017/06/04/learning-about-kubernetes/ |
| IA30 | https://kubernetes-on-aws.readthedocs.io/en/latest/admin-guide/kubernetes-inproduction.html |
| IA31 | https://kubernetes-security.info/ |
| IA32 | https://kubernetes.cn/docs/setup/production-environment/windows/intro-windows-inkubernetes/ |
| IA33 | https://kubernetes.io/blog/2018/08/03/out-of-the-clouds-onto-the-ground-how-to-makekubernetes-production-grade-anywhere/ |
| IA34 | https://kubernetes.io/blog/2019/04/17/the-future-of-cloud-providers-in-kubernetes/ |

**Table A3 – continued from previous page**

| | |
|---|---|
| IA35 | https://kubernetes.io/case-studies/ |
| IA36 | https://kubernetes.io/docs/concepts/architecture/cloud-controller/ |
| IA37 | https://kubernetes.io/docs/concepts/overview/what-is-kubernetes/ |
| IA38 | https://kubernetes.io/docs/concepts/overview/working-with-objects/object-management/ |
| IA39 | https://kubernetes.io/docs/concepts/security/ |
| IA40 | https://kubernetes.io/docs/concepts/security/overview/ |
| IA41 | https://kubernetes.io/docs/reference/issues-security/security/ |
| IA42 | https://kubernetes.io/docs/setup/production-environment/tools/kubeadm/highavailability/ |
| IA43 | https://kubernetes.io/docs/tasks/administer-cluster/securing-a-cluster/ |
| IA44 | https://kubernetes.io/docs/tasks/debug-application-cluster/debug-application/ |
| IA45 | https://kubernetes.io/docs/tasks/debug-application-cluster/resource-usage-monitoring/ |
| IA46 | https://kubernetes.io/docs/tasks/manage-kubernetes-objects/declarative-config/ |
| IA47 | https://kubernetes.io/docs/tutorials/kubernetes-basics/ |

Continued on next page

**Table A3 – continued from previous page**

| Index | URL |
|---|---|
| IA48 | https://kublr.com/industry-info/docker-and-kubernetes-survey/ |
| IA49 | https://learnk8s.io/blog/kubernetes-chaos-engineering-lessons-learned |
| IA50 | https://learnk8s.io/production-best-practices |
| IA51 | https://logz.io/blog/kubernetes-challenges-at-scale/ |
| IA52 | https://logz.io/blog/kubernetes-security/ |
| IA53 | https://logz.io/blog/resources-learn-kubernetes/ |
| IA54 | https://logz.io/blog/what-are-the-hardest-parts-of-kubernetes-to-learn/ |
| IA55 | https://medium.com/@alexho140/kubernetes-best-practices-lessons-learned-e7437c158bb2 |
| IA56 | https://medium.com/@jain.sm/security-challenges-with-kubernetes-818fad4a89f2 |
| IA57 | https://medium.com/@srikanth.k/kubernetes-what-is-it-what-problems-does-it-solve-howdoes-it-compare-with-its-alternatives-937fe80b754f |
| IA58 | https://medium.com/better-programming/3-years-of-kubernetes-in-production-heres-whatwe-learned-44e77e1749c8 |
| IA59 | https://medium.com/faun/35-advanced-tutorials-to-learn-kubernetes-dae5695b1f18 |
| IA60 | https://medium.com/platformer-blog/benefits-of-kubernetes-e6d5de39bc48 |
| IA61 | https://medium.com/swlh/hurdles-and-challenges-hindering-mass-kubernetes-adoption9a7134f581a1 |
| IA62 | https://neuvector.com/container-security/kubernetes-security-guide/ |
| IA63 | https://platform9.com/blog/kubernetes-networking-challenges-at-scale/ |
| IA64 | https://platform9.com/blog/kubernetes-security-what-and-what-not-to-expect/ |
| IA65 | https://platform9.com/blog/kubernetes-use-cases/ |
| IA66 | https://pythonspeed.com/articles/dont-need-kubernetes/ |
| IA67 | https://redmondmag.com/articles/2018/12/05/critical-kubernetes-flaws.aspx |
| IA68 | https://rx-m.com/kubernetes/the-kubernetes-learning-journey-for-developers/ |
| IA69 | https://scaffold.digital/kubernetes-lessons-learned/ |
| IA70 | https://searchitoperations.techtarget.com/news/252454267/Kubernetes-security-issuesraise-concerns-for-enterprise-shops |
| IA71 | https://securityboulevard.com/2020/03/security-concerns-remain-with-containers-andkubernetes-per-new-report/ |
| IA72 | https://securityboulevard.com/2020/05/security-in-kubernetes-environment/ |
| IA73 | https://siliconangle.com/2019/08/06/34-vulnerabilities-uncovered-security-auditkubernetes-code/ |
| IA74 | https://snyk.io/blog/secure-your-kubernetes-applications-with-snyk-container/ |
| IA75 | https://snyk.io/learn/kubernetes-security/ |
| IA76 | https://softwareengineeringdaily.com/2018/01/13/the-gravity-of-kubernetes/ |
| IA77 | https://stackify.com/kubernetes-security-best-practices-you-must-know/ |
| IA78 | https://stackoverflow.com/questions/35900435/what-is-a-good-use-case-for-kubernetes-pod |
| IA79 | https://stfalcon.com/en/blog/post/kubernetes |
| IA80 | https://strategicfocus.com/2020/07/16/kubernetes-security-challenges-risks-and-attackvectors/ |
| IA81 | https://sysdig.com/blog/monitoring-kubernetes/ |
| IA82 | https://tanzu.vmware.com/content/blog/five-key-decisions-to-make-before-runningkubernetes-in-production |
| IA83 | https://techbeacon.com/enterprise-it/top-5-container-adoption-management-challenges-itops |
| IA84 | https://techolution.com/kubernetes-challenges/ |
| IA85 | https://the-report.cloud/benefits-of-kubernetes |
| IA86 | https://thenewstack.io/7-key-considerations-for-kubernetes-in-production/ |
| IA87 | https://thenewstack.io/guide-for-2019-what-to-consider-about-vms-and-kubernetes/ |
| IA88 | https://thenewstack.io/kubernetes-deep-dive-and-use-cases/ |
| IA89 | https://thenewstack.io/laying-the-groundwork-for-kubernetes-security-across-workloadspods-and-users/ |
| IA90 | https://thenewstack.io/learning-kubernetes-the-need-for-a-realistic-playground/ |
| IA91 | https://thenewstack.io/top-challenges-kubernetes-users-face-deployment/ |
| IA92 | https://threatpost.com/kubernetes-bugs-authentication-bypass-dos/149265/ |

Continued on next page

| Index | URL |
|---|---|
| IA93 | https://vexxhost.com/blog/address-these-4-kubernetes-security-challenges-now/ |
| IA94 | https://virtualizationreview.com/articles/2020/05/13/state-of-kubernetes.aspx |

**Table A3 – continued from previous page**

| | |
|---|---|
| IA95 | https://www.aquasec.com/solutions/kubernetes-container-security/ |
| IA96 | https://www.brighttalk.com/webcast/18009/392616/do-you-know-your-kubernetesruntime-vulnerabilities |
| IA97 | https://www.brighttalk.com/webcast/18009/413396/kubernetes-security-7-things-youshould-consider |
| IA98 | https://www.brighttalk.com/webcast/6793/435999/how-to-accelerate-kubernetesdeployment-in-the-enterprise |
| IA99 | https://www.cio.com/article/3411994/kubernetes-security-best-practices-for-enterprisedeployment.html |
| IA100 | https://www.cisco.com/c/dam/en/us/solutions/data-center/managing-kubernetesperformance-scale.pdf |
| IA101 | https://www.cncf.io/blog/2019/01/14/9-kubernetes-security-best-practices-everyone-mustfollow/ |
| IA102 | https://www.cncf.io/blog/2020/08/21/kubernetes-troubleshooting-7-essential-steps-fordelivering-reliable-applications/ |
| IA103 | https://www.criticalcase.com/blog/kubernetes-features-and-benefits.html |
| IA104 | https://www.cvedetails.com/vulnerability-list/vendor id-15867/product id34016/Kubernetes-Kubernetes.html |
| IA105 | https://www.darkreading.com/vulnerabilities—threats/kubernetes-shows-built-inweakness/d/d-id/1336956 |
| IA106 | https://www.deployhub.com/kubernetes-pipeline/ |
| IA107 | https://www.developintelligence.com/blog/2017/02/kubernetes-actually-use/ |
| IA108 | https://www.ericsson.com/en/blog/2020/3/benefits-of-kubernetes-on-bare-metal-cloudinfrastructure |
| IA109 | https://www.exabeam.com/information-security/kubernetes-security-monitoring/ |
| IA110 | https://www.fairwinds.com/the-guide-to-kubernetes-adoption?hsLang=en |
| IA111 | https://www.freecodecamp.org/news/learn-kubernetes-in-under-3-hours-a-detailed-guideto-orchestrating-containers-114ff420e882/ |
| IA112 | https://www.helpnetsecurity.com/2020/01/21/kubernetes-security-challenges/ |
| IA113 | https://www.hitechnectar.com/blogs/pros-cons-kubernetes/ |
| IA114 | https://www.hyscale.io/blog/kubernetes-in-production-five-challenges-youre-likely-to-faceand-how-to-approach-them/ |
| IA115 | https://www.informationsecuritybuzz.com/articles/what-are-the-top-5-kubernetes-securitychallenges-and-risks/ |
| IA116 | https://www.infoworld.com/article/3173266/4-reasons-you-should-use-kubernetes.html |
| IA117 | https://www.infoworld.com/article/3268073/what-is-kubernetes-your-next-applicationplatform.html |
| IA118 | https://www.infoworld.com/article/3545797/how-kubernetes-tackles-it-s-scalingchallenges.html |
| IA119 | https://www.instana.com/blog/problems-solved-and-problems-created-by-kubernetes/ |
| IA120 | https://www.itprotoday.com/hybrid-cloud/8-problems-kubernetes-architecture |
| IA121 | https://www.leverege.com/iot-ebook/lessons-learned |
| IA122 | https://www.linkbynet.com/what-are-the-real-benefits-of-kubernetes |
| IA123 | https://www.mobilise.cloud/kubernetes-deployment-strategies/ |
| IA124 | https://www.networkcomputing.com/data-centers/kubernetes-challenges-enterprises |
| IA125 | https://www.newsbreak.com/news/1558016691912/4-kubernetes-security-challenges-andhow-to-address-them |
| IA126 | https://www.paloaltonetworks.com/prisma/environments/kubernetes |
| IA127 | https://www.portshift.io/blog/challenges-adopting-k8s-production/ |
| IA128 | https://www.portshift.io/blog/what-kubernetes-does-for-security/ |
| IA129 | https://www.redhat.com/en/topics/containers/what-is-a-kubernetes-cluster |
| IA130 | https://www.redhat.com/en/topics/containers/what-is-kubernetes |
| IA131 | https://www.replex.io/blog/announcing-state-of-kubernetes-report-replex |

Continued on next page

| Index | URL |
|---|---|
| IA132 | https://www.replex.io/blog/the-state-of-cloud-native-challenges-culture-and-technology |
| IA133 | https://www.sans.org/webcasts/state-kubernetes-security-110230 |
| IA134 | https://www.sdxcentral.com/articles/news/cloud-native-security-remains-a-complexorganism/2020/06/ |
| IA135 | https://www.sdxcentral.com/articles/news/kubernetes-opportunities-challenges-escalatedin-2019/2019/12/ |

**Table A3 – continued from previous page**

| | |
|---|---|
| IA136 | https://www.sdxcentral.com/articles/news/latest-kubernetes-security-flaw-linked-toincomplete-patch-of-past-flaw/2019/06/ |
| IA137 | https://www.securitymagazine.com/articles/91755-container-and-kubernetes-securityconcerns-are-inhibiting-business-innovation |
| IA138 | https://www.sentinelone.com/blog/kubernetes-security-challenges-risks-and-attack-vectors/ |
| IA139 | https://www.simplilearn.com/tutorials/kubernetes-tutorial/kubernetes-architecture |
| IA140 | https://www.sitecore.com/knowledge-center/getting-started/should-my-team-adopt-docker |
| IA141 | https://www.stackrox.com/kubernetes-adoption-and-security-trends-and-market-share-forcontainers/ |
| IA142 | https://www.stackrox.com/post/2019/05/how-to-build-production-ready-kubernetesclusters-and-containers/ |
| IA143 | https://www.stackrox.com/post/2020/02/top-7-container-security-use-cases-for-kubernetesenvironments/ |
| IA144 | https://www.stackrox.com/post/2020/05/kubernetes-security-101/ |
| IA145 | https://www.threatstack.com/blog/3-things-to-know-about-kubernetes-security |
| IA146 | https://www.trendmicro.com/vinfo/de/security/news/vulnerabilities-and-exploits/kubernetes-vulnerability-cve-2019-11246-discovered-due-to-incomplete-updatesfrom-a-previous-flaw |
| IA147 | https://www.trendmicro.com/vinfo/us/security/news/virtualization-and-cloud/guidanceon-kubernetes-threat-modeling |
| IA148 | https://www.weave.works/blog/6-business-benefits-of-kubernetes |
| IA149 | https://www.weave.works/blog/aws-and-kubernetes-networking-options-and-trade-offspart-3 |
| IA150 | https://www.weave.works/technologies/the-journey-to-kubernetes/ |
| IA151 | http://blog.kubecost.com/blog/requests-and-limits/ |
| IA152 | https://about.gitlab.com/blog/2020/09/16/year-of-kubernetes/ |
| IA153 | https://acotten.com/post/1year-kubernetes |
| IA154 | https://acquisitiontalk.com/2020/01/f-16-running-on-kubernetes-and-the-challenges-of-adisconnected-environment/ |
| IA155 | https://assets.ext.hpe.com/is/content/hpedam/documents/a000390009999/a00039700/a00039700enw.pdf |
| IA156 | https://banzaicloud.com/blog/cert-management-on-kubernetes/ |
| IA157 | https://banzaicloud.com/blog/hybrid-cloud-kubernetes/ |
| IA158 | https://blog.alcide.io/kubernetes-security |
| IA159 | https://blog.cloudticity.com/five-benefits-kubernetes-healthcare |
| IA160 | https://blog.container-solutions.com/answers-to-11-big-questions-about-kubernetesversioning |
| IA161 | https://blog.container-solutions.com/riding-the-tiger-lessons-learned-implementing-istio |
| IA162 | https://blog.jetstack.io/blog/k8s-getting-started-part1/ |
| IA163 | https://blog.quasardb.net/quasardb-on-kubernetes |
| IA164 | https://blog.sonatype.com/kubesecops-kubernetes-security-practices-you-should-follow |
| IA165 | https://blog.styra.com/blog/why-rbac-is-not-enough-for-kubernetes-api-security |
| IA166 | https://blog.thundra.io/do-you-really-need-kubernetes |
| IA167 | https://blogs.vmware.com/load-balancing/2020/08/21/overcoming-application-deliverychallenges-for-kubernetes/ |
| IA168 | https://builders.intel.com/docs/networkbuilders/cpu-pin-and-isolation-in-kubernetes-appnote.pdf |
| IA169 | https://cloud.gov/docs/ops/runbook/troubleshooting-kubernetes/ |

Continued on next page

| Index | URL |
|---|---|
| IA170 | https://cloud.netapp.com/blog/gcp-cvo-blg-multicloud-kubernetes-centralizing-multicloudmanagement |
| IA171 | https://cloud.netapp.com/kubernetes-hub |
| IA172 | https://cloudacademy.com/blog/kubernetes-the-current-and-future-state-of-k8s-in-theenterprise/ |
| IA173 | https://cloudacademy.com/blog/what-is-kubernetes/ |
| IA174 | https://cloudowski.com/articles/10-differences-between-openshift-and-kubernetes/ |
| IA175 | https://cloudowski.com/articles/4-ways-to-manage-kubernetes-resources/ |
| IA176 | https://cloudplex.io/blog/microservices-orchestration-with-kubernetes/ |
| IA177 | https://cloudplex.io/blog/top-10-kubernetes-tools/ |
| IA178 | https://conferences.oreilly.com/velocity/vl-eu-2018/public/schedule/detail/71360.html |
| IA179 | https://coreos.com/blog/pitfalls-of-diy-kubernetes |

**Table A3 – continued from previous page**

| Index | URL |
|---|---|
| IA180 | https://developer.squareup.com/blog/kubernetes-pod-security-policies/ |
| IA181 | https://devopscon.io/blog/kubernetes-is-not-an-afterthought/ |
| IA182 | https://diginomica.com/kubernetes-and-misconception-multi-cloud-portability |
| IA183 | https://diginomica.com/kubernetes-evolving-enterprise-friendly-platform-challenges-remain |
| IA184 | https://discuss.hashicorp.com/t/what-are-advantages-use-consul-in-kubernetes-use-caseswithout-service-mesh/9901 |
| IA185 | https://docs.cloud.oracle.com/iaas/Content/ContEng/Concepts/contengoverview.htm |
| IA186 | https://docs.docker.com/docker-for-windows/kubernetes/ |
| IA187 | https://docs.gitlab.com/ee/user/project/clusters/ |
| IA188 | https://docs.honeycomb.io/getting-data-in/integrations/kubernetes/usecases/ |
| IA189 | https://docs.influxdata.com/platform/integrations/kubernetes/ |
| IA190 | https://docs.mattermost.com/install/install-kubernetes.html |
| IA191 | https://docs.mirantis.com/mcp/q4-18/mcp-ref-arch/kubernetes-cluster-plan/cloudprovider-overview.html |
| IA192 | https://docs.openstack.org/developer/performance-docs/issues/scale testing _issues.html |
| IA193 | https://docs.wavefront.com/kubernetes.html |
| IA194 | https://dzone.com/articles/aws-and-kubernetes-networking-options-and-trade-of |
| IA195 | https://dzone.com/articles/container-and-kubernetes-security-a-2020-update |
| IA196 | https://dzone.com/articles/how-big-companies-are-using-kubernetes |
| IA197 | https://dzone.com/articles/kubernetes-benefits-microservices-architecture-for |
| IA198 | https://dzone.com/articles/kubernetes-security-best-practices |
| IA199 | https://dzone.com/articles/the-challenges-of-adopting-k8s-for-production-and |
| IA200 | https://engineering.bitnami.com/articles/running-helm-in-production.html |
| IA201 | https://enterprisersproject.com/article/2017/10/how-explain-kubernetes-plain-english |
| IA202 | https://enterprisersproject.com/article/2017/10/how-make-case-kubernetes |
| IA203 | https://enterprisersproject.com/article/2018/11/kubernetes-production-4-myths-debunked |
| IA204 | https://enterprisersproject.com/article/2019/1/kubernetes-security-4-areas-focus |
| IA205 | https://enterprisersproject.com/article/2019/11/kubernetes-3-ways-get-started |
| IA206 | https://enterprisersproject.com/article/2019/8/multi-cloud-statistics |
| IA207 | https://enterprisersproject.com/article/2020/1/kubernetes-trends-watch-2020 |
| IA208 | https://enterprisersproject.com/article/2020/2/kubernetes-6-secrets-success |
| IA209 | https://enterprisersproject.com/article/2020/5/kubernetes-managing-7-tips |
| IA210 | https://enterprisersproject.com/article/2020/5/kubernetes-migrations-5-mistakes |
| IA211 | https://grafana.com/blog/2019/05/08/kubernetes-co-creator-brendan-burns-lessonslearned-monitoring-cloud-native-systems/ |
| IA212 | https://grafana.com/blog/2019/07/24/how-a-production-outage-was-caused-usingkubernetes-pod-priorities/ |
| IA213 | https://gruntwork.io/guides/kubernetes/how-to-deploy-production-grade-kubernetescluster-aws/ |
| IA214 | https://info.roundtower.com/hubfs/app/pdf/roundtower-rancherhow to build an enterprise kubernetes strategy.pdf |
| IA215 | https://jamesdefabia.github.io/docs/user-guide/production-pods/ |

Continued on next page

| Index | URL |
|---|---|
| IA216 | https://jfrog.com/whitepaper/the-jfrog-journey-to-kubernetes-best-practices-for-takingyour-containers-all-the-way-to-production/ |
| IA217 | https://linkerd.io/2020/02/25/multicluster-kubernetes-with-service-mirroring/ |
| IA218 | https://newrelic.com/platform/kubernetes/monitoring-guide |
| IA219 | https://nickjanetakis.com/blog/docker-swarm-vs-kubernetes-which-one-should-you-learn |
| IA220 | https://opensource.com/article/19/6/reasons-kubernetes |
| IA221 | https://opensource.com/article/20/6/kubernetes-garbage-collection |
| IA222 | https://phoenixnap.com/blog/kubernetes-monitoring-best-practices |
| IA223 | https://phoenixnap.com/blog/kubernetes-vs-openstack |
| IA224 | https://phoenixnap.com/kb/kubernetes-security-best-practices |
| IA225 | https://phoenixnap.com/kb/understanding-kubernetes-architecture-diagrams |
| IA226 | https://phoenixnap.com/kb/what-is-kubernetes |
| IA227 | https://qconnewyork.com/ny2018/presentation/cri-runtimes-deep-dive-whos-running-mykubernetes-pod |
| IA228 | https://query.prod.cms.rt.microsoft.com/cms/api/am/binary/RE36AY2 |
| IA229 | https://softchris.github.io/pages/kubernetes-one.html |

**Table A3 – continued from previous page**

| | |
|---|---|
| IA230 | https://spark.apache.org/docs/latest/running-on-kubernetes.html |
| IA231 | https://srcco.de/posts/web-service-on-kubernetes-production-checklist-2019.html |
| IA232 | https://tech.target.com/2018/08/08/running-cassandra-in-kubernetes-across-1800stores.html |
| IA233 | https://techbeacon.com/devops/one-year-using-kubernetes-production-lessons-learned |
| IA234 | https://techbeacon.com/enterprise-it/4-kubernetes-security-challenges-how-address-them |
| IA235 | https://techbeacon.com/enterprise-it/hackers-guide-kubernetes-security |
| IA236 | https://techbeacon.com/security/lessons-kubernetes-flaw-why-you-should-shift-yoursecurity-upstream |
| IA237 | https://techcloudlink.com/wp-content/uploads/2019/10/Operating-Kubernetes-Clustersand-Applications-Safely.pdf |
| IA238 | https://www.ansible.com/blog/how-useful-is-ansible-in-a-cloud-native-kubernetesenvironment |
| IA239 | https://www.atlassian.com/continuous-delivery/microservices/kubernetes |
| IA240 | https://www.bleepingcomputer.com/news/security/severe-flaws-in-kubernetes-expose-allservers-to-dos-attacks/ |
| IA241 | https://www.bluematador.com/blog/iam-access-in-kubernetes-the-aws-security-problem |
| IA242 | https://www.bluematador.com/blog/kubernetes-log-management-the-basics |
| IA243 | https://www.bluematador.com/blog/kubernetes-on-aws-eks-vs-kops |
| IA244 | https://www.bluematador.com/blog/kubernetes-security-essentials |
| IA245 | https://www.capitalone.com/tech/cloud/why-kubernetes-alone-wont-solve-enterprisecontainer-needs/ |
| IA246 | https://www.capitalone.com/tech/software-engineering/conquering-statefulness-onkubernetes/ |
| IA247 | https://www.capitalone.com/tech/software-engineering/create-and-deploy-kubernetesclusters/ |
| IA248 | https://www.checkpoint.com/downloads/products/checkpoint-cloud-native-security.pdf |
| IA249 | https://www.cloudmanagementinsider.com/google-what-kubernetes-best-practices/ |
| IA250 | https://www.crn.com/news/data-center/nutanix-cto-new-kubernetes-paas-bests-vmwarevia-simplicity- |
| IA251 | https://www.datadoghq.com/container-report/ |
| IA252 | https://www.dellemc.com/en-in/collaterals/unauth/briefs-handouts/solutions/h18141dellemc-dpd-kubernetes.pdf |
| IA253 | https://www.dellemc.com/en-us/collaterals/unauth/white-papers/products/convergedinfrastructure/dellemc-hci-for-kubernetes.pdf |
| IA254 | https://www.devopsdigest.com/the-kubernetes-security-paradox |
| IA255 | https://www.dragonspears.com/blog/powering-kubernetes-benefits-and-drawbacks-ofazure-vs-aws |

Continued on next page

| Index | URL |
|---|---|
| IA256 | https://www.dynatrace.com/support/help/technology-support/cloudplatforms/kubernetes/monitoring/monitor-kubernetes-openshift-clusters/ |
| IA257 | https://www.enterprisedb.com/blog/gartner-report-best-practices-running-containers-andkubernetes-production |
| IA258 | https://www.esecurityplanet.com/applications/tips-for-container-and-kubernetessecurity.html |
| IA259 | https://www.esecurityplanet.com/products/top-container-and-kubernetes-securityvendors.html |
| IA260 | https://www.fairwinds.com/blog/heroku-vs.-kubernetes-the-big-differences-you-shouldknow |
| IA261 | https://www.fairwinds.com/blog/how-we-learned-to-stop-worrying-and-love-clusterupgrades |
| IA262 | https://www.fairwinds.com/blog/kubernetes-best-practice-efficient-resource-utilization |
| IA263 | https://www.fairwinds.com/blog/kubernetes-best-practices-for-security |
| IA264 | https://www.fairwinds.com/blog/what-problems-does-kubernetes-solve |
| IA265 | https://www.fairwinds.com/why-kubernetes |
| IA266 | https://www.fingent.com/blog/5-reasons-to-adopt-kubernetes-into-your-business-it/ |
| IA267 | https://www.fortinet.com/content/dam/fortinet/assets/alliances/sb-xtending-enterprisesecurity-into-kubernetes-environments.pdf |
| IA268 | https://www.guru99.com/kubernetes-vs-docker.html |
| IA269 | https://www.heliossolutions.co/blog/kubernetes-security-defined-explained-and-explored/ |
| IA270 | https://www.ibm.com/cloud/learn/kubernetes |

**Table A3 – continued from previous page**

| Index | URL |
|---|---|
| IA271 | https://www.inovex.de/blog/kubernetes-security-tools/ |
| IA272 | https://www.itproportal.com/features/kubernetes-as-a-cloud-native-operating-system/ |
| IA273 | https://www.jeffgeerling.com/blog/2018/kubernetes-complexity |
| IA274 | https://www.jeffgeerling.com/blog/2019/everything-i-know-about-kubernetes-i-learnedcluster-raspberry-pis |
| IA275 | https://www.jeffgeerling.com/blog/2019/monitoring-kubernetes-cluster-utilization-andcapacity-poor-mans-way |
| IA276 | https://www.jeffgeerling.com/blog/2019/running-drupal-kubernetes-docker-production |
| IA277 | https://www.linode.com/docs/kubernetes/kubernetes-use-cases/ |
| IA278 | https://www.magalix.com/blog/deploying-an-application-on-kubernetes-from-a-to-z |
| IA279 | https://www.magalix.com/blog/the-best-kubernetes-tools-for-managing-large-scale-projects |
| IA280 | https://www.magalix.com/blog/understanding-kubernetes-objects |
| IA281 | https://www.magalix.com/blog/what-we-learned-from-running-fully-containerized-serviceson-kubernetes-part-i |
| IA282 | https://www.magalix.com/blog/why-teams-adopting-kubernetes-fight-over-capacitymanagement |
| IA283 | https://www.metricfire.com/blog/aws-ecs-vs-kubernetes/ |
| IA284 | https://www.metricfire.com/blog/kubernetes-security-secrets-from-the-trenches/ |
| IA285 | https://www.nutanix.com/blog/enterprise-kubernetes-done-right-nutanix-cloud-native |
| IA286 | https://www.nutanix.com/content/dam/nutanix/resources/solution-briefs/sb-karbon.pdf |
| IA287 | https://www.openshift.com/blog/kubernetes-1.18-strengthens-networking-and-storagewhile-getting-ready-for-the-next-big-adventure |
| IA288 | https://www.openshift.com/blog/kubernetes-adoption-challenges-solved |
| IA289 | https://www.openshift.com/blog/red-hat-chose-kubernetes-openshift |
| IA290 | https://www.openshift.com/learn/topics/kubernetes/ |
| IA291 | https://www.ovhcloud.com/en/public-cloud/kubernetes/ |
| IA292 | https://www.paladion.net/blogs/kubernetes-introduction-and-security-aspects |
| IA293 | https://www.presslabs.com/blog/kubernetes-cloud-providers-2019/ |
| IA294 | https://www.prweb.com/releases/new nirmata study more than half of kubernetes users cite lack of expertise prevents wider adoption across the organization/prweb16055189.htm |
| IA295 | https://www.pulumi.com/blog/crosswalk-kubernetes/ |
| IA296 | https://www.pulumi.com/docs/intro/cloud-providers/kubernetes/ |
| IA297 | https://www.rackspace.com/blog/kubernetes-explained-for-business-leaders |
| IA298 | https://www.rackspace.com/managed-kubernetes |
| IA299 | https://www.rackspace.com/solve/how-kubernetes-has-changed-face-hybrid-cloud |
| IA300 | https://www.rapidvaluesolutions.com/azure-kubernetes-service-aks-simplifying-deploymentwith-stateful-applications/ |
| IA301 | https://www.redapt.com/blog/a-critical-aspect-of-day-2-kubernetes-operations |
| IA302 | https://www.replex.io/blog/kubernetes-in-production |
| IA303 | https://www.replex.io/blog/kubernetes-in-production-best-practices-for-governance-costmanagement-and-security-and-access-control |
| IA304 | https://www.replex.io/blog/kubernetes-in-production-the-ultimate-guide-to-monitoringresource-metrics |
| IA305 | https://www.replex.io/blog/the-ultimate-kubernetes-cost-guide-aws-vs-gce-vs-azure-vsdigital-ocean |
| IA306 | https://www.spectrocloud.com/blog/kubernetes-multi-tenant-vs-single-tenant-clusters/ |
| IA307 | https://www.splunk.com/en us/blog/it/kubernetes-navigator-real-time-monitoring-and-aidriven-analytics-for-kubernetes-environments-now-generally-available.html |
| IA308 | https://www.splunk.com/en us/blog/it/monitoring-kubernetes.html |
| IA309 | https://www.splunk.com/en us/blog/it/strategies-for-monitoring-docker-andkubernetes.html |
| IA310 | https://www.splunk.com/en us/blog/security/approaching-kubernetes-security-detectingkubernetes-scan-with-splunk.html |
| IA311 | https://www.sumologic.com/blog/kubernetes-vs-docker/ |
| IA312 | https://www.sumologic.com/blog/troubleshooting-kubernetes/ |
| IA313 | https://www.sumologic.com/blog/why-use-kubernetes/ |
| IA314 | https://www.sumologic.com/kubernetes/security/ |
| IA315 | https://www.tutorialspoint.com/kubernetes/kubernetes kubectl commands.htm |
| IA316 | https://www.vmware.com/topics/glossary/content/kubernetes |

Continued on next page

**Table A3 – continued from previous page**

| | |
|---|---|
| IA317 | https://www.vmware.com/topics/glossary/content/kubernetes-security |
| IA318 | https://www.xenonstack.com/insights/kubernetes-security-tools/ |
| IA319 | https://www.xenonstack.com/use-cases/cloud-native-devops/ |
| IA320 | https://www.zdnet.com/article/corporate-culture-complicates-kubernetes-and-containercollaboration/ |
| IA321 | https://www.zdnet.com/article/red-hat-takes-kubernetes-to-the-clouds-edge/ |

Table A4: Quality Assessments of 321 Internet Artifacts

| Index | Q1 | Q2 | Q3 | Q4 | Q5 | Q6 | Q7 | Q8 | Q9 | Q10 | Q11 | Q12 |
|---|---|---|---|---|---|---|---|---|---|---|---|---|
| IA1 | 1 | 1 | 0.75 | 1 | 0.5 | 0.75 | 0.0 | 0 | 0.5 | 0.25 | 0 | 1 |
| IA2 | 0.25 | 0.25 | 0.75 | 1 | 1 | 1 | 0.0 | 1 | 0.5 | 0.75 | 0 | 0 |
| IA3 | 0.25 | 0.25 | 0.75 | 0 | 0.25 | 1 | 0.0 | 1 | 0 | 0.5 | 0 | 0 |
| IA4 | 0.25 | 0.25 | 0.5 | 0.5 | 0.25 | 1 | 0.0 | 1 | 0.5 | 0.5 | 0 | 0.25 |
| IA5 | 1 | 1 | 0.25 | 1 | 0 | 1 | 0.0 | 0 | 0.75 | 0.75 | 0 | 1 |
| IA6 | 1 | 1 | 1 | 1 | 0.75 | 1 | 0.5 | 0.75 | 1 | 0.25 | 0 | 1 |
| IA7 | 1 | 1 | 1 | 1 | 0.5 | 1 | 0.5 | 0.75 | 1 | 0.75 | 0 | 1 |
| IA8 | 1 | 1 | 1 | 1 | 0.5 | 1 | 0.0 | 1 | 1 | 0.75 | 0 | 1 |
| IA9 | 1 | 1 | 1 | 1 | 1 | 0.5 | 0.0 | 0 | 1 | 0.75 | 0 | 1 |
| IA10 | 1 | 1 | 1 | 1 | 0.5 | 0.75 | 0.0 | 0.5 | 1 | 0.75 | 0 | 1 |
| IA11 | 1 | 1 | 1 | 1 | 0.5 | 1 | 0.5 | 0.75 | 1 | 0.75 | 0 | 1 |
| IA12 | 1 | 1 | 0.5 | 1 | 0.5 | 0.75 | 0.5 | 1 | 1 | 0.75 | 0 | 1 |
| IA13 | 1 | 1 | 0.75 | 1 | 0.5 | 0.75 | 0.5 | 1 | 0.75 | 0.5 | 0 | 1 |
| IA14 | 0.25 | 0.25 | 1 | 0.5 | 0.5 | 0.75 | 0.5 | 1 | 0.75 | 0.5 | 0 | 0.25 |
| IA15 | 0.25 | 0 | 1 | 0.5 | 0.75 | 0.5 | 0 | 1 | 0.75 | 0.5 | 0 | 0 |
| IA16 | 0.25 | 0 | 0.5 | 0 | 0.75 | 0.5 | 0 | 1 | 0.5 | 0.25 | 0 | 0 |

Continued on next page

**Table A4 – continued from previous page**

| Index | Q1 | Q2 | Q3 | Q4 | Q5 | Q6 | Q7 | Q8 | Q9 | Q10 | Q11 | Q12 |
|---|---|---|---|---|---|---|---|---|---|---|---|---|
| IA17 | 0.25 | 0.25 | 0.5 | 0.25 | 0.5 | 0.75 | 0 | 1 | 0.5 | 0.75 | 0 | 0.25 |
| IA18 | 0.25 | 0.25 | 0.5 | 0.5 | 0.25 | 0.75 | 0 | 1 | 0.25 | 0.25 | 0 | 0.25 |
| IA19 | 0.25 | 0.25 | 0.5 | 0.5 | 0.25 | 0.75 | 0 | 1 | 0.5 | 0.5 | 0 | 0.25 |
| IA20 | 0 | 0 | 0.75 | 0.25 | 0.25 | 0.75 | 0 | 1 | 0.5 | 0.75 | 0 | 0 |
| IA21 | 0.25 | 0.25 | 1 | 0.5 | 0.5 | 1 | 0 | 1 | 0.5 | 0.5 | 0.5 | 0.25 |
| IA22 | 0.25 | 0.25 | 1 | 1 | 0.75 | 0.75 | 0 | 1 | 1 | 0.75 | 0 | 0.25 |
| IA23 | 0.25 | 0.25 | 1 | 1 | 0.75 | 1 | 0 | 1 | 1 | 0.75 | 0.25 | 0.25 |
| IA24 | 0.75 | 0.75 | 1 | 1 | 0.75 | 1 | 0 | 1 | 1 | 0.5 | 0 | 0.5 |
| IA25 | 0.5 | 0 | 0.5 | 0.5 | 0.25 | 0.5 | 0 | 0.5 | 0.5 | 0 | 0.25 | 0.25 |
| IA26 | 0.5 | 0 | 0.75 | 0.25 | 0.5 | 1 | 0 | 1 | 0.5 | 0.5 | 1 | 0 |
| IA27 | 0.5 | 0.75 | 0.75 | 0 | 0.75 | 1 | 0 | 1 | 0.25 | 1 | 1 | 0 |
| IA28 | 0.25 | 0.25 | 1 | 0 | 0.25 | 0.5 | 0 | 1 | 0.25 | 0.25 | 0 | 0 |
| IA29 | 0 | 0 | 1 | 0.75 | 0.75 | 0.75 | 0 | 0.5 | 1 | 0 | 0 | 0 |
| IA30 | 0.25 | 0.25 | 1 | 1 | 0.25 | 1 | 0 | 0 | 1 | 0.5 | 0 | 0.25 |
| IA31 | 0.5 | 0.5 | 0.5 | 0.5 | 0.5 | 0.5 | 0 | 0 | 0.5 | 0.25 | 0 | 0.5 |
| IA32 | 0.5 | 0.5 | 0.5 | 0.5 | 0 | 0.5 | 0 | 0 | 0.5 | 0 | 0 | 0.5 |
| IA33 | 1 | 1 | 1 | 1 | 1 | 1 | 0 | 1 | 1 | 0.5 | 0 | 0.5 |
| IA34 | 1 | 1 | 0.5 | 0.75 | 0.5 | 1 | 0 | 1 | 0.25 | 0 | 0 | 0.5 |
| IA35 | 0.5 | 0.5 | 0.5 | 0.5 | 0 | 0.25 | 0 | 0 | 0.5 | 0 | 0 | 0.5 |
| IA36 | 1 | 1 | 0.5 | 1 | 0.5 | 1 | 0 | 0.5 | 1 | 0 | 0 | 1 |
| IA37 | 1 | 1 | 0.75 | 0.75 | 1 | 0.75 | 0 | 0.5 | 0.75 | 0 | 0 | 1 |
| IA38 | 1 | 1 | 0.75 | 1 | 0 | 1 | 0 | 0.5 | 1 | 0.5 | 0 | 1 |
| IA39 | 0.5 | 0.5 | 0 | 0.5 | 0 | 0.5 | 0 | 0.5 | 0.5 | 0 | 0 | 0.5 |
| IA40 | 1 | 1 | 1 | 1 | 0.75 | 1 | 0 | 1 | 1 | 0.5 | 0 | 1 |
| IA41 | 1 | 1 | 1 | 1 | 0 | 0.5 | 0.75 | 1 | 1 | 0.5 | 0 | 1 |
| IA42 | 1 | 1 | 1 | 1 | 0.25 | 1 | 0.0 | 1 | 1 | 0 | 0 | 1 |
| IA43 | 1 | 1 | 1 | 1 | 0.5 | 1 | 0.0 | 1 | 1 | 0.5 | 0 | 1 |
| IA44 | 1 | 1 | 0.75 | 1 | 0.5 | 1 | 0.0 | 1 | 1 | 0.25 | 0 | 1 |
| IA45 | 1 | 1 | 0.5 | 1 | 0.25 | 0.75 | 0.0 | 1 | 1 | 0 | 0 | 0.5 |
| IA46 | 1 | 1 | 0.5 | 1 | 0.5 | 0.75 | 0.0 | 1 | 1 | 0.5 | 0 | 1 |
| IA47 | 1 | 1 | 1 | 1 | 0.75 | 0.75 | 0.0 | 1 | 1 | 0.25 | 0 | 1 |
| IA48 | 0 | 0 | 1 | 0 | 0 | 1 | 0 | 0 | 0 | 0.5 | 0 | 0.25 |
| IA49 | 0.25 | 0 | 0.75 | 0.25 | 0 | 1 | 0 | 0.75 | 0.25 | 0.5 | 0 | 0.25 |
| IA50 | 0.25 | 0 | 1 | 0 | 0 | 1 | 0.0 | 0 | 0 | 0.75 | 0 | 0.25 |
| IA51 | 0.25 | 0 | 0.75 | 1 | 0.5 | 0.75 | 0.0 | 1 | 0.75 | 0.5 | 0 | 0 |
| IA52 | 0.25 | 0 | 1 | 1 | 0 | 1 | 0 | 1 | 0.5 | 0.75 | 0 | 0 |
| IA53 | 0.25 | 0 | 1 | 0.5 | 0 | 0.5 | 0 | 1 | 1 | 0.25 | 0 | 0 |
| IA54 | 0.25 | 0 | 1 | 0.5 | 0.5 | 0.75 | 0 | 1 | 1 | 0.25 | 0 | 0 |
| IA55 | 0.25 | 0 | 1 | 1 | 0 | 0.75 | 0 | 1 | 1 | 0.75 | 0.75 | 0.25 |
| IA56 | 0.25 | 0 | 1 | 1 | 0 | 1 | 0 | 1 | 1 | 0.5 | 0.5 | 0.25 |
| IA57 | 0.25 | 0 | 1 | 1 | 1 | 1 | 0 | 1 | 1 | 0.25 | 1 | 0.25 |
| IA58 | 0.25 | 0 | 0.75 | 0.75 | 0.25 | 0.75 | 0 | 0.5 | 0.5 | 0 | 1 | 0.25 |
| IA59 | 0.25 | 0 | 1 | 1 | 0 | 0.5 | 0 | 0.5 | 1 | 0.25 | 0.75 | 0.25 |
| IA60 | 0.25 | 0 | 1 | 0.25 | 0.25 | 0.5 | 0 | 1 | 0.25 | 0.5 | 0.75 | 0 |
| IA61 | 0.25 | 0 | 0.5 | 0.25 | 0 | 0.5 | 0 | 1 | 0.25 | 0.75 | 0.75 | 0 |
| IA62 | 0 | 0 | 0.75 | 0.5 | 0.75 | 0.75 | 0.5 | 1 | 0.75 | 0.5 | 0 | 0 |
| IA63 | 0 | 0 | 1 | 0.5 | 0.5 | 0.75 | 0 | 1 | 0.5 | 0.5 | 0 | 0 |
| IA64 | 0 | 0 | 1 | 0.75 | 0.5 | 0.75 | 0 | 1 | 0.25 | 0.25 | 0 | 0 |
| IA65 | 0 | 0.5 | 0.75 | 0.75 | 0.25 | 1 | 0 | 1 | 0.25 | 0.5 | 0 | 0.25 |
| IA66 | 0 | 0 | 0.5 | 0.5 | 0.25 | 1 | 0 | 1 | 1 | 0.75 | 0 | 0 |
| IA67 | 0 | 0 | 0.5 | 0.5 | 0.25 | 0.5 | 0 | 1 | 0.5 | 0.75 | 0 | 0 |
| IA68 | 0 | 0 | 1 | 0 | 0.75 | 0.25 | 0.5 | 0 | 0.25 | 0.25 | 0 | 0 |
| IA69 | 0 | 0 | 0.75 | 0.75 | 0.5 | 0.5 | 0.0 | 0.5 | 0.25 | 0.75 | 0 | 0 |
| IA70 | 0 | 0 | 0.5 | 0.5 | 0 | 0.25 | 0 | 1 | 0.5 | 0.5 | 0.25 | 0 |
| IA71 | 0 | 0 | 0.5 | 0.25 | 0 | 0.5 | 0 | 1 | 0.5 | 0.5 | 0 | 0 |

Continued on next page

| Index | Q1 | Q2 | Q3 | Q4 | Q5 | Q6 | Q7 | Q8 | Q9 | Q10 | Q11 | Q12 |
|---|---|---|---|---|---|---|---|---|---|---|---|---|
| IA72 | 0 | 0 | 1 | 0.25 | 0 | 0.5 | 0 | 1 | 0.5 | 0 | 0 | 0 |

**Table A4 – continued from previous page**

| Index | Q1 | Q2 | Q3 | Q4 | Q5 | Q6 | Q7 | Q8 | Q9 | Q10 | Q11 | Q12 |
|---|---|---|---|---|---|---|---|---|---|---|---|---|
| IA73 | 0 | 0 | 0.5 | 0.75 | 0.25 | 1 | 0 | 1 | 0.5 | 0.5 | 0 | 0 |
| IA74 | 0 | 0 | 0.5 | 0.25 | 0.25 | 0.5 | 0.5 | 1 | 0.25 | 0.75 | 0 | 0 |
| IA75 | 0 | 0 | 0.75 | 0.75 | 0.5 | 0.75 | 0 | 1 | 0.75 | 0.25 | 0 | 0 |
| IA76 | 0 | 0 | 0.75 | 0.75 | 0.5 | 1 | 0 | 1 | 1 | 0.75 | 0 | 0.25 |
| IA77 | 0 | 0 | 1 | 0.5 | 0.75 | 0.5 | 0 | 1 | 0.5 | 0.75 | 0 | 0 |
| IA78 | 0.5 | 0 | 0.5 | 0.25 | 1 | 0.75 | 0 | 1 | 0.75 | 0.25 | 0.25 | 0.5 |
| IA79 | 0 | 0 | 1 | 0.5 | 1 | 0.75 | 0 | 1 | 0.75 | 0.25 | 0 | 0 |
| IA80 | 0 | 0 | 0 | 0 | 0 | 0 | 0 | 0 | 0 | 0 | 0 | 0 |
| IA81 | 0 | 0 | 1 | 0.5 | 0.75 | 0.75 | 0 | 1 | 1 | 0.75 | 0 | 0 |
| IA82 | 0.75 | 1 | 0.5 | 0.75 | 0.25 | 1 | 0 | 1 | 1 | 0.75 | 0 | 0.5 |
| IA83 | 0 | 0 | 0.5 | 0.75 | 0.5 | 0.5 | 0 | 0 | 0.5 | 0.25 | 0 | 0 |
| IA84 | 0 | 0 | 1 | 0.75 | 0.25 | 0.5 | 0 | 1 | 0.75 | 0.75 | 0 | 0 |
| IA85 | 0 | 0 | 1 | 0.25 | 0.25 | 0.25 | 0 | 0 | 0.25 | 0.5 | 0 | 0 |
| IA86 | 0.25 | 0.25 | 0.5 | 0.75 | 0.25 | 0.5 | 0 | 1 | 0.25 | 0.5 | 0 | 0.25 |
| IA87 | 0.25 | 0.25 | 1 | 0.5 | 0.5 | 0.5 | 0 | 1 | 0.5 | 0.25 | 0 | 0.25 |
| IA88 | 0.25 | 0.25 | 1 | 1 | 0.25 | 0.5 | 0 | 1 | 0.5 | 0.5 | 0 | 0.25 |
| IA89 | 0.25 | 0 | 0.5 | 1 | 0.25 | 1 | 0 | 1 | 0.75 | 0.5 | 0 | 0.25 |
| IA90 | 0.25 | 0.25 | 0.5 | 0.75 | 0.25 | 0.5 | 0 | 1 | 1 | 0.25 | 0 | 0.25 |
| IA91 | 0.25 | 0 | 0.75 | 1 | 0 | 0.25 | 0 | 1 | 1 | 0.75 | 0 | 0.25 |
| IA92 | 0 | 0 | 0.75 | 1 | 0.25 | 0.75 | 0 | 1 | 1 | 0.5 | 0 | 0 |
| IA93 | 0 | 0 | 1 | 0.25 | 0.25 | 0.25 | 0 | 0 | 0.25 | 0.25 | 0 | 0.5 |
| IA94 | 0 | 0 | 0.25 | 0.25 | 0 | 0.5 | 0 | 1 | 0.25 | 0.75 | 0 | 0 |
| IA95 | 0 | 0 | 0.5 | 0.5 | 0 | 0.25 | 0.5 | 0.5 | 0 | 0.25 | 0 | 0 |
| IA96 | 0 | 0.25 | 0.75 | 0 | 0.5 | 0.75 | 0 | 1 | 0 | 0 | 0 | 0 |
| IA97 | 0 | 0.25 | 0.75 | 0.5 | 0 | 0.75 | 0 | 1 | 0.5 | 0.75 | 0 | 0 |
| IA98 | 0 | 0 | 1 | 0.25 | 0.25 | 0.75 | 0 | 1 | 0.25 | 0.25 | 0 | 0 |
| IA99 | 0.25 | 0.25 | 0.75 | 0.5 | 0 | 0.25 | 0 | 0.5 | 0.5 | 0.25 | 0.5 | 0.25 |
| IA100 | 1 | 1 | 1 | 1 | 0 | 0.5 | 0 | 0.5 | 0.5 | 0.25 | 0 | 1 |
| IA101 | 0.75 | 0 | 1 | 0.75 | 0 | 0.5 | 0 | 1 | 0.25 | 0.75 | 0 | 0 |
| IA102 | 0.75 | 0.75 | 1 | 1 | 0 | 1 | 0 | 0.5 | 0.5 | 0.75 | 0 | 0.5 |
| IA103 | 0 | 0 | 1 | 0.5 | 0 | 0.5 | 0 | 1 | 0 | 0.5 | 0 | 0 |
| IA104 | 0.25 | 0.25 | 0.25 | 0.25 | 0 | 0.5 | 0 | 0.5 | 0.5 | 0 | 0 | 0.25 |
| IA105 | 0 | 0 | 0.75 | 0.5 | 0.5 | 0.25 | 0 | 1 | 0.5 | 0.75 | 0.75 | 0 |
| IA106 | 0 | 0 | 0.75 | 0.5 | 0 | 0.25 | 0 | 1 | 0.75 | 0.75 | 0.25 | 0 |
| IA107 | 0 | 0 | 1 | 0.25 | 1 | 0.5 | 0 | 0 | 0.25 | 0.25 | 0 | 0 |
| IA108 | 0.25 | 0.25 | 0.5 | 0.25 | 0 | 0.25 | 0 | 1 | 0.25 | 0.25 | 0 | 0 |
| IA109 | 0 | 0.5 | 1 | 0.75 | 0.25 | 0.75 | 0 | 1 | 0.5 | 0.75 | 0 | 0 |
| IA110 | 0 | 0 | 0.5 | 0 | 0 | 0 | 0 | 0 | 0 | 0 | 0 | 0 |
| IA111 | 0.25 | 0 | 1 | 0.75 | 0.5 | 1 | 0 | 1 | 0.5 | 0 | 0 | 0 |
| IA112 | 0 | 0 | 0.5 | 0.75 | 0.25 | 0.25 | 0 | 0.5 | 0.5 | 0.25 | 0 | 0 |
| IA113 | 0 | 0.25 | 0.5 | 0 | 0.75 | 0.25 | 0 | 0 | 0.25 | 0.25 | 0 | 0 |
| IA114 | 0 | 0 | 1 | 0.75 | 0.25 | 0.5 | 0 | 1 | 0.5 | 0.75 | 0 | 0 |
| IA115 | 0 | 0.25 | 0.5 | 0.75 | 1 | 0.75 | 0 | 1 | 0.75 | 0.5 | 0.5 | 0.25 |
| IA116 | 0 | 0.25 | 0.5 | 0.25 | 0 | 0.25 | 0 | 1 | 0.25 | 0.25 | 0 | 0 |
| IA117 | 0 | 0 | 0.5 | 1 | 0.25 | 0.5 | 0 | 1 | 0.75 | 0 | 0 | 0 |
| IA118 | 0 | 0.25 | 0.5 | 0.25 | 0.75 | 0.5 | 0 | 1 | 0.25 | 0 | 0 | 0 |
| IA119 | 0 | 0 | 0.75 | 0.75 | 0 | 1 | 0 | 1 | 0.75 | 0.25 | 0 | 0 |
| IA120 | 0 | 0 | 0.5 | 0.75 | 0 | 0.75 | 0 | 1 | 1 | 0.5 | 0 | 0 |
| IA121 | 0 | 0 | 0.5 | 0.25 | 0 | 0.75 | 0 | 0 | 0 | 0.25 | 0 | 0.5 |
| IA122 | 0 | 0 | 0.5 | 0.75 | 0.75 | 0.75 | 0 | 0 | 0.5 | 0.75 | 0 | 0 |
| IA123 | 0 | 0 | 1 | 0.5 | 0.5 | 1 | 0 | 1 | 0 | 0.25 | 0.75 | 0 |
| IA124 | 0 | 0 | 1 | 0.5 | 0 | 0.5 | 0 | 1 | 0.75 | 0.5 | 0 | 0 |
| IA125 | 0 | 0 | 0 | 0 | 0 | 0 | 0 | 0 | 0 | 0 | 0 | 0 |
| IA126 | 0 | 0 | 0.25 | 0 | 0 | 0.25 | 0.5 | 0 | 0 | 0.25 | 0 | 0 |
| IA127 | 0 | 0 | 0.5 | 0.5 | 0 | 0.75 | 0 | 1 | 0.5 | 0.75 | 0 | 0 |
| IA128 | 0 | 0 | 0.5 | 0.5 | 0.25 | 0.75 | 0 | 1 | 0 | 0.5 | 0 | 0 |
| IA129 | 1 | 0.5 | 0.5 | 1 | 0.5 | 1 | 0 | 0 | 1 | 0 | 0 | 1 |
| IA130 | 1 | 0.5 | 0.5 | 1 | 0.5 | 1 | 0 | 0 | 1 | 0 | 0 | 1 |

Continued on next page

**Table A4 – continued from previous page**

| | | | | | | | | | | | | |
|---|---|---|---|---|---|---|---|---|---|---|---|---|
| IA131 | 0.5 | 0 | 0.5 | 0.5 | 0 | 1 | 0 | 1 | 0 | 0.5 | 0 | 0 |
| IA132 | 0 | 0 | 0.5 | 0.5 | 0 | 0.5 | 0 | 1 | 0.5 | 0 | 0 | 0 |
| IA133 | 0.5 | 0.25 | 0.5 | 0.5 | 0 | 1 | 0 | 0.5 | 0.25 | 0 | 0 | 0 |
| IA134 | 0 | 0 | 0.75 | 0.5 | 0 | 0.25 | 0 | 1 | 0.5 | 0.25 | 0.5 | 0.25 |
| IA135 | 0 | 0 | 0.5 | 0.75 | 0 | 0.25 | 0 | 1 | 1 | 0.25 | 0 | 0.25 |
| IA136 | 0 | 0 | 1 | 0.75 | 0 | 0.75 | 0 | 1 | 0.5 | 0 | 0 | 0.25 |
| IA137 | 0 | 0 | 0.5 | 0.25 | 0 | 0.25 | 0 | 1 | 0 | 0.5 | 0 | 0 |
| IA138 | 0 | 0 | 1 | 0.5 | 0 | 1 | 0 | 0.5 | 0.5 | 0 | 0 | 0 |
| IA139 | 0 | 0 | 1 | 0.25 | 0 | 1 | 0 | 1 | 0.5 | 0 | 0 | 0 |
| IA140 | 0 | 0 | 0.5 | 0.25 | 1 | 0.25 | 0 | 0 | 0 | 0.25 | 0 | 0 |
| IA141 | 0.25 | 0.25 | 1 | 0 | 0.5 | 0.5 | 0 | 0.5 | 0 | 0 | 0 | 0.25 |
| IA142 | 0.25 | 0.25 | 1 | 0.75 | 0.75 | 1 | 0 | 0.5 | 0.25 | 0.5 | 0 | 0.25 |
| IA143 | 0.25 | 0.25 | 1 | 0.5 | 0 | 0.75 | 0 | 1 | 0.75 | 0.75 | 0 | 0 |
| IA144 | 0.25 | 0.25 | 1 | 0 | 0 | 0.75 | 0 | 0.5 | 0.75 | 0.25 | 0 | 0.5 |
| IA145 | 0 | 0 | 1 | 0.5 | 0.5 | 0.25 | 0 | 1 | 1 | 0.75 | 0 | 0 |
| IA146 | 0 | 0 | 1 | 1 | 0 | 1 | 0 | 1 | 1 | 0.5 | 0 | 0.25 |
| IA147 | 0 | 0 | 1 | 0.75 | 0 | 0.25 | 0 | 1 | 1 | 0.25 | 0 | 0.25 |
| IA148 | 0 | 0 | 1 | 0.5 | 0.5 | 0.25 | 0 | 1 | 0.5 | 0.5 | 0 | 0 |
| IA149 | 0 | 0.25 | 0.5 | 0.75 | 0 | 0.5 | 0.5 | 1 | 0.5 | 0.25 | 0 | 0 |
| IA150 | 0 | 0 | 1 | 1 | 0.25 | 0.5 | 0 | 0.5 | 1 | 0 | 0 | 0 |
| IA151 | 0 | 0 | 1 | 0.5 | 0 | 1 | 0 | 1 | 0.5 | 0 | 0 | 0 |
| IA152 | 0.75 | 0.5 | 1 | 0.5 | 0 | 0.5 | 0 | 1 | 0.75 | 0.25 | 0.5 | 0.5 |
| IA153 | 0.25 | 0 | 1 | 1 | 0 | 0.25 | 0 | 0.5 | 1 | 0 | 0.25 | 0 |
| IA154 | 0 | 0 | 0.5 | 0.5 | 0 | 0.5 | 0 | 1 | 0.25 | 0.5 | 0 | 0 |
| IA155 | 1 | 0.75 | 0.5 | 0.5 | 0 | 0.5 | 0 | 0 | 0.5 | 0.75 | 0 | 0 |
| IA156 | 0 | 0 | 1 | 1 | 0 | 1 | 0 | 1 | 1 | 0.5 | 0 | 0 |
| IA157 | 0.25 | 0 | 1 | 0.5 | 0 | 0.25 | 0 | 0.5 | 0.5 | 0.5 | 0 | 0 |
| IA158 | 0 | 0 | 0.75 | 0.75 | 0.5 | 0.5 | 0 | 1 | 0.5 | 0.5 | 0 | 0 |
| IA159 | 0 | 0 | 0.75 | 0.75 | 0 | 0.25 | 0 | 1 | 0.75 | 0 | 0 | 0 |
| IA160 | 0.5 | 0.5 | 0.75 | 1 | 0.75 | 0.25 | 0 | 1 | 0.75 | 0 | 0.5 | 0 |
| IA161 | 0.5 | 0.5 | 0.5 | 0.75 | 0 | 0.25 | 0 | 0.5 | 0.5 | 0 | 0 | 0.5 |
| IA162 | 0.5 | 0.5 | 0.75 | 1 | 1 | 0.5 | 0 | 0.5 | 0.25 | 0.25 | 0 | 0.5 |
| IA163 | 0.5 | 0.5 | 1 | 0.5 | 0 | 1 | 0 | 0.5 | 0.5 | 0.25 | 0 | 0.5 |
| IA164 | 0.5 | 0.25 | 0.5 | 1 | 0 | 0.75 | 0.25 | 0.5 | 0.5 | 0.5 | 0 | 0 |
| IA165 | 0 | 0 | 1 | 1 | 0.25 | 0.25 | 0 | 0.5 | 0.5 | 0.25 | 0 | 0 |
| IA166 | 0 | 0 | 1 | 1 | 0.5 | 0.75 | 0 | 0.5 | 0.5 | 0.25 | 0 | 0.5 |
| IA167 | 1 | 0 | 0.5 | 0.25 | 0 | 0.5 | 0 | 0.5 | 0 | 0 | 0 | 0 |
| IA168 | 1 | 0.5 | 0.5 | 1 | 0 | 1 | 0 | 0.5 | 1 | 0.5 | 0 | 0.5 |
| IA169 | 0.5 | 0.5 | 1 | 1 | 0 | 1 | 0 | 0 | 1 | 0 | 0 | 0 |
| IA170 | 0.25 | 0.25 | 1 | 0.5 | 0 | 0.75 | 0 | 1 | 0.75 | 0 | 0.5 | 0.5 |
| IA171 | 0.25 | 0.25 | 1 | 0.25 | 0 | 0.75 | 0 | 1 | 0.5 | 0 | 0.5 | 0.5 |
| IA172 | 0 | 0 | 1 | 0.75 | 0.5 | 0.5 | 0 | 1 | 0.75 | 0.75 | 0.75 | 0.25 |
| IA173 | 0 | 0.25 | 0.75 | 0.75 | 0.5 | 0.5 | 0 | 1 | 0.5 | 0 | 0.75 | 0.25 |
| IA174 | 0 | 0.25 | 1 | 0.75 | 0.5 | 0.75 | 0 | 1 | 0.5 | 1 | 0.75 | 0.25 |
| IA175 | 0 | 0.25 | 1 | 0.75 | 0 | 0.5 | 0 | 1 | 0.5 | 0.5 | 0.75 | 0.25 |
| IA176 | 0 | 0 | 1 | 0.5 | 0 | 1 | 0.5 | 0.5 | 0.5 | 0.5 | 0 | 0 |
| IA177 | 0 | 0 | 0.5 | 0 | 0.5 | 0.5 | 0 | 0.5 | 0.5 | 0.25 | 0.5 | 0 |
| IA178 | 0.5 | 0.25 | 1 | 0 | 0 | 0.5 | 0 | 0.5 | 0 | 0 | 0 | 0 |
| IA179 | 0.5 | 0.25 | 0.5 | 0.5 | 0.5 | 0.25 | 0 | 0.5 | 0.5 | 0.75 | 0 | 0 |
| IA180 | 0 | 0 | 1 | 0.75 | 0 | 1 | 0 | 1 | 0.5 | 0.5 | 0 | 0 |
| IA181 | 0.25 | 0.5 | 1 | 0.5 | 0 | 0.5 | 0 | 0.25 | 0.5 | 1 | 0 | 0 |

Continued on next page

| Index | Q1 | Q2 | Q3 | Q4 | Q5 | Q6 | Q7 | Q8 | Q9 | Q10 | Q11 | Q12 |
|---|---|---|---|---|---|---|---|---|---|---|---|---|
| IA182 | 0 | 0 | 0.75 | 0.75 | 0 | 0.75 | 0 | 1 | 0.5 | 0.75 | 0.25 | 0 |
| IA183 | 0 | 0 | 0.75 | 0.75 | 0 | 0.5 | 0 | 1 | 0.75 | 0.25 | 0.25 | 0 |
| IA184 | 0.5 | 0.5 | 0.75 | 0.75 | 1 | 0.5 | 0 | 0.75 | 0.5 | 0 | 0.75 | 0 |
| IA185 | 1 | 1 | 0.5 | 0.75 | 0 | 0.5 | 1 | 0 | 0.5 | 0.25 | 0 | 1 |
| IA186 | 1 | 1 | 0.75 | 1 | 0.5 | 0.5 | 1 | 0 | 0.5 | 0.25 | 0 | 1 |
| IA187 | 1 | 1 | 1 | 1 | 0 | 1 | 0 | 0 | 1 | 0.25 | 0 | 1 |
| IA188 | 0 | 0 | 0.5 | 1 | 0 | 1 | 1 | 0 | 0.5 | 0.5 | 0 | 0 |

**Table A4 – continued from previous page**

| Index | Q1 | Q2 | Q3 | Q4 | Q5 | Q6 | Q7 | Q8 | Q9 | Q10 | Q11 | Q12 |
|---|---|---|---|---|---|---|---|---|---|---|---|---|
| IA189 | 0 | 0 | 0.75 | 1 | 0 | 0.5 | 0 | 0 | 0.5 | 0 | 0 | 0 |
| IA190 | 0 | 0 | 1 | 1 | 0 | 1 | 0.5 | 0 | 0.5 | 0 | 0 | 0.5 |
| IA191 | 0 | 0 | 0.5 | 0 | 0 | 0.5 | 0.5 | 0 | 0 | 0 | 0 | 0 |
| IA192 | 0.75 | 0.25 | 0.5 | 0.75 | 0 | 1 | 0 | 0 | 1 | 0.5 | 0 | 0.75 |
| IA193 | 0.25 | 0.25 | 1 | 1 | 0 | 1 | 0.5 | 0 | 1 | 0 | 0 | 0.25 |
| IA194 | 0 | 0 | 1 | 0.5 | 0 | 0.75 | 0 | 1 | 0.75 | 0 | 0.5 | 0.25 |
| IA195 | 0 | 0 | 1 | 0.5 | 0 | 0.75 | 0 | 1 | 0.25 | 0 | 0.5 | 0.25 |
| IA196 | 0 | 0 | 1 | 0.75 | 0.5 | 0.75 | 0 | 1 | 0.75 | 0.5 | 0.5 | 0.25 |
| IA197 | 0 | 0 | 0.75 | 0.5 | 0 | 0.25 | 0 | 1 | 0.25 | 0.25 | 0.5 | 0.25 |
| IA198 | 0 | 0 | 1 | 0.75 | 0.25 | 0.75 | 0 | 1 | 0.5 | 0.5 | 0.5 | 0.25 |
| IA199 | 0 | 0 | 1 | 0.25 | 0 | 0.5 | 0 | 1 | 0 | 0.25 | 0.5 | 0.25 |
| IA200 | 0 | 0 | 1 | 1 | 0 | 1 | 0 | 1 | 1 | 0.5 | 0 | 0 |
| IA201 | 0.25 | 0.25 | 0.5 | 0.5 | 0.75 | 0.75 | 0 | 1 | 1 | 0 | 0.75 | 0.25 |
| IA202 | 0.25 | 0.25 | 0.5 | 0.5 | 0.25 | 0.5 | 0 | 0.5 | 1 | 0.25 | 0.75 | 0.25 |
| IA203 | 0.25 | 0.25 | 0.5 | 0.5 | 0 | 0.25 | 0 | 1 | 0.5 | 0.25 | 0.75 | 0.25 |
| IA204 | 0.25 | 0.25 | 0.5 | 0.75 | 0.5 | 0.5 | 0 | 1 | 0.75 | 0.5 | 0.75 | 0.25 |
| IA205 | 0.25 | 0.25 | 1 | 0.75 | 0.5 | 0.75 | 0 | 1 | 0.5 | 0 | 0.75 | 0.25 |
| IA206 | 0.25 | 0.25 | 0.5 | 0.75 | 0 | 0.5 | 0 | 1 | 0.5 | 0.5 | 0.75 | 0.25 |
| IA207 | 0.25 | 0.25 | 1 | 1 | 0 | 0.75 | 0 | 1 | 0.5 | 0.5 | 0.75 | 0.25 |
| IA208 | 0.25 | 0.25 | 1 | 0.75 | 0 | 0.25 | 0 | 1 | 0.75 | 0.5 | 0.75 | 0.25 |
| IA209 | 0.25 | 0.25 | 0.75 | 1 | 0 | 0.5 | 0 | 1 | 0.75 | 0.75 | 0.75 | 0.25 |
| IA210 | 0.25 | 0.25 | 0.5 | 0.5 | 0.5 | 0.25 | 0 | 1 | 0.5 | 0.25 | 0.75 | 0.25 |
| IA211 | 0 | 0 | 1 | 0.5 | 0 | 1 | 0 | 1 | 0 | 0.5 | 0 | 0 |
| IA212 | 0 | 0 | 1 | 0.25 | 0 | 0.5 | 0 | 1 | 0.25 | 0.5 | 0 | 0 |
| IA213 | 0 | 0 | 1 | 0.5 | 0.75 | 0.75 | 0 | 0 | 0.5 | 0.5 | 0.75 | 0 |
| IA214 | 0.25 | 0.25 | 1 | 1 | 0.75 | 0.75 | 0.5 | 0 | 1 | 0 | 0 | 0.75 |
| IA215 | 0.25 | 0.25 | 1 | 1 | 0 | 1 | 0 | 0 | 1 | 0.5 | 0.5 | 0.5 |
| IA216 | 0 | 0 | 1 | 1 | 0 | 0.75 | 0 | 0 | 0.5 | 0.75 | 0 | 0.5 |
| IA217 | 0 | 0 | 1 | 1 | 0 | 1 | 0 | 1 | 1 | 0.5 | 0 | 0 |
| IA218 | 0.25 | 0 | 0.75 | 1 | 0.25 | 0.75 | 0.5 | 0 | 1 | 0.75 | 0 | 0.25 |
| IA219 | 0 | 0 | 1 | 1 | 1 | 0.5 | 0 | 0.5 | 0.5 | 0.25 | 0.25 | 0 |
| IA220 | 0.25 | 0.25 | 1 | 1 | 0 | 0.25 | 0 | 1 | 0.5 | 0.75 | 0.75 | 0.25 |
| IA221 | 0.25 | 0.25 | 1 | 1 | 0.25 | 0.25 | 0 | 1 | 1 | 0.75 | 0.75 | 0.25 |
| IA222 | 0 | 0 | 0.5 | 1 | 0 | 1 | 0.5 | 1 | 1 | 0.5 | 0 | 0 |
| IA223 | 0 | 0 | 1 | 0.75 | 1 | 0.75 | 0.5 | 1 | 1 | 0.25 | 0 | 0 |
| IA224 | 0 | 0 | 1 | 1 | 0 | 0.75 | 0 | 1 | 0.5 | 0.75 | 0 | 0 |
| IA225 | 0 | 0 | 1 | 0.75 | 0 | 1 | 0 | 1 | 0.75 | 0 | 0 | 0 |
| IA226 | 0 | 0 | 0.75 | 0.75 | 0.5 | 0.75 | 0 | 1 | 0.5 | 0.5 | 0 | 0 |
| IA227 | 0.5 | 1 | 1 | 0.5 | 0.5 | 0.75 | 0 | 1 | 0.5 | 0.25 | 0 | 0.75 |
| IA228 | 0.25 | 0.25 | 1 | 0.5 | 0.75 | 1 | 0 | 0 | 0.5 | 0.25 | 0 | 0.75 |
| IA229 | 0.5 | 0 | 1 | 1 | 0 | 1 | 0 | 0 | 1 | 0 | 0 | 0 |
| IA230 | 1 | 1 | 1 | 1 | 0 | 1 | 1 | 0 | 1 | 0.75 | 0 | 1 |
| IA231 | 0 | 0 | 1 | 0.5 | 0 | 0.75 | 0 | 1 | 1 | 0.5 | 0 | 0 |
| IA232 | 0.5 | 0 | 0.75 | 0.5 | 0 | 1 | 0 | 1 | 0.75 | 0 | 0 | 0.5 |
| IA233 | 0 | 1 | 0.75 | 0.75 | 0 | 1 | 0 | 0 | 1 | 0.75 | 0 | 0.25 |
| IA234 | 0 | 1 | 1 | 0.75 | 0 | 0.5 | 0 | 0 | 1 | 0.75 | 0 | 0.25 |
| IA235 | 0 | 0.5 | 0.5 | 1 | 0 | 1 | 0 | 0 | 1 | 0.75 | 0 | 0.25 |
| IA236 | 0 | 1 | 0.5 | 0.5 | 0 | 0.5 | 0 | 0 | 1 | 0.75 | 0 | 0.25 |
| IA237 | 1 | 1 | 1 | 1 | 0 | 1 | 0 | 0.5 | 1 | 0.75 | 0 | 1 |
| IA238 | 1 | 0.25 | 0.5 | 0.5 | 0.5 | 1 | 0.25 | 1 | 1 | 0.25 | 0 | 0.5 |
| IA239 | 1 | 0.5 | 0.5 | 0.75 | 1 | 0.25 | 0.25 | 0 | 0.25 | 0 | 0 | 1 |
| IA240 | 0 | 0.25 | 0.5 | 1 | 0 | 1 | 0 | 1 | 1 | 0.5 | 0.25 | 0.25 |
| IA241 | 0 | 0 | 1 | 1 | 0 | 0.5 | 0 | 0.5 | 1 | 0.5 | 0 | 0 |
| IA242 | 0 | 0 | 1 | 1 | 0 | 1 | 0 | 1 | 1 | 0 | 0 | 0 |
| IA243 | 0 | 0 | 1 | 1 | 0 | 0.75 | 0 | 1 | 1 | 0.5 | 0 | 0 |
| IA244 | 0 | 0 | 1 | 1 | 0 | 0.5 | 0 | 0.5 | 1 | 0.25 | 0 | 0 |
| IA245 | 1 | 0.5 | 0.5 | 1 | 0 | 0.75 | 0 | 1 | 1 | 0.25 | 0 | 0.5 |
| IA246 | 1 | 1 | 0.5 | 1 | 0 | 1 | 0 | 1 | 1 | 0.25 | 0 | 0.75 |

Continued on next page

**Table A4 – continued from previous page**

| Index | Q1 | Q2 | Q3 | Q4 | Q5 | Q6 | Q7 | Q8 | Q9 | Q10 | Q11 | Q12 |
|---|---|---|---|---|---|---|---|---|---|---|---|---|
| IA247 | 1 | 0.75 | 1 | 0.75 | 0.75 | 1 | 0 | 1 | 0.75 | 0.25 | 0 | 0.5 |
| IA248 | 0 | 0 | 1 | 0.75 | 0 | 1 | 0 | 0 | 0.75 | 0.25 | 0 | 0.5 |
| IA249 | 0 | 0 | 1 | 0.25 | 1 | 0.25 | 0 | 1 | 0.75 | 0.25 | 0.25 | 0 |
| IA250 | 0 | 0 | 1 | 0 | 0 | 0.5 | 0 | 1 | 0.75 | 0.5 | 0 | 0.25 |
| IA251 | 0.25 | 0 | 0.5 | 0.5 | 0 | 0.5 | 0 | 0.5 | 0.25 | 0 | 0 | 0 |
| IA252 | 1 | 1 | 1 | 0.75 | 0 | 0.75 | 1 | 0 | 0.25 | 0.25 | 0 | 0.5 |
| IA253 | 1 | 1 | 1 | 1 | 0 | 0.75 | 0.5 | 0 | 0.75 | 0.75 | 0 | 1 |
| IA254 | 0 | 0 | 1 | 0 | 0.5 | 0.25 | 0 | 1 | 0 | 0.25 | 0 | 0 |
| IA255 | 0 | 0 | 1 | 0.5 | 0 | 0.25 | 0 | 0.5 | 1 | 0 | 0 | 0 |
| IA256 | 0 | 0 | 0.5 | 0.5 | 0 | 1 | 0 | 0 | 0.5 | 0 | 0 | 0 |
| IA257 | 0 | 0 | 1 | 0.25 | 0 | 0.25 | 0 | 1 | 0.25 | 0.25 | 0 | 0 |
| IA258 | 0 | 0 | 1 | 0.5 | 0 | 0.5 | 0 | 1 | 0.75 | 0.5 | 0 | 0 |
| IA259 | 0 | 0 | 1 | 0.5 | 0 | 0.5 | 0 | 1 | 1 | 0.75 | 0 | 0 |
| IA260 | 0 | 0 | 0.75 | 0 | 0 | 0.25 | 0 | 1 | 0.25 | 0 | 0 | 0 |
| IA261 | 0 | 0 | 1 | 0.25 | 0.5 | 0.75 | 0 | 0.5 | 0.25 | 0.25 | 0 | 0 |
| IA262 | 0 | 0 | 1 | 0.25 | 0 | 0.5 | 0 | 1 | 0.25 | 0.75 | 0 | 0 |
| IA263 | 0 | 0 | 1 | 0.25 | 0 | 0.25 | 0 | 1 | 0.25 | 0.75 | 0 | 0 |
| IA264 | 0 | 0 | 1 | 0.75 | 1 | 1 | 0 | 1 | 0.75 | 0 | 0 | 0 |
| IA265 | 0 | 0 | 1 | 0.25 | 0.25 | 0.25 | 0.5 | 0 | 0 | 0.25 | 0 | 0 |
| IA266 | 0 | 0 | 1 | 0 | 0 | 0.25 | 0 | 0.5 | 0 | 0 | 0 | 0 |
| IA267 | 0 | 0 | 1 | 0.5 | 0 | 0.75 | 0.25 | 0 | 0.5 | 0.75 | 0 | 0.5 |
| IA268 | 0 | 0 | 1 | 0.25 | 0.25 | 0.5 | 0 | 0 | 0 | 0.25 | 0 | 0 |
| IA269 | 0 | 0 | 1 | 0.5 | 0 | 1 | 0 | 1 | 0.25 | 0 | 0 | 0 |
| IA270 | 1 | 1 | 1 | 1 | 0.75 | 1 | 0.5 | 0.75 | 0.5 | 0 | 0 | 0.75 |
| IA271 | 0 | 0 | 0.5 | 1 | 0 | 1 | 0 | 1 | 1 | 0.5 | 0 | 0 |
| IA272 | 0 | 0 | 0.5 | 0 | 0 | 0.5 | 0 | 1 | 0 | 0 | 0 | 0 |
| IA273 | 0 | 0 | 1 | 1 | 0 | 1 | 0 | 1 | 1 | 0.75 | 0.75 | 0 |
| IA274 | 0 | 0 | 0.5 | 0.75 | 0 | 1 | 0 | 1 | 0.5 | 0 | 0.75 | 0.25 |
| IA275 | 0 | 0 | 0.5 | 0.25 | 0 | 0.75 | 0 | 1 | 0.25 | 0 | 0.25 | 0 |
| IA276 | 0 | 0 | 1 | 0.5 | 0 | 1 | 0 | 1 | 0.5 | 0 | 0.5 | 0 |
| IA277 | 0.25 | 0.25 | 0.5 | 1 | 0 | 0.75 | 0 | 0.5 | 0.5 | 0.75 | 0 | 0.5 |
| IA278 | 0 | 0 | 1 | 1 | 0 | 1 | 0 | 1 | 1 | 0 | 0.25 | 0 |
| IA279 | 0 | 0 | 1 | 0.5 | 0.5 | 0.25 | 0 | 1 | 1 | 0.25 | 0 | 0 |
| IA280 | 0 | 0 | 1 | 1 | 0 | 0.5 | 0 | 1 | 0.5 | 0 | 0.25 | 0 |
| IA281 | 0 | 0 | 0.75 | 0.5 | 1 | 0.75 | 0 | 1 | 0.5 | 0 | 0 | 0 |
| IA282 | 0 | 0 | 0.5 | 0.5 | 1 | 0.25 | 0 | 1 | 0.5 | 0.25 | 0 | 0 |
| IA283 | 0 | 0 | 1 | 0.25 | 0 | 0.75 | 0 | 1 | 0.5 | 0.5 | 0 | 0 |
| IA284 | 0 | 0 | 1 | 0.5 | 0.25 | 0.75 | 0 | 0.75 | 1 | 0 | 0.25 | 0 |
| IA285 | 0.25 | 0.25 | 0.5 | 0 | 0 | 0.25 | 0.75 | 0 | 0.25 | 0.5 | 0 | 0 |
| IA286 | 0.25 | 0.25 | 0.5 | 0 | 0.25 | 0.25 | 1 | 0 | 0 | 0.25 | 0 | 0.25 |
| IA287 | 0.5 | 0.25 | 0.5 | 0.5 | 0 | 0.75 | 0 | 1 | 1 | 0 | 0 | 0 |
| IA288 | 0.5 | 0 | 1 | 0 | 0 | 0.5 | 0 | 1 | 0.25 | 0.25 | 0 | 0 |
| IA289 | 0.5 | 0 | 1 | 0.5 | 0.5 | 0.75 | 0 | 1 | 0.75 | 0.75 | 0 | 0 |
| IA290 | 0.5 | 0 | 0.5 | 0.25 | 0 | 0.25 | 0 | 0 | 0 | 0.5 | 0 | 0 |
| IA291 | 0 | 0 | 1 | 0.25 | 0 | 0.25 | 0.5 | 0 | 0 | 0.25 | 0 | 0 |
| IA292 | 0 | 0 | 1 | 0.25 | 0 | 0.5 | 0 | 0.5 | 0 | 0.5 | 0 | 0 |
| IA293 | 0 | 0 | 0.5 | 0.75 | 0.25 | 0.75 | 0 | 1 | 0.75 | 0 | 0 | 0 |
| IA294 | 0 | 0 | 1 | 0 | 0 | 0.5 | 0 | 1 | 0.25 | 0.5 | 0 | 0 |
| IA295 | 0 | 0 | 0.75 | 0.75 | 0 | 1 | 0 | 1 | 0.75 | 0 | 0 | 0 |
| IA296 | 0 | 0 | 0.25 | 0.5 | 0 | 0.75 | 0.5 | 0.5 | 0.5 | 0.25 | 0 | 0 |
| IA297 | 0 | 0 | 0.75 | 0.5 | 0.5 | 0.5 | 0.25 | 1 | 0.5 | 0.5 | 0 | 0 |
| IA298 | 0 | 0 | 0.5 | 0 | 0 | 0.25 | 1 | 0 | 0 | 0.25 | 0 | 0 |
| IA299 | 0 | 0.75 | 1 | 1 | 0.75 | 0.75 | 0 | 1 | 1 | 0.75 | 0 | 0 |
| IA300 | 0 | 0 | 0.5 | 0.25 | 0 | 0.75 | 0 | 1 | 0 | 0.5 | 0 | 0 |
| IA301 | 0 | 0 | 1 | 0.25 | 0 | 0.25 | 0 | 1 | 1 | 0.75 | 0 | 0 |
| IA302 | 0 | 0 | 1 | 0.5 | 0.5 | 0.75 | 0 | 1 | 1 | 0.75 | 0 | 0 |
| IA303 | 0 | 0 | 1 | 1 | 0.5 | 1 | 0 | 1 | 1 | 0.75 | 0 | 0 |
| IA304 | 0 | 0 | 1 | 0.5 | 0.5 | 1 | 0 | 1 | 0.75 | 0.25 | 0 | 0 |
| IA305 | 0 | 0 | 1 | 0.25 | 0.5 | 1 | 0 | 1 | 0.25 | 0.25 | 0 | 0 |

Continued on next page

**Table A4 – continued from previous page**

| | | | | | | | | | | | | |
|---|---|---|---|---|---|---|---|---|---|---|---|---|
| IA306 | 0 | 0 | 1 | 0 | 0.75 | 0.5 | 0 | 1 | 0.5 | 0.5 | 0 | 0 |
| IA307 | 0.25 | 0.25 | 0.75 | 0.5 | 0.25 | 1 | 1 | 1 | 0.75 | 0 | 0 | 0 |
| IA308 | 0.25 | 0 | 1 | 0.75 | 0.5 | 1 | 0 | 1 | 1 | 0 | 0 | 0 |
| IA309 | 0.25 | 0.5 | 1 | 0.75 | 0.5 | 0.5 | 0.5 | 1 | 0.75 | 0 | 0 | 0.25 |
| IA310 | 0.25 | 0.25 | 0.5 | 1 | 0.25 | 1 | 0.5 | 1 | 1 | 0.5 | 0 | 0 |
| IA311 | 0 | 0 | 1 | 0.75 | 1 | 0.75 | 0 | 1 | 0.5 | 0 | 0 | 0 |
| IA312 | 0 | 0 | 0.75 | 0.25 | 0 | 0.5 | 0 | 1 | 0.25 | 0 | 0 | 0 |
| IA313 | 0 | 0.25 | 1 | 0.5 | 0.5 | 0.5 | 0.5 | 1 | 0.5 | 0.75 | 0 | 0 |
| IA314 | 0 | 0 | 1 | 0 | 0.5 | 0.75 | 0.25 | 1 | 0 | 0 | 0 | 0 |
| IA315 | 0.25 | 0 | 0.75 | 0 | 0 | 1 | 0 | 0 | 0 | 0 | 0.25 | 0.25 |
| IA316 | 1 | 1 | 0.75 | 1 | 0.5 | 0.75 | 0.25 | 0 | 1 | 0.25 | 0 | 1 |
| IA317 | 1 | 1 | 0.75 | 0.25 | 0.75 | 0.5 | 0 | 0 | 0.75 | 0.5 | 0 | 1 |
| IA318 | 0 | 0 | 1 | 0.75 | 0.5 | 0.25 | 0 | 1 | 0.75 | 0.75 | 0 | 0 |
| IA319 | 0 | 0 | 1 | 0 | 0.5 | 0.75 | 0 | 1 | 0.25 | 0.5 | 0 | 0 |
| IA320 | 0.25 | 0 | 0.5 | 0.5 | 0.25 | 1 | 0 | 1 | 0.5 | 0.75 | 0 | 0 |
| IA321 | 0.25 | 0 | 0.5 | 0.75 | 0 | 0.75 | 0 | 1 | 0.75 | 0.5 | 0 | 0 |
| Avg. | 0.25 | 0.22 | 0.79 | 0.61 | 0.27 | 0.67 | 0.08 | 0.72 | 0.59 | 0.37 | 0.11 | 0.22 |

Table A5: Mapping Between Benefits and Internet Artifacts

| Benefits | Internet Artifacts | Count |
|---|---|---|
| Deployment of Software & Services | IA1, IA2, IA3, IA4, IA5, IA9, IA10, IA11, IA12, IA13, IA14, IA15, IA16, IA22, IA23, IA24, IA27, IA28, IA30, IA33, IA35, IA37, IA46, IA47, IA48, IA52, IA53, IA55, IA56, IA57, IA58, IA60, IA61, IA62, IA65, IA68, IA69, IA72, IA76, IA77, IA79, IA81, IA85, IA86, IA87, IA88, IA107, IA111, IA113, IA116, IA117, IA120, IA129, IA130, IA139, IA150, IA152, IA153, IA155, IA159, IA160, IA172, IA173, IA183, IA185, IA189, IA196, IA197, IA201, IA203, IA213, IA214, IA215, IA218, IA219, IA220, IA223, IA224, IA226, IA233, IA234, IA252, IA264, IA266, IA268, IA270, IA272, IA277, IA279, IA282, IA283, IA284, IA285, IA286, IA287, IA288, IA289, IA290, IA291, IA292, IA293, IA294, IA295, IA296, IA297, IA298, IA299, IA300, IA301, IA302, IA303, IA304, IA305, IA306, IA307, IA308, IA309, IA310, IA311, IA312, IA313, IA314, IA315, IA316, IA317, IA318, IA319, IA320, IA321 | 126 |

Continued on next page

Table A5 – continued from previous page

| Benefits | Internet Artifacts | Count |
|---|---|---|
| Ease in Cloudbased Interfacing | IA1, IA2, IA4, IA5, IA6, IA7, IA9, IA10, IA11, IA12, IA13, IA14, IA15, IA18, IA22, IA23, IA27, IA28, IA30, IA33, IA34, IA35, IA36, IA37, IA40, IA42, IA48, IA51, IA52, IA53, IA54, IA57, IA59, IA60, IA61, IA62, IA67, IA72, IA76, IA77, IA79, IA85, IA87, IA88, IA103, IA107, IA109, IA111, IA113, IA114, IA115, IA116, IA117, IA118, IA119, IA120, IA121, IA122, IA123, IA124, IA125, IA126, IA127, IA128, IA129, IA130, IA131, IA140, IA148, IA150, IA152, IA159, IA162, IA173, IA174, IA176, IA185, IA190, IA191, IA196, IA197, IA206, IA213, IA214, IA219, IA220, IA223, IA238, IA246, IA250, IA255, IA260, IA267, IA272, IA277, IA280, IA281, IA282, IA283, IA285, IA286, IA288, IA289, IA290, IA291, IA293, IA295, IA297, IA298, IA299, IA300, IA301, IA304, IA305, IA306, IA307, IA308, IA309, IA310, IA311, IA312, IA313, IA316, IA319, IA320, IA321 | 111 |
| Configuration Management | IA2, IA4, IA5, IA9, IA10, IA11, IA13, IA14, IA22, IA23, IA24, IA30, IA33, IA36, IA37, IA38, IA40, IA42, IA43, IA46, IA50, IA51, IA52, IA55, IA56, IA57, IA59, IA60, IA68, IA69, IA74, IA75, IA77, IA86, IA88, IA143, IA144, IA155, IA160, IA165, IA168, IA171, IA172, IA176, IA180, IA181, IA187, IA207, IA213, IA215, IA220, IA225, IA239, IA246, IA253, IA260, IA261, IA268, IA270, IA277, IA278, IA279, IA280, IA283, IA284, IA286, IA287, IA290, IA291, IA292, IA293, IA295, IA296, IA298, IA300, IA302, IA303, IA304, IA306, IA308, IA310, IA314, IA315, IA317 | 86 |
| Community Support | IA1, IA2, IA9, IA10, IA13, IA14, IA20, IA22, IA23, IA24, IA27, IA28, IA29, IA30, IA34, IA37, IA41, IA46, IA47, IA48, IA51, IA52, IA55, IA56, IA57, IA58, IA59, IA60, IA61, IA67, IA73, IA75, IA76, IA85, IA94, IA127, IA134, IA135, IA141, IA159, IA172, IA179, IA201, IA206, IA207, IA213, IA214, IA219, IA229, IA233, IA251, IA265, IA268, IA270, IA277, IA279, IA283, IA284, IA287, IA288, IA289, IA290, IA292, IA293, IA294, IA295, IA296, IA297, IA298, IA299, IA301, IA303, IA308, IA309, IA316, IA318, IA320, IA321 | 75 |
| SLO-based Scalability | IA1, IA2, IA6, IA9, IA11, IA12, IA14, IA22, IA23, IA24, IA27, IA28, IA30, IA37, IA45, IA47, IA49, IA50, IA51, IA52, IA55, IA57, IA59, IA60, IA61, IA65, IA68, IA77, IA85, IA88, IA111, IA116, IA121, IA129, IA130, IA139, IA151, IA152, IA154, IA155, IA159, IA160, IA164, IA165, IA172, IA173, IA186, IA187, IA197, IA214, IA220, IA223, IA225, IA226, IA239, IA240, IA247, IA251, IA252, IA253, IA264, IA268, IA270, IA277, IA281, IA283, IA286, IA290, IA291, IA292, IA296, IA300, IA302, IA303, IA306, IA307, IA309, IA311, IA313, IA315, IA316, IA319, IA320 | 75 |
| Resource Limit Specification | IA1, IA2, IA4, IA6, IA9, IA10, IA11, IA12, IA13, IA14, IA15, IA22, IA23, IA27, IA29, IA33, IA37, IA42, IA45, IA49, IA51, IA52, IA54, IA55, IA57, IA60, IA79, IA85, IA88, IA94, IA103, IA113, IA121, IA122, IA127, IA131, IA140, IA152, IA155, IA159, IA168, IA173, IA182, IA197, IA213, IA215, IA223, IA226, IA233, IA239, IA264, IA265, IA266, IA268, IA270, IA282, IA283, IA285, IA292, IA293, IA297, IA300, IA304, IA307, IA309, IA310, IA311, IA313, IA315, IA316, IA318, IA320 | 72 |
| Availability of Software | IA1, IA2, IA6, IA9, IA11, IA12, IA13, IA14, IA16, IA22, IA23, IA24, IA27, IA30, IA33, IA37, IA42, IA51, IA55, IA57, IA58, IA60, IA79, IA102, IA121, IA130, IA150, IA155, IA165, IA182, IA201, IA251, IA253, IA265, IA272, IA281, IA283, IA285, IA286, IA291, IA293, IA298, IA302, IA303, IA306, IA307, IA309, IA311, IA313, IA316 | 53 |
| Self-healing Containers & Pods | IA9, IA11, IA27, IA33, IA37, IA57, IA58, IA60, IA77, IA130, IA159, IA198, IA213, IA214, IA253, IA264, IA265, IA268, IA270, IA280, IA283, IA284, IA292, IA298, IA300, IA311 | 26 |

Continued on next page

Table A6: Mapping Between Challenges and Internet Artifacts

| Challenges | Internet Artifacts | Count |
|---|---|---|
| Lack of Security Practices & Tools | IA1, IA2, IA3, IA4, IA5, IA7, IA8, IA10, IA11, IA12, IA13, IA16, IA17, IA18, IA19, IA20, IA21, IA23, IA27, IA30, IA33, IA36, IA39, IA40, IA41, IA43, IA48, IA50, IA51, IA52, IA55, IA56, IA59, IA62, IA63, IA64, IA65, IA67, IA70, IA71, IA72, IA73, IA74, IA75, IA77, IA82, IA87, IA89, IA93, IA95, IA96, IA97, IA99, IA101, IA105, IA112, IA114, IA115, IA125, IA126, IA127, IA128, IA133, IA138, IA142, IA143, IA144, IA145, IA146, IA147, IA158, IA160, IA164, IA165, IA166, IA170, IA171, IA173, IA174, IA175, IA176, IA177, IA180, IA181, IA195, IA198, IA199, IA200, IA203, IA204, IA213, IA214, IA216, IA217, IA222, IA224, IA228, IA230, IA231, IA234, IA235, IA237, IA240, IA241, IA244, IA248, IA249, IA253, IA257, IA258, IA263, IA269, IA271, IA272, IA279, IA284, IA289, IA290, IA295, IA298, IA301, IA302, IA303, IA306, IA307, IA310, IA313, IA314, IA317, IA318 | 132 |
| Attack Surface Reduction | IA1, IA3, IA4, IA5, IA8, IA10, IA11, IA13, IA16, IA17, IA18, IA19, IA20, IA21, IA31, IA38, IA40, IA52,IA54, IA55, IA56, IA57, IA59, IA62, IA64, IA66, IA67, IA70, IA71, IA72, IA73, IA75, IA76, IA77, IA83, IA84, IA86, IA89, IA91, IA92, IA93, IA99, IA101, IA105, IA109, IA110, IA115, IA124, IA125, IA127, IA128, IA133, IA134, IA135, IA136, IA137, IA138, IA141, IA143, IA144, IA145, IA146, IA147, IA149, IA150, IA154, IA158, IA160, IA164, IA165, IA170, IA174, IA177, IA179, IA180, IA181, IA187, IA195, IA198, IA199, IA200, IA203, IA204, IA207, IA214, IA216, IA224, IA227, IA228, IA230, IA234, IA235, IA236, IA240, IA241, IA243, IA244, IA245, IA248, IA254, IA257, IA258, IA259, IA263, IA267, IA269, IA271, IA279, IA281, IA284, IA292, IA294, IA302, IA310, IA314, IA317, IA318, IA319 | 126 |

Continued on next page

Table A6 – continued from previous page

| Challenges | Internet Artifacts | Count |
|---|---|---|
| Lack of Diag-nostics Tools | IA1, IA2, IA5, IA8, IA9, IA10, IA11, IA12, IA13, IA14, IA15, IA18, IA20, IA21, IA22, IA23, IA30, IA33, IA37, IA43,IA45, IA50, IA51, IA52, IA53, IA54, IA55, IA57, IA59, IA62, IA63, IA65, IA66, IA67, IA68, IA70, IA75, IA76, IA77, IA81, IA82, IA84, IA86, IA91, IA109, IA115, IA119, IA120, IA121, IA127, IA130, IA142, IA143, IA152, IA155, IA157, IA158, IA161, IA162, IA170, IA174, IA177, IA178, IA181, IA189, IA190, IA192, IA193, IA198, IA199, IA203, IA204, IA208, IA209, IA211, IA213, IA214, IA215, IA216, IA218, IA222, IA228, IA231, IA232, IA233, IA237, IA242, IA251, IA256, IA257, IA269, IA271, IA275, IA282, IA283, IA284, IA285, IA286, IA288, IA290, IA291, IA292, IA293, IA295, IA297, IA298, IA300, IA301, IA302, IA303, IA304, IA306, IA307, IA308, IA309, IA310, IA311, IA312, IA313, IA314, IA317 | 121 |
| Maintenancerelated Challenges | IA1, IA2, IA5, IA10, IA11, IA14, IA18, IA20, IA21, IA22, IA23, IA24, IA26, IA27, IA46, IA51, IA52, IA53, IA54, IA57, IA58, IA61, IA62, IA63, IA64, IA65, IA66, IA69, IA70, IA71, IA72, IA73, IA74, IA77, IA81, IA82, IA83, IA84, IA86, IA89, IA100, IA102, IA106, IA108, IA113, IA114, IA115, IA119, IA120, IA121, IA124, IA125, IA131, IA138, IA140, IA142, IA150, IA154, IA156, IA158, IA159, IA160, IA166, IA170, IA171, IA175, IA179, IA182, IA183, IA192, IA198, IA199, IA214, IA217, IA218, IA222, IA233, IA235, IA236, IA238, IA243, IA245, IA249, IA254, IA255, IA261, IA265, IA269, IA273, IA281, IA282, IA283, IA286, IA288, IA290, IA291, IA292, IA294, IA295, IA297, IA298, IA300, IA301, IA302, IA306, IA308, IA312, IA315, IA317 | 109 |
| Learning Curve | IA1, IA2, IA3, IA10, IA15, IA17, IA18, IA19, IA20, IA22, IA25, IA26, IA27, IA29, IA30, IA48, IA51, IA53, I54, IA55, IA56, IA57, IA58, IA59, IA60, IA61, IA62, IA63, IA64, IA65, IA66, IA70, IA71, IA72, IA75, IA83, IA86, IA94, IA119, IA120, IA122, IA127, IA131, IA134, IA137, IA139, IA140, IA141, IA147, IA159, IA164, IA166, IA174, IA178, IA182, IA203, IA205, IA207, IA208, IA209, IA210, IA214, IA221, IA222, IA225, IA231, IA234, IA236, IA241, IA253, IA257, IA260, IA265, IA268, IA273, IA279, IA283, IA286, IA288, IA290, IA292, IA293, IA294, IA295, IA297, IA301, IA306, IA307, IA309, IA311, IA312, IA313, IA314, IA317, IA318, IA320 | 91 |

Continued on next page

Table A6 – continued from previous page

| Challenges | Internet Artifacts | Count |
|---|---|---|
| Networking | IA1, IA2, IA4, IA5, IA7, IA8, IA10, IA11, IA12, IA13, IA15, IA19, IA29, IA30, IA31, IA33, IA36, IA40, IA42, IA43, IA44, IA49, IA51, IA52, IA55, IA56, IA62, IA63, IA66, IA70, IA72, IA73, IA74, IA75, IA76, IA77, IA78, IA79, IA91, IA99, IA128, IA130, IA139, IA149, IA154, IA164, IA173, IA177, IA181, IA182, IA183, IA184, IA194, IA198, IA214, IA222, IA245, IA251, IA253, IA257, IA260, IA265, IA267, IA272, IA273, IA279, IA280, IA281, IA283, IA284, IA285, IA287, IA289, IA290, IA292, IA293, IA301, IA302, IA303, IA306, IA307, IA308, IA309, IA314, IA316, IA317, IA318, IA321 | 84 |
| Migration Cost | IA2, IA6, IA10, IA11, IA13, IA14, IA16, IA18, IA21, IA22, IA23, IA24, IA25, IA26, IA30, IA48, IA51, IA52, IA53, IA55, IA56, IA57, IA58, IA64, IA66, IA70, IA75, IA79, IA82, IA86, IA91, IA92, IA94, IA100, IA106, IA107, IA108, IA109, IA110, IA111, IA112, IA113, IA114, IA124, IA126, IA131, IA137, IA140, IA141, IA150, IA151, IA159, IA165, IA166, IA167, IA182, IA183, IA194, IA195, IA205, IA210, IA214, IA233, IA236, IA239, IA245, IA253, IA259, IA260, IA265, IA268, IA272, IA273, IA282, IA284, IA286, IA287, IA288, IA293, IA294, IA297, IA303, IA304, IA305, IA306, IA312, IA313, IA319, IA320 | 82 |
| Storage | IA1, IA2, IA3, IA5, IA6, IA8, IA10, IA11, IA12, IA13, IA30, IA33, IA37, IA38, IA52, IA53, IA54, IA55, IA56, IA57, IA69, IA76, IA78, IA79, IA82, IA83, IA91, IA130, IA131, IA139, IA163, IA171, IA237, IA253, IA272, IA276, IA278, IA280, IA285, IA286, IA287, IA289, IA290, IA291, IA292, IA293, IA297, IA302, IA303, IA306, IA308, IA309, IA310, IA314 | 52 |
| System Environment Configurations | IA4, IA6, IA7, IA9, IA10, IA11, IA12, IA13, IA23, IA30, IA33, IA37, IA38, IA40, IA48, IA52, IA55, IA56, IA57, IA58, IA63, IA64, IA66, IA68, IA72, IA75, IA78, IA81, IA86, IA89, IA98, IA121, IA130, IA156, IA201, IA281, IA283, IA292, IA294, IA295, IA303, IA306, IA314, IA317, IA319 | 46 |
| Testing | IA1, IA2, IA3, IA10, IA11, IA13, IA16, IA20, IA21, IA24, IA26, I53, IA55, IA57, IA63, IA66, IA71, IA72, IA82, IA83, IA127, IA128, IA142, IA199, IA216, IA248, IA273, IA279, IA283, IA285, IA294, IA307, IA310, IA314, IA317, IA318 | 36 |
| Failure Troubleshooting | IA8, IA13, IA18, IA19, IA20, IA26, IA29, IA30, IA38, IA44,IA3, I54, IA63, IA81, IA84, IA86, IA102, IA113, IA167, IA169, IA215, IA216, IA218, IA222, IA231, IA242, IA255, IA272, IA301, IA307, IA308, IA312 | 32 |
| Performance | IA1, IA2, IA4, IA8, IA13, IA16, IA26, IA30, IA31, IA33, IA37, IA45, IA54, IA57, IA58, IA63, IA66, IA100, IA123, IA151, IA152, IA153, IA218, IA222, IA249, IA256, IA262, IA279, IA281, IA282, IA285, IA290, IA295, IA304, IA306, IA307, IA308, IA309, IA313, IA321 | 26 |
| Cultural Change | IA2, IA19, IA27, IA29, IA58, IA61, IA72, IA84, IA110, IA125, IA128, IA132, IA140, IA164, IA209, IA234, IA253, IA257, IA281, IA282, IA285, IA288, IA320 | 23 |

Continued on next page

Table A6 – continued from previous page

| **Challenges** | **Internet Artifacts** | **Count** |
|---|---|---|
| Idempotency | IA6, IA17, IA19, IA20, IA23, IA30, IA77, IA119, IA120, IA164, IA192, IA203, IA230, IA288, IA306, IA321 | 16 |
| Hardware Compatibility | IA8, IA15, IA20, IA274 | 4 |

## 8 Declarations

**Funding**: The research was partially funded by the U.S. National Science Foundation (NSF) award # 2026869.

**Conflict of Interests/Competing Interests**: The third author is an employee at IBM. None of the authors have financial or proprietary interests in any material discussed in this article.